\documentclass[a4paper,12pt,oneside]{article}
\usepackage[latin1]{inputenc}
\usepackage{amsmath}
\usepackage{amsfonts}
\usepackage{amssymb}
\usepackage{graphicx}
\usepackage{footnote}
\usepackage{float}
\usepackage{subcaption}
\DeclareGraphicsExtensions{.bmp,.png,.pdf,.jpg,.eps}
\usepackage{physics}
\usepackage{mathrsfs}

\usepackage[colorlinks]{hyperref}
\hypersetup{
    colorlinks=true,
    linkcolor=blue,
    filecolor=magenta,  
    urlcolor=[rgb]{0.00,0.00,0.50},    
    citecolor=red,
    }

\usepackage{url}

\advance \textheight by \topskip
\usepackage{amsmath}
\usepackage[english]{babel}
\makeatletter
 \@addtoreset{equation}{section}
\makeatother
\usepackage[latin1]{inputenc}
\usepackage{amsmath}
\numberwithin{equation}{section}

\advance \textheight by \topskip
\usepackage{amsmath}
\usepackage[english]{babel}
\DeclareMathAlphabet\mathbfcal{OMS}{cmsy}{b}{n}
\DeclareMathAlphabet{\boldmathe}{T1}{cmr}{bx}{it}
\newcommand{\mbf}[1]{\boldmathe{#1}}
\newcommand{\mbfgr}[1]{\textit{\mbox{\boldmath$#1$}}}

\def\vA{\mbf{A}}
\def\vB{\mbf{B}}

\def\vV{\mbf{V}}
\def\vG{\mbf{G}}

\def\vJ{\mbf{J}}

\def\va{\mbf{a}}

\def\vx{\mbf{x}}

\def\vn{\mbf{n}}
\def\vp{\mbf{p}}

\def\vr{\mbf{r}}

\def\vpi{\mbfgr{\pi}}

\def\be{\begin{equation}}
\def\ee{\end{equation}}

\def\A{\mathbb A}

\def\R{\mathbb R}

\def\C{\mathbb C}

\def\be{\begin{equation}}
\def\ee{\end{equation}}

\def\A{\mathbb A}

\def\R{\mathbb R}

\def\C{\mathbb C}

\def\vA{\mbf{A}}
\def\vB{\mbf{B}}

\def\vV{\mbf{V}}

\def\vG{\mbf{G}}

\def\vJ{\mbf{J}}

\def\va{\mbf{a}}

\def\vn{\mbf{n}}
\def\vp{\mbf{p}}
\def\vr{\mbf{r}}

\def\vpi{\mbfgr{\pi}}

\def\be{\begin{equation}}
\def\ee{\end{equation}}

\def\A{\mathbb A}

\def\Z{\mathbb Z}

\def\R{\mathbb R}

\def\C{\mathbb C}

\usepackage{todonotes}

\begin{document}
%%%%%%%%%%%%%%%%%%%%%%%%%%%%%

\title{
{\bf Conformal bridge transformation\\ and\\
 $\mathcal{PT}$ symmetry}}

\author{{\bf  Luis Inzunza and Mikhail S. Plyushchay} 
 \\
[8pt]
{\small \textit{Departamento de F\'{\i}sica,
Universidad de Santiago de Chile, Casilla 307, Santiago,
Chile  }}\\
[4pt]
 \sl{\small{E-mails:   
\textcolor{blue}{luis.inzunza@usach.cl},
\textcolor{blue}{mikhail.plyushchay@usach.cl}
}}
}
\date{}
\maketitle

\begin{abstract}The conformal bridge transformation (CBT) is reviewed in the light of  
the $ \mathcal{PT}$ symmetry. Originally, the CBT was presented as a 
non-unitary transformation (a complex canonical transformation in the classical case) 
that relates two different forms of dynamics in the sense of Dirac.
Namely, it maps the asymptotically free form into  the  harmonically confined 
form of dynamics associated with the $\mathfrak{so}(2,1) \cong \mathfrak{sl}(2, \R) $ conformal symmetry. 
However, as the transformation relates the non-Hermitian operator $ i \hat{D} $, where $ \hat{D}$ 
is the generator of dilations, with the compact Hermitian generator 
$ \hat{\mathcal{J}}_0 $
of the $ \mathfrak{sl}(2,\R) $ algebra,   
 the  CBT generator can be associated with  a 
 $\mathcal{PT}$-symmetric metric. In this work we review the applications of this transformation 
for one- and two-dimensional systems, as well as for systems on a 
cosmic string background, and for a conformally extended
charged particle in the field of  Dirac monopole. 
We also compare and unify the CBT with
the Darboux transformation.
The latter  is used to construct 
$\mathcal{PT}$-symmetric solutions of the  equations of the KdV hierarchy
with the properties of extreme waves.
As a new result, 
by using a modified CBT we relate the 
  one-dimensional 
$ \mathcal{PT}$-regularized asymptotically free conformal mechanics model with 
the $ \mathcal{PT}$-regularized version of the de Alfaro, Fubini and Furlan system. 

\vskip.5cm\noindent
\end{abstract}

\section{Introduction}
The very fact that the properties of various  complex systems can be 
related with the properties of a free particle and obtained  
from it  in elegant ways is just amazing.
A good example of this is the connection between the  free particle and the KdV hierarchy, 
based on the covariance of the Lax representation with respect to 
Darboux transformations \cite{MatSal}. 
The stationary Schr\"odinger equation for a one-dimensional free particle 
enters the game when  the operators of the auxiliary spectral problem in the  
Lax pair representation  are taken with a zero  potential  identified as a trivial  
solution of the KdV equation. 
 Then, the iterative application of the  Darboux transformation 
 to the  linear equations  associated with the Lax representation 
 combined with the Darboux dressing of the Lax operator 
allows ones
 to generate multi-soliton solutions of the equations of the KdV hierarchy.
A Schr\"odinger  system of the auxiliary spectral problem  with the obtained 
multi-soliton potential is reflectionless being  almost isospectral to the  
free particle, and its states are generated from the eigenstates of the free 
particle Hamiltonian operator by the Darboux transformation 
\cite{AMGP,AranPly}.
At least  some  of reflectionless 
systems  are  converted  by  periodization  into the finite-gap  quantum  
systems \cite{DunneRao,CorPlyJA,CJNP},  
 and their  potentials  can be promoted to the cnoidal type solutions  
 by using, again, the Darboux covariance of the Lax representation
  \cite{AranPly}.
 One can  introduce soliton defects propagating in  a crystalline  background 
 with the help of the  same Darboux transformations
 \cite{AranPly}. 
 By the  Miura transformation, intimately related to
 supersymmetry, one also can relate the free particle system with the 
 modified KdV equation \cite{AGM1}. 
 Using the same methods, one can construct 
 $\mathcal{PT}$-regularized Calogero type quantum models
 with the exotic properties, whose potentials can be transformed 
into  complex  $\mathcal{PT}$-symmetric 
solutions of the equations of  the KdV  hierarchy
 \cite{JuanMP1,JuanMP2}.

On the other hand, 
an  important  characteristic of the free particle
in arbitrary case of $d$-dimensional Euclidean space 
 is that this non-relativistic system is described   by the 
 $\mathfrak{so}(2,1)$ conformal symmetry. 
The 
 non-relativistic conformal symmetry 
 occurs naturally in a wide variety of physical phenomena, 
 and attracted recently a lot of attention in 
the context of  non-relativistic AdS/CFT correspondence
\cite{LeiPly,BarFue,DuvHasHor,Jack},
black hole physics
 \cite{ConformalBH1,ConformalBH2,ConformalBH3},
 cosmology \cite{DGH,ConformalBH0, PioWal,CGGH},
 AdS/CDM correspondence \cite{GAdS1,GAdS2,App2,App3} and  
 QCD confinement \cite{App1,Brod2}, to name a few.
 The  non-relativistic conformal symmetry of the free particle 
 and its generalizations  lie in the base  of  the so-called conformal bridge transformation (CBT) 
\cite{InzPlyWipf1,InzPlyWipf2,InzPly7,InzPly8} by which 
 the dynamics and symmetries 
 of  an asymptomatically free  conformally invariant  system  can be mapped into  
 those  of the  associated in a certain way  harmonically trapped system. 
 This corresponds to the picture described  in  the Dirac seminal article   
 \cite{Dirac}, where different forms of dynamics are studied by choosing, in the general case, 
 a linear combination of the  
 generators of a given symmetry  as the Hamiltonian of the system. 
 In the original work \cite{InzPlyWipf1}, 
 it is shown that in one dimension  the CBT  relates 
 the  free particle  and the two-body Calogero model 
 with, respectively,  the harmonic oscillator 
 and  the conformal 
 mechanical model of de Alfaro, Fubini and Furlan (AFF) \cite{AFF}.
 In two dimensions,  the free particle system can be  related 
 with a variety of systems such as the isotropic and anisotropic harmonic oscillators, 
 the Landau 
 problem, and the exotic family of rotationally invariant harmonic oscillators \cite{InzPlyWipf1,InzPly8}. 
 As the transformation is based on the  algebraic arguments, 
 it can be applied to  systems in any conformally-invariant space-time 
 and gauge backgrounds.
 In this way, it was employed to study the dynamics and hidden symmetries 
 in backgrounds of the Dirac monopole \cite{InzPlyWipf2}  and cosmic string
 \cite{InzPly7}. One of the goals  of this article  is to 
 review how this transformation works and the scope of its applications. 

In comparison with the Darboux transformation, which relates almost isospectral one-dimensional 
systems by means of   finite order differential 
operators, the CBT is generated by a non-local operator,  whose realization
is not restricted by a space dimension, 
and it relates the systems with essentially different  spectra   in a non-trivial way. 
In fact, it is expected that the possibilities to  connect  the systems by the CBT
expand  with increasing the number of 
 dimensions and with the  conformally invariant change of the space-time
 metric.  Additionally, 
some hints on a possible  close relationship of the CBT with 
$\mathcal{PT}$-symmetric systems 
\cite{Bender,BBJo,Mostafazadeh,Bender2007} 
were indicated in the original works \cite{InzPlyWipf1,InzPlyWipf2,InzPly7,InzPly8}. 
They are based on the  fact that at the quantum level, 
the transformation is realized by a non-unitary operator 
$\hat{\mathfrak{S}}$ that transforms a non-Hermitian operator 
$i\hat{D}$, where $\hat{D}$ is a generator of dilations,   
into the Hermitian  compact generator of  the $\mathfrak{sl}(2,\R)$ symmetry
which has  a real discrete spectrum. 
In this work, we show that the connection between the one-dimensional 
free particle and the harmonic oscillator corresponds to a 
particular example of the $\mathcal{PT}$-symmetric Swanson models 
studied  in \cite{PTmetric1,MFFr,PTmetric2,PEGAAF}. 
In this way our Hermitian  generator  $\hat{\mathfrak{S}}$  of the CBT
can be related to a $\mathcal{PT}$-symmetric metric operator. 
Furthermore, since the applications of our CBT touch the systems in
the spaces $\R^d$  with  $d\geq 1$,  and the models in different 
geometric and gauge 
backgrounds,  new possibilities are opened for  connecting 
$\mathcal{PT}$-symmetric systems with models that reveal   interesting 
physical properties such as quantum anomalies, Bose-Einstein condensation, 
gauge symmetries, etc. 

The paper  is organized as follows.  In Sec. \ref{SecCBT} we present the basic 
properties of the CBT at the classical and quantum levels,
and  establish its connection with  $\mathcal{PT}$-symmetric systems. 
In Sec. \ref{SecAFF1d} we consider  the explicit applications of the CBT 
to one-dimensional systems. 
As a new result we present the connection between a one-parametric family of
the $\mathcal{PT}$-regularized perfectly invisible zero-gap Calogero type systems  
 with a $\mathcal{PT}$-symmetric version of the AFF conformal mechanics.
  We use the relation of the former family with  
  the free-particle  by means of Darboux transformations
  based on a scale-invariant higher-order differential equation
   to build the complete set of the spectrum-generating ladder operators for the latter 
   system. We also consider there  a reinterpretation of the  CBT from the point of view of the 
   $\mathcal{PT}$-symmetric Swanson model. 
In Sec. \ref{SecCBT2d} we consider the isotropic and anisotropic CBT in 
$d$-dimensions, as well as the generation 
of the exotic rotationally  invariant harmonic oscillator in two dimensions. In Sec. \ref{SecCBT3d}
we study  the application of the CBT for systems in two different backgrounds which correspond to 
the cosmic string and Dirac monopole. 
In Sec. \ref{SecDar} we show how the Darboux transformation applied 
appropriately 
to the $\mathcal{PT}$-regularized 
Calogero type systems allows us to produce 
complex $\mathcal{PT}$-symmetric 
solutions of the equations of the KdV hierarchy which reveal the properties 
typical for extreme waves.
In Sec. \ref{SecDis} we  conclude with discussion of 
 some  interesting open problems and 
 further generalizations of the CBT in the light of the 
 $\mathcal{PT}$ symmetry. 

\section{Conformal bridge transformation}
\label{SecCBT}

\subsection{Classical case}
\label{SecClassicalCBT}
Consider the classical $\mathfrak{so}(2,1)$ algebra 
\begin{eqnarray}\label{so21alg}
&\{D_0,H_0\}=H_0\,,\qquad\{D_0,K_0\}=-K_0\,,
\qquad \{K_0,H_0\}=2D_0\,,&
\end{eqnarray}
without specifying  the concrete  form of the generators.
Identifying  $H_0$ as a Hamiltonian of a particular classical system, one sees that
$D_0$ and $K_0$ cannot be true, not depending explicitly on time,   integrals 
of motion. They, however, can easily be promoted to the 
dynamical, explicitly depending on time integrals of motion 
in the sense of the evolution equation  $\dot{A}=\{A,H\}+\frac{\partial A}{\partial t}=0$,
\begin{eqnarray}\label{KD(t)}
&K_0\, \rightarrow \, K=K(t)=T_{H_0}(t)(K_0)\,,\qquad D_0\, \rightarrow \,
D=D(t)=T_{H_0}(t)(D_0)\,.&
\end{eqnarray}
Here, $T_{H_0}(t)$ indicates the Hamiltonian flux in a  phase space,
  \begin{eqnarray}&\label{TransCan}
\exp(\gamma F)\star f(q,p):=f(q,p)+\sum_{n=1}^\infty 
\frac{\gamma^n}{n!}\{F,\{\ldots,\{F,f\underbrace{\}\ldots\}\}}_{n}:=T_F(\gamma)(f)\,,&
\end{eqnarray}
that is  a canonical transformation.
Obviously, $T_{H_0}(t)(H_0)=H_0$, and $H_0$, $D$ and $K$ 
satisfy the same $\mathfrak{so}(2,1)$ algebra, in which
$D$ and $K$ are identified as  generators of dilations and 
special conformal transformations, respectively.

The real and complex linear combinations
  \begin{eqnarray}
\label{sl2Rgen}
&\mathcal{J}_0=\frac{1}{2}(\omega^{-1}H_0+\omega K_0)\,,\qquad
\mathcal{J}_\pm=\mathcal{J}_1\pm i\mathcal{J}_2=-
\frac{1}{2}\left( \omega^{-1}H_0- \omega K_0 \pm 2i D_0\right)\,,&
\end{eqnarray}
satisfy the classical $\mathfrak{sl}(2,\R) $ algebra, 
  \begin{eqnarray}
&\{\mathcal{J}_0,\mathcal{J}_\pm\}=\mp i \mathcal{J}_\pm\,,\qquad
\{\mathcal{J}_-,\mathcal{J}_+\}=-2i\mathcal{J}_{0} \,.&
\end{eqnarray}
A constant $ \omega $ of dimension of frequency 
is introduced   to compensate the dimensions of the generators $H_0$ and $K_0$. 
$H_0$ has a nature of a non-compact  (parabolic) generator of the 
$\mathfrak{so}(2,1)$ algebra \cite{Barg,SL2R}, and so, the corresponding classical system 
can be asymptotically 
free  (like, e.g.  a  free particle, or  conformal mechanics model). 
 $\mathcal{J}_0$ is a compact (elliptic) generator of  $\mathfrak{sl}(2,\R)$, 
and  represents a  harmonically trapped (confined)  version
of the system $H_0$. 
The  quantity  $2\omega \mathcal{J}_0$ can be considered 
as a Hamiltonian of such a system, which corresponds to another 
form of dynamics with respect to the same conformal 
symmetry  $\mathfrak{so}(2,1)\cong\mathfrak{sl}(2,\R)$. 
Then $\mathcal{J}_\pm$ can be promoted to the dynamical 
integrals of motion for the harmonically trapped system
by the analog of the  transformation given by Eqs.  (\ref{KD(t)}) and (\ref {TransCan})
with $H_0$ and $t$ changed for $2\omega \mathcal{J}_0$ and $\tau$.
The quantities $2\omega \mathcal{J}_0$ and 
$\mathcal{J}_\pm$  
generate  the Newton-Hooke symmetry of the harmonically trapped system
\cite{NH1,NH2,NH3,Papan,NH4}. 

Consider now  the transformation 
\begin{eqnarray}
&\label{GenConBr+}
\mathfrak{S}:\,(H_0,D,K)\,\rightarrow \, (-\omega \mathcal{J}_-,-i\mathcal{J}_0,
{\omega}^{-1}
{\mathcal{J}}_+)\,,&
\end{eqnarray}
where we assume that $\mathcal{J}_\pm$ are the dynamical integrals of motion
with respect to the evolution generated by $2\omega \mathcal{J}_0$.
It is an internal automorphism of the conformal algebra 
$\mathfrak{so}(2,1)\cong\mathfrak{sl}(2,\R)$
generated by the composition of the canonical transformations 
\begin{eqnarray}
\label{ClasicalBrige}
&\mathscr{T}(\tau,\beta,\gamma,\delta,t)= T_{2\omega 
\mathcal{J}_0}(\tau)\circ T_{\beta\gamma\delta} \circ T_H(-t)\,,&
\end{eqnarray}
where 
\begin{eqnarray}
\label{Tabg0}
&T_{\beta\gamma\delta }:=T_{\omega K_0}(\beta)\circ 
T_{\omega^{-1}H_0}(\frac{1}{2}\delta) \circ T_{D_0}(\gamma)=
 T_{\omega K_0}(\beta)\circ T_{D_0}(\gamma) \circ T_{\omega^{-1}H_0}(\delta)\,,
&\\&
 \beta=-i \,,\qquad
\gamma=-\ln 2 \,,\qquad  \delta=i\,.
&\label{betadeltagamma}
\end{eqnarray}
In this composition, the first transformation $T_{H}(-t)$ removes
dependence on  $t$  in the dynamical integrals $D$ and $K$.
The second transformation
relates the $t=0$ generators  with the generators 
$\mathcal{J}_0$ 
and $\mathcal{J}_\pm$
of the $\mathfrak{sl}(2,\R)$ algebra,  taken at $\tau=0$.  
The last transformation  $T_{2\omega \mathcal{J}_0}(\tau)$ 
restores the $\tau$ dependence. 
The independent of the  evolution parameters 
transformation $T_{\beta\gamma \delta}$ 
is equivalent to 
\begin{eqnarray}\label{Tequiv}
&T_{\beta\gamma \delta} =T_{G_1}( \varepsilon)\,,\qquad
G_1:=\omega^{-1}H_0-\omega K_0\,,\quad \varepsilon=\frac{\pi}{4}i\,.&
\end{eqnarray}
Since the parameters $\beta$, $\delta$ and $\varepsilon$ are
pure imaginary,  this canonical transformation is of an unusual,
complex form from the point of view of the conventional classical mechanics.
It transforms, particularly,  the  $\mathfrak{so}(2,1)$ 
hyperbolic \cite{Barg,SL2R} generator $D_0$ 
multiplied by $2i\omega$ 
into the compact  real 
 $\mathfrak{sl}(2,\R)$ generator 
$\mathcal{J}_0$ multiplied by $2\omega$. 
This picture 
 corresponds to the change of the form of dynamics in the sense of Dirac
 \cite{Dirac}. 
 Both,  the asymptotically free and the harmonically 
confined, forms of dynamics are  associated to the 
conformal symmetry, and  are related  one to another by the described 
classical conformal bridge transformation \cite{InzPlyWipf1,InzPly7}. 

One can easily check that for the complex $\mathfrak{so}(2,1)$ automorphism (\ref{GenConBr+}) 
the following relation 
\begin{eqnarray}\label{S2}
&\mathfrak{S}^2=\mathfrak{S} \circ \mathfrak{S}\,:\, (\mathcal{J}_0,\mathcal{J}_1,\mathcal{J}_2)\,\rightarrow\,
 (-\mathcal{J}_0,\mathcal{J}_1,-\mathcal{J}_2)&
 \end{eqnarray}
 is valid in terms  of the $\mathfrak{sl}(2,\R)$ generators,
that is a rotation by $\pi$ about $\mathcal{J}_1$.
Therefore, (\ref{GenConBr+}) 
is the fourth order root of the  $\mathfrak{so}(2,1)\cong\mathfrak{sl}(2,\R)$
identity automorphism, $\mathfrak{S}^4=1$.
In terms of the $\mathfrak{so}(2,1)$ generators its eigenelement of eigenvalue
$1$ is
$G_1=\omega^{-1}H_0-\omega K_0$, see Eq.  (\ref{Tequiv}),  
while linear combinations $G_{+i}=-2D_0+\omega^{-1}H_0+\omega K_0$ and 
$G_{-i}=2D_0+\omega^{-1}H_0+\omega K_0$ are eigenelements of eigeinvalues 
$+i$ and $-i$, respectively. 
There is no eigenelement of
the automorphism  (\ref{GenConBr+})  of eigenvalue $-1$ to be 
 linear in the $\mathfrak{so}(2,1)$ generators.
The quadratic eigenelement of eigenvalue $-1$ is 
$a(G_{+i})^2+b(G_{-i})^2$ with arbitrary coefficients $a$ and $b$.
The  $\mathfrak{so}(2,1)$ classical Casimir 
$C=H_0K_0-D_0^2=\frac{1}{4}G_{+i}G_{-i}$
is an eigenelement of automorphism $\mathfrak{S}$ of eigenvalue $1$.

Given a particular system described by some symmetry algebra (that can be of a 
non-linear,  $W$ type), 
in which conformal symmetry 
$\mathfrak {so} (2,1) $ appears as a subalgebra, we can always apply to this system the  
 classical (and quantum, see below) conformal bridge transformation to relate it to 
 its harmonically confined version. As a result, 
 the integrals of one system can be mapped into integrals of another. 
 Explicit examples of this are discussed in 
Secs. \ref{SecCBT2d} and \ref{SecCBT3d}. 
Note that in this picture, the classical dynamics of the harmonically trapped system,
which is generated by the compact Hamiltonian 
$2\omega\mathcal{J}_0$, corresponds to the Hamiltonian flow  generated
by the complex quantity $2i\omega  D_0$  in the asymptotically 
free system described by the Hamiltonian $H_0$.

\subsection{Quantum version}
\label{SecQuantumCBT}

Consider now  the quantum   $\mathfrak{so}(2,1)$ algebra 
 \begin{eqnarray}
&[\hat{D},\hat{H}_0]=i\hbar\hat{H}_0\,,\qquad
[\hat{D},\hat{K}]=-i\hbar\hat{K}\,,\qquad
[\hat{K},\hat{H}_0]=2i\hbar \hat{D}\,.&
\label{so(2,1)}
\end{eqnarray}
Here 
\begin{eqnarray}
&\hat{D}=e^{-i\frac{\hat{H}_0}{\hbar}t}\hat{D}_0e^{i\frac{\hat{H}_0}{\hbar}t}\,,\qquad
\hat{K}=e^{-i\frac{\hat{H}_0}{\hbar}t}\hat{K}_0e^{i\frac{\hat{H}_0}{\hbar}t}\,,&
\end{eqnarray}
and we just note  that acting on a solution of the time-dependent Schr\"odinger equation 
$\Psi(t)=e^{-i\frac{\hat{H}_0}{\hbar}t}\Psi(0)$, a generic 
 dynamical integral operator
$\hat{A}=e^{-i\frac{\hat{H}_0}{\hbar}t}\hat{A}_0e^{i\frac{\hat{H}_0}{\hbar}t}$ produces 
$
\hat{A}(t)\Psi(t)=e^{-i\frac{\hat{H}_0}{\hbar}t}\hat{A}_{0}\Psi(0)\,.
$
In this work we  consider only stationary eigenstates and linear combinations of them, 
and for this reason, at the quantum level we suppose $\hat{D}=\hat{D}_0$ and $\hat{K}=\hat{K}_0$, 
bearing in mind that  the time dependence can be reconstructed by
 application of  the corresponding time-evolution operator.   
 
By introducing the rescaled by $\hbar$
quantum analogs of the  linear combinations (\ref{sl2Rgen}), 
\begin{eqnarray}\label{JJJ}
&\hat{\mathcal{J}}_0=\frac{1}{2\omega\hbar}(\hat{H}_0+\omega^2\hat{K})\,,\qquad
\hat{\mathcal{J}}_\pm=\hat{\mathcal{J}}_1\pm i \hat{\mathcal{J}}_2=
-\frac{1}{2\omega\hbar}(\hat{H}_0-\omega^2\hat{K}\pm 2i\omega\hat{D})\,,&
\end{eqnarray}
we produce the quantum  $\mathfrak{sl}(2,\R)$ algebra 
\begin{eqnarray}
&[\hat{\mathcal{J}}_0,\hat{\mathcal{J}}_\pm]=\pm \hat{\mathcal{J}}_\pm\,,\qquad
[\hat{\mathcal{J}}_-,\hat{\mathcal{J}}_+]=2\hat{\mathcal{J}}_0\,.&
\end{eqnarray}
The quantum conformal bridge transformation (CBT)  
\cite{InzPlyWipf1,InzPly7,InzPly8} is a similarity transformation 
\begin{eqnarray}
&\label{GenConBr}
\hat{\mathfrak{S}}(\hat{H}_0)\hat{\mathfrak{S}}^{-1}=-\omega\hbar \hat{\mathcal{J}}_-\,,\qquad
\hat{\mathfrak{S}}(i\hat{D})\hat{\mathfrak{S}}^{-1}=\hbar \hat{\mathcal{J}}_0\,,\qquad
\hat{\mathfrak{S}}(\hat{K})\hat{\mathfrak{S}}^{-1}={\hbar}{\omega}^{-1}
\hat{\mathcal{J}}_+\,,&
\end{eqnarray}
generated by
the non-unitary, Hermitian operator 
\begin{eqnarray}
\label{QCB}
& \hat{\mathfrak{S}}
=
e^{-\frac{\omega}{\hbar} \hat{K}}e^{\frac{i}{\hbar}\ln(2)\hat{D}} e^{\frac{\hat{H}_0}{\hbar \omega}}
=\exp\left[\frac{\pi}{4\hbar}\left({\omega}^{-1}\hat{H}_0 - \omega\hat{K}\right)\right]\,.
&
\end{eqnarray}
Notice that all the operators $\hat{H}_0$,  $\hat{K}$ and $i\hat{D}$,
to which the CBT (\ref{GenConBr}) is applied, as well as  the 
CBT generator $\hat{\mathfrak{S}}$ itself commute with the $\mathcal{PT}$ operator, i.e. 
they are 
$\mathcal{PT}$-symmetric~\footnote{Here $\mathcal{P}$ is a space reflection operator, 
$\mathcal{P}x = -x\mathcal{P}$, 
$\mathcal{P}^2=1$, and
a complex conjugation operator $\mathcal{T}$ is defined by 
$\mathcal{T}z = -z^* \mathcal{T}$, $\mathcal{T}^2=1$, 
where $z\in \C$ is an arbitrary complex number.}.

Relations (\ref{GenConBr}) imply that 
\begin{eqnarray}
&\label{EigenEstates}
\hat{D} \ket{\lambda} = i\hbar \lambda \ket{\lambda}\quad \Rightarrow\quad \hat{\mathcal{J}}_0
(\hat{\mathfrak{S}}\ket{\lambda}) = \lambda\hat{\mathfrak{S}}\ket{\lambda}\,,
&\\&
\label{CoherentEstates}
\hat{H}_0\ket{E} = E\ket{E}\quad \Rightarrow\quad \hat{\mathcal{J}}_-(\hat{\mathfrak{S}}\ket{E})
= -\frac{E}{\hbar 
\omega}\hat{\mathfrak{S}}\ket{E}\,.&
\end{eqnarray}
Then, 
to have a physical eigenstate $ \hat{\mathfrak{S}}\ket{\lambda} $ of $ \hat{\mathcal{J}}_0 $, the 
formal state $\ket{\lambda}$ must obey  the following conditions:
\begin{itemize}
\item[I.] The series  $ \exp(\frac{\hat{H}_0}{2\hbar \omega})\ket{\lambda} =
\sum_{n=0}^{\infty}\frac{1}{n!(2\hbar \omega)^{n}} (\hat{H}_0)^n\ket{\lambda} 
$
 has  to  reduce to  a finite number of terms; this means that  $\ket{\lambda}$ should 
be a Jordan 
state
 of the operator $ \hat{H}_0 $ corresponding to zero energy \footnote{
The wave functions of generalized Jordan states 
corresponding to energy $ \lambda $ 
satisfy relations of the form $ P(\hat{H})\Omega_{\lambda}=\psi_{\lambda} $, 
where $ \hat{H}\psi_{\lambda} = \lambda\psi_{\lambda}$ and $P(\eta) $ 
 is a polynomial 
\cite{CJP,CarPlyJ,InzPly3}.
Here
we consider  the Jordan states satisfying  
the relations $ (\hat{H})^\ell\Omega_{\lambda} = \lambda\psi_{\lambda} $ 
with $ \lambda=0 $ for a certain 
natural number $\ell$.}. 
\item[II.] If a  wave function $ \Omega_\lambda=\bra{\vx}\ket{\lambda} $ is a rank 
$n$ Jordan state of $\hat{H}$ 
of  zero energy,  
$\hat{H}^n \Omega_\lambda=0$, 
then $\Omega_\lambda$,  
as well as $(\hat{H}_0)^{k}\Omega_\lambda$,  $k=1,\ldots, n-1$, 
 must not have poles and 
have to be single-valued.
\end{itemize}
On the other hand, the
 eigenvectors $\ket{E}$ 
(physical, or  non-physical,
with complex eigenvalues in general case)
 of $\hat{H}_0$  
 are transformed into eigenvectors of 
the lowering operator 
$\hat{\mathcal{J}}_-$ of  the $ \mathfrak{sl}(2,\R)$  algebra. 
Therefore, the resulting 
eigenstates in (\ref{CoherentEstates}) 
are the coherent states of $ \mathfrak{sl}(2,\R)$ in the sense of Perelomov 
\cite{Perelomov}.

The formalism related to the quantum CBT admits a reinterpretation 
in the context of the  $\mathcal{PT}$ symmetry  \cite{PTmetric1,PTmetric2}.
Indeed,  the second 
 relation in (\ref{GenConBr}) can be written as 
\begin{eqnarray}\label{PTHC}
&\hat{\mathfrak{S}}\hat{H}_{\mathcal{PT}}=\hat{H}_{HC}\hat{\mathfrak{S}}\,,&
\end{eqnarray}
where 
$\hat{H}_{\mathcal{PT}}=2i\omega\hbar \hat{D}$ is a  $\mathcal{PT}$-invariant operator, while
$\hat{H}_{HC}=2\omega \hbar \hat{\mathcal{J}}_0$ 
is a Hermitian operator.
From here we see that the Hermitian operator $\hat{\mathfrak{S}}$,
being the generator of the complex automorphism of the conformal 
algebra $\mathfrak{so}(2,1)\cong \mathfrak{sl}(2,\R)$,   intertwines 
a non-Hermitian, but  $\mathcal{PT}$-invariant Hamiltonian with a Hermitian one.  
By multiplying this relation from the 
left by $\hat{\mathfrak{S}}$ we obtain $\hat{\Theta}\hat{H}_{\mathcal{PT}}=
\hat{H}_{\mathcal{PT}}^\dagger \hat{\Theta}$, 
where $\hat{\Theta}=\hat{\mathfrak{S}}^2$,
and we have taken into account the
relation  $\hat{D}=-\hbar \hat{\mathcal{J}}_2$ 
and the quantum analog of (\ref{S2}).
 This implies that   if $\hat{H}_{HC}$ represents 
a well defined quantum system with real eigenvalues and normalizable eigenfunctions, then 
$\hat{H}_{\mathcal{PT}}$ has a real spectrum with  corresponding eigenstates of finite but not positive definite
norm 
under the indefinite scalar  product
$\bra{\lambda_1}\hat{\Theta}\ket{\lambda_2}$. 
With respect to this inner product, operator  $\hat{H}_{\mathcal{PT}}$
is pseudo-Hermitian \cite{Mostafazadeh}.

\section{Applications of CBT to one-dimensional systems}
\label{SecAFF1d}

In this section, based on  \cite{InzPlyWipf1},  we apply 
the quantum CBT to one-dimensional systems.  For the sake of simplicity, we use 
here the units $\hbar=m=1$. 
In each of the examples,  we consider  the symmetry operators, the eigenstates, and 
the rank $n$ Jordan states of zero energy corresponding to an asymptotically  free system. 
By applying  the CBT, we  get  the symmetry operators, the eigenstates and the coherent states 
of the corresponding  harmonically confined models. 

In subsection \ref{SecNon-HermitianCal}, we give a reinterpretation of the 
CBT  by comparing its construction  with  the Swanson
 $\mathcal{PT}$-symmetric system \cite{PTmetric1,PTmetric2}
  in correspondence with Eq. (\ref{PTHC}) and related comments there. 
  This will allow us, particularly, to generalize the 
  construction of ref. \cite{PTmetric1,PTmetric2} to the case of  
 the AFF conformal mechanics model.

\subsection{Example 1: The 
free particle - harmonic oscillator relation}
\label{CBTExp1}

Let us start with the one-dimensional free particle symmetry  generators  
\begin{eqnarray}
&\label{1dfreeparticlegen}
\hat{H}=-\frac{1}{2}\frac{d^2}{dx^2}
\,,\qquad
\hat{D}=-\frac{i}{2}\left[x\frac{d}{dx}+\frac{1}{2}\right]\,,
\qquad 
\hat{K}=\frac{1}{2}x^2\,,
&\\&
\hat{p}=-i\frac{d}{dx}\,,\qquad
\hat{x}=x\,.\label{1dfreeparticlegen2}&
\end{eqnarray}
They produce the one-dimensional Schr\"odinger symmetry \cite{Nied1}.
The 
eigenstates and eigenvalues 
of $\hat{H}$
 are 
\begin{eqnarray}
&
\label{freePeig}
\psi_{\kappa}=e^{i\kappa x}\,,\qquad
E=\frac{1}{2}\hbar^2\kappa^2\,,\qquad \kappa \in \R\,.
&
\end{eqnarray} 
The 
functions $\bra{x}\ket{\lambda}$ that satisfy   
the two conditions specified  in Sec. \ref{SecQuantumCBT} 
correspond to 
\be
\label{JordanOsci}
\Omega_{n}(x)=\bra{x}\ket{\lambda}=x^{n}\,,\qquad n=0,1,2,\ldots.
\ee
The set of states (\ref {JordanOsci}) as a whole is invariant under the action of the symmetry generators 
(\ref{1dfreeparticlegen}), (\ref{1dfreeparticlegen2}), 
\begin{eqnarray}
&\label{HosinOm}
\hat{H}\Omega_{n}=-\frac{1}{2}n(n-1)\Omega_{n-2}\,,\qquad
\hat{K}\Omega_{n}=\frac{1}{2}\Omega_{n+2}\,,\qquad
2i\hat{D}\Omega_{n}=(n+\frac{1}{2})\Omega_{n}\,,&\\&
\hat{p}\Omega_{n}=-in\Omega_{n-1}\,,\qquad
\hat{x}\Omega_{n}=\Omega_{n+1}\,.\label{HosinOm2}
&
\end{eqnarray}
Via the 
repeated application 
of $\hat{H}$ to $\Omega_{n}$, we arrive at the functions $\Omega_{0}=1$ (if $n$ is even) or $\Omega_{1}=x$ (if $n$ is odd), which 
are the (physical and non-physical) 
zero energy solutions of the free particle stationary Schr\"odinger equation. So, 
$\Omega_{n}(x)$ are the rank $[n/2]+1$ Jordan 
states of zero energy of the free particle, 
$(\hat{H})^{[n/2]+1}\Omega_{n}=0$, where $[.]$ denotes an integer part. 
We see that $\Omega_{n}$ are common formal eigenfunctions
of the operators $2i\hat{D}$  and $(\hat{H})^{[n/2]+1}$.

According to Eq.  (\ref{GenConBr}), 
the application of the  CBT to the free particle's 
$\mathfrak{so}(2,1)$ generators 
gives us the $\mathfrak{sl}(2,\R)$ generators 
of the one-dimensional harmonic oscillator system,
\begin{eqnarray}
&
\hat{H}_{\text{os}}=2\omega\hat{\mathcal{J}}_0= -\frac{1}{2}\frac{d^2}{dx^2}+\frac{\omega^2}{2}x^2\,,
\qquad 
\hat{\mathcal{J}}_\pm=\frac{1}{4\omega}\left[\frac{d^2}{dx^2}+\omega^2 x^2 \mp \omega(x\frac{d}{dx}+\frac{1}{2})\right]\,.
&
\end{eqnarray}
Along with them, we obtain the  Heisenberg generators (ladder operators)
\begin{eqnarray}
&\label{d-lader1}
\hat{a}^\pm=\sqrt{\frac{\omega}{2}}\left(x \mp \frac{1}{\omega}\frac{d}{d x}\right)\,,\qquad
[\hat{a}^-,\hat{a}^+]=1\,, 
&
\end{eqnarray}
according to 
the relations 
$\hat{\mathfrak{S}}(\hat{p})\hat{\mathfrak{S}}^{-1}=
-i\sqrt{\omega}\,\hat{a}^-$ and $ 
\hat{\mathfrak{S}}(\hat{x})\hat{\mathfrak{S}}^{-1}=\sqrt{\frac{1}{\omega}}\,\hat{a}^+
$. 
From here one deduces  that 
this non-unitary (similarity) transformation 
can be identified as the
fourth order root of the space reflection operator $\mathcal{P}$,  
\begin{eqnarray}\label{S4=1}
& \hat{\mathfrak{S}}: (x,\hat{p},\hat{a}^+,\hat{a}^-){\rightarrow}(\hat{a}^+,-i\hat{a}^-,-i\hat{p}, x)\,,\qquad
 \hat{\mathfrak{S}}^2: (x,\hat{p},\hat{a}^+,\hat{a}^-){\rightarrow}(-i\hat{p}, -ix, -\hat{a}^-,\hat{a}^+)\,,&\nonumber\\
& \hat{\mathfrak{S}}^4: (x,\hat{p},\hat{a}^+,\hat{a}^-){\rightarrow}(-x, -\hat{p}, -\hat{a}^+,-\hat{a}^-)\,.&
\end{eqnarray}
For the sake of simplicity, we set $\omega=1$ in (\ref{S4=1}).
Notice here that the action of the CBT on generators of the Heisenberg algebra
corresponds to  the  eighth order  root of the identity automorphism, 
$\hat{\mathfrak{S}}^8=1$~\footnote{See the comments  related to Eq. (\ref{scalhol}) below,
and  refs. \cite{Howe,Wiki}  where 
the automorphism of the one-dimensional Heisenberg 
group is discussed  in the context of the
Stone-von Neumann theorem and the properties of the Fourier transform.}.

 The application of the operator $\hat{\mathfrak{S}}$ to the functions (\ref{JordanOsci}) gives us 
\begin{eqnarray}
&
\label{JordanTransformed}
\hat{\mathfrak{S}}\Omega_n=\mathcal{N}_{n}\psi_{n}\,,\qquad 
\psi_{n}=\frac{1}{\sqrt{2^{n}n!}}(\frac{\omega}{\pi})^{\frac{1}{4}}H_{n}(\sqrt{\omega}x)e^{-\frac{\omega x^2}{2}}\,,\qquad
\mathcal{N}_{n}=\frac{(2\pi)^{\frac{1}{4}}}{\omega^{\frac{n}{2}+\frac{1}{4}}}\sqrt{n}\,,&
\end{eqnarray}
where we have used the Weierstrass transformation \cite{Bilo,InzPlyWipf1}. 
Additionally, by acting  from the left by the operator $\hat{\mathfrak{S}}$ on both sides of  Eqs.  
(\ref{HosinOm}) and (\ref{HosinOm2}) one gets the well known relations
\begin{eqnarray}
&\hat{H}_{\text{os}}\psi_{n}=\omega (n+\frac{1}{2})\psi_{n}\,,\quad 
\hat{\mathcal{J}}_\pm\psi_{n}=\sqrt{(n\pm \beta_\pm)(n+\beta_\pm \pm 1)}\psi_{n\pm 2}\,,\quad
\beta_\pm=\frac{1\pm 1}{2}\,,
&\\&
\hat{a}^-\psi_{n}=\sqrt{n}\psi_{n-1}\,,\qquad
\hat{a}^+\psi_{n}=\sqrt{n+1}\psi_{n+1}\,.
&
\end{eqnarray}
Finally, the action of the operator 
$\hat{\mathfrak{S}}$ on the free particle eigenstates of the form (\ref{freePeig}),
being simultaneously  eigenfunctions of $\hat{p}$,  produces 
\begin{eqnarray}
&\phi(x,\kappa)=
\hat{\mathfrak{S}}e^{i\frac{\kappa x}{\sqrt{2}}}=2^{\frac{1}{4}}
\exp(-\frac{\omega^2}{2} x^2 +\frac{\kappa^2}{4\omega}+i\kappa x)=
(\frac{2\pi}{\omega})^{\frac{1}{4}}\sum_{n=0}^{\infty}(\frac{ik}{\sqrt{2\omega}})^{n}
\frac{\psi_n}{\sqrt{n!}}\,.
&
\end{eqnarray}
These are the coherent states \cite{Gazeau} that satisfy the relation
$\hat{a}^-\phi(x,\kappa)=\frac{i\kappa}{\sqrt{2\omega}}\phi(x,\kappa)$.
Under the time evolution 
these functions take the form 
\be
\phi(x,\kappa,t)=e^{-i\hat{H}_{\text{os}}t}\phi(x,\kappa)=e^{-i\frac{\omega t}{2}}\phi(x,\kappa e^{-i\omega t})\,.
\ee
To have the over-complete set of coherent states  of the quantum harmonic oscillator,
we allow the parameter $\kappa$  to take complex values.
In the same vein  
one can show that the 
free particle  Gaussian wave packets are mapped into the squeezed states of 
the harmonic oscillator, see ref. \cite{InzPlyWipf1}. 

Finally, we note that the described  CBT formalism is close
with the unitary transformation 
between the coordinate and the Fock-Bargmann representations. 
In fact, this last representation  
can be obtained 
if we formally replace the spacial variable $x$ with the complex  variable  $z$ in operators 
(\ref{1dfreeparticlegen}), (\ref{1dfreeparticlegen2}) as well as in the Jordan states (\ref{JordanOsci}), 
and  as an additional step,  substitute  the usual $L^2(\R)$ scalar product 
for  the inner product
\begin{eqnarray}\label{scalhol}
&(\psi_1, \psi_2)=\frac{1}{\pi}\int_{\R^2}\overline{\psi_1(z)}
\psi_2(z)e^{-\bar{z}z}d^2z,\qquad 
d^2z=d(\text{Re}\,z)d(\text{Im}\,z)\,.&
\end{eqnarray}
The kernel of the integral transformation, which is a unitary transformation
 from  the $L^2(\R)$ Hilbert space to  the Fock-Bargmann space,  
 can be related to the  considered CBT, for further details see ref. \cite{InzPlyWipf1}.

\subsection{Example 2: The one-dimensional Calogero model - AFF model relation}
\label{SecCalAFF}
The two-particle Calogero model admits a separation of variables in terms of the relative coordinate 
and the coordinate of the center of mass, which has a  free dynamics.
The corresponding  $\mathfrak{so}(2,1)$ symmetry generators associated with the relative coordinate $x>0$
  are 
defined  on  the positive real half-line $\R^+$, and they correspond to 
\begin{eqnarray}
&\label{HermitianCal}
\hat{H}_{\nu}=-\frac{1}{2}\frac{d^2}{dx^2}+\frac{\nu(\nu+1)}{2x^2}
\,,\qquad
\hat{D}=-\frac{i}{2}\left[x\frac{d}{dx}+\frac{1}{2}\right]\,,
\qquad 
\hat{K}=\frac{1}{2}x^2\,,
&
\end{eqnarray}
where we assume  that $\nu>-\frac{1}{2}$ \cite{InzPly3}.
The  
eigenstates and eigenvalues of $\hat{H}_{\nu}$,
are given by 
\begin{eqnarray}
\label{Calogerostates}
&\psi_{\kappa,\nu}(x)=\sqrt{x}\,{J}_{\nu+1/2}(\kappa x)\,,
\qquad
E=\frac{1}{2}\hbar^2\kappa^2\,,\qquad \kappa>0\,,&
\end{eqnarray}
where $J_\alpha$ are the Bessel functions of the first kind. 
Besides,  
the functions $\bra{x}\ket{\lambda}=\Omega_{n,\nu}$, that are well defined on $\R^+$
and are the zero energy Jordan states of $\hat{H}_{\nu}$ 
 correspond to
\begin{eqnarray}
\label{Omega1}
&\Omega_{n,\nu}=x^{2n+\nu+1}\,.&
\end{eqnarray}
These functions satisfy the following set of equations, 
\begin{eqnarray}
\label{actionHKAFF}
&\hat{H}_{\nu}\Omega_{n,\nu}=-n(2n+2\nu+1)\Omega_{n-1,\nu}\,,\qquad
\hat{K}\Omega_{n,\nu}=\frac{1}{2}\Omega_{n+1,\nu}\,,&\\&
2i\hat{D}\Omega_{n,\nu}=(2n+\nu+\frac{3}{2})\Omega_{n,\nu}\,.\label{actionHKAFF2}
&
\end{eqnarray}
In the same vein  as in the previous subsection, one can see that after the repeated application
of $\hat{H}_{\nu}$ one gets the zero energy solution $\Omega_{0,\nu}=x^{\nu+1}$, 
which is a regular function 
on $\R^+$. 

On the other hand, the  conformal symmetry generators  of the AFF model  \cite{AFF},
\begin{eqnarray}
&\label{AFFsl2Rgen}
\hat{H}_{\nu}^{AFF}=2\omega\hat{\mathcal{J}}_0=
-\frac{1}{2}\frac{d^2}{dx^2}+\frac{\nu(\nu+1)}{2x^2}+\frac{\omega^2}{2}x^2\,,
&\\&
\hat{\mathcal{J}}_\pm=\frac{1}{4\omega}\left[\frac{d^2}{dx^2}-\frac{\nu(\nu+1)}{x^2}+\omega^2 x^2 \mp 
\omega(x\frac{d}{dx}+\frac{1}{2})\right]\,,
&
\end{eqnarray}
are obtained by applying the CBT to   generators (\ref{HermitianCal}). 
In the same way, the normalized eigenstates of $\hat{H}_\nu^{AFF}$ correspond to 
\begin{eqnarray}
&
\hat{\mathfrak{S}}\Omega_{n,\nu}=\mathcal{N}_{n,\nu} \psi_{n,\nu}\,,\qquad
\psi_{n,\nu}=\sqrt{\frac{2\omega^{\nu+\frac{3}{2}}n!}{\Gamma(n+\nu+\frac{3}{2})}}
x^{\nu+1}L_{n}^{(\nu+\frac{1}{2})}(\omega x^2)e^{-\frac{\omega x^2}{2}}\,, &
\end{eqnarray}
where $\mathcal{N}_{n,\nu} =
(-1)^{n}\left(\frac{2}{\omega}\right)^{\frac{\nu}{2}+n}
\sqrt{\omega^{-1}n!\Gamma(n+\nu+\frac{3}{2})}\,.$
They satisfy equations 
\begin{eqnarray}
&\hat{{H}}_{\nu}^{AFF}\psi_{n,\nu}=E_{n,\nu}\psi_{n,\nu}\,,\qquad E_{n,\nu}=\omega(2n+\nu+\frac{3}{2})\,, &\\&
\hat{\mathcal{J}}_{-}\psi_{n,\nu}=-\sqrt{n(n+\nu+\frac{1}{2})}\psi_{n-1,\nu}\,,\quad
\hat{\mathcal{J}}_{+}\psi_{n,\nu}=-\sqrt{(n+1)(n+\nu+\frac{3}{2})}\psi_{n+1,\nu}\,,&
\end{eqnarray}
that are obtained directly from the application of $\hat{\mathfrak{S}}$ to 
 equations (\ref{actionHKAFF}), (\ref{actionHKAFF2}).

On the other hand, the application of the operator
$\hat{\mathfrak{S}}$ to  eigenstates (\ref{Calogerostates})
of the system $\hat{H}_\nu$ 
yields 
\begin{eqnarray}
\label{coherent0}
&\hat{\mathfrak{S}}\psi_{\kappa,\nu}(\frac{1}{\sqrt{2}}x)=2^{\frac{1}{4}}
e^{-\frac{1}{2}x^2+\frac{1}{4}\kappa^2}
\sqrt{x}\,{J}_{\nu+1/2}(\kappa x):=\phi_\nu(x,\kappa)\,,&
\end{eqnarray} 
that are the states satisfying the relation
$
\hat{\mathcal{J}}_-\phi_\nu(x,\kappa)
=-\frac{1}{4}\kappa^2\phi_\nu(x,\kappa)\,.
$
By changing the parameter $\kappa$ for the complex parameter $z$,
one obtains coherent states that are eigenstates of operator  $\hat{\mathcal{J}}_-$ with complex 
eigenvalue $-\frac{1}{4}z^2$.
By using  the evolution operator $\exp(-it\hat{H}^{\text{AFF}}_\nu)$, the 
time-dependent coherent states are obtained, 
\begin{eqnarray}
\phi_{\nu}(x,z,t)=2^{1/4}\sqrt{x}\,{J}_{\nu+1/2}(z(t) x)e^{-x^2/2+z^2(t)/4-it}\,,\qquad
z(t)=z e^{-it}\,.
\end{eqnarray}

\subsection{Example 3: CBT and $\mathcal{PT}$-regularized conformal systems}
\label{SecNon-HermitianCal}

The conformal symmetry operators of the 
$\mathcal{PT}$-regularized Calogero systems, which are defined for 
$x\in \R$, 
are given by \cite{JuanMP1,JuanMP2}
\begin{eqnarray}
&
\hat{H}_{\alpha,\nu}=-\frac{1}{2}\frac{d^2}{dx^2}+\frac{\nu(\nu+1)}{2(x+i\alpha)^2}
\,,\qquad
\hat{D}_\alpha=-\frac{i}{2}\left[(x+i\alpha)\frac{d}{dx}+\frac{1}{2}\right]\,,
&\\&
\hat{K}_\alpha=\frac{1}{2}(x+i\alpha)^2\,.
&
\end{eqnarray}
Notice that $\hat{H}_{\alpha,\nu}$ and $\hat{K}_\alpha$ are  $\mathcal{PT}$-symmetric,
$[\mathcal{PT},\hat{H}_{\alpha,\nu}]=[\mathcal{PT},\hat{K}_\alpha]=0$, while $\hat{D}_\alpha$ is 
the $\mathcal{PT}$-odd, $\mathcal{PT} \hat{D}_\alpha=-\hat{D}_\alpha \mathcal{PT}$, operators.
They are obtained by application of the 
complex translation $x\rightarrow x+i\alpha$, $\alpha\in \R$, $\alpha\neq 0$, 
generated by the Hermitian operator
\be\hat{I}_\alpha=e^{-\alpha \hat{p}}=e^{i\alpha\frac{d}{dx}}\,, 
\ee 
to the generators of the 
Hermitian Calogero system  (\ref{HermitianCal}),
\begin{eqnarray}
&\hat{I}_\alpha (\hat{H}_{0,\nu})\hat{I}_{-\alpha}=\hat{H}_{\alpha,\nu}\,,\qquad
\hat{I}_\alpha (\hat{D}_0)\hat{I}_{-\alpha}=\hat{D}_{\alpha}\,,\qquad
\hat{I}_\alpha (\hat{K}_0)\hat{I}_{-\alpha}=\hat{K}_{\alpha}\,,&
\end{eqnarray}
supplemented  by extension of the domain for the position variable to the 
entire real line, that we imply in the rest of this subsection. 
 In the same way, 
the eigenstates of $\hat{H}_{\alpha,\nu}$ are formally obtained by  application of operator $\hat{I}_{\alpha}$ 
to the  states (\ref{Calogerostates}), and to  their corresponding linearly independent partners, 
which are given by
the Neumann functions $Y_\nu(\kappa x)$. 
On the other hand, the rank $[n/2]+1$ Jordan states of zero energy are given by 
\begin{eqnarray}
\label{JordanNonH}
&\Omega_{n,\nu}^{\alpha}= \hat{I}_{\alpha}\Omega_{n,\nu}=\left(x+i\alpha\right)^{2n+\nu+1}\,,\qquad
\Xi_{n,\nu}^{\alpha}= \hat{I}_{\alpha}\Xi_{n,\nu}=(x+i\alpha)^{2n-\nu}\,,&
\end{eqnarray}
where $\Xi_{n,\nu}=x^{2n-\nu}$ is obtained by  the transformation 
$\rho: \nu\rightarrow -\nu-1$
 (with respect to which generators (\ref{HermitianCal}) are invariant) 
 over   functions 
(\ref{Omega1}) \cite{InzPly3}.  Relations analogous to (\ref{actionHKAFF}), (\ref{actionHKAFF2}) 
for functions $\Omega_{n,\nu}^{\alpha}$ are generated by  applying the
operator $\hat{I}_\alpha$. Additional  transformation $\rho$ then 
yields 
\begin{eqnarray}
&
\hat{H}_{\alpha,\nu}\Xi_{n,\nu}^{\alpha}=-n(2n-2\nu-1)\Xi_{n-1,\nu}^{\alpha}\,,\qquad
\hat{K}_{\alpha}\Xi_{n,\nu}^{\alpha}=\frac{1}{2}\Xi_{n+1,\nu}^{\alpha}\,,&\\&
2i\hat{D}_{\alpha}\Xi_{n,\nu}^{\alpha}=(2n-\nu+\frac{1}{2})\Xi_{n,\nu}^{\alpha}\,.
&
\end{eqnarray}

In the special case $\nu=m$ with $m=1,2,\ldots$, the system 
$\hat{H}_{\alpha,m}$ can be obtained from the free particle 
system on $\R$ by the Darboux transformation of the order $m$ 
\cite{CorOlMP,JuanMP1}. As a consequence, each such system
possesses a hidden symmetry described by the Darboux-dressed 
generators of the translations and Galilean boosts of the free particle, 
\begin{eqnarray}\label{LacNoviIN}
&\hat{\mathcal{P}}_{\alpha,m}=\hat{\A}_{\alpha,m}^-\hat{p} \hat{\A}_{\alpha,m}^+\,,\qquad
\hat{\mathcal{X}}_{\alpha,m}=\hat{\A}_{\alpha,m}^-(\hat{x}+i\alpha) \hat{\A}_{\alpha,m}^+\,,
&\\
&\label{Am operaor}
\hat{\A}_{\alpha,m}^-=\hat{A}_{\alpha,m}^-\ldots \hat{A}_{\alpha,1}^-\,,\quad
\hat{\A}_{\alpha,m}^+=\hat{A}_{\alpha,1}^+\ldots \hat{A}_{\alpha,m}^+\,,\quad 
\hat{A}_{\alpha,m}^\pm=\mp\frac{1}{\sqrt{2}}\left(\frac{d}{dx}\pm\frac{m}{x+i\alpha}\right)\,.&
\end{eqnarray}
The order $(2m+1)$ differential operator $\hat{\mathcal{P}}_{\alpha,m}$
is the  analog of the Lax-Novikov integral in the quantum reflectionless and finite-gap systems
whose potentials are  snapshots of the corresponding multi-soliton 
and cnoidal-type solutions to the KdV equation 
\cite{AGM1,AranPly}.
Here the operators $\hat{\A}_{\alpha,m}^\pm$ are the 
higher order intertwining operators that connect the free particle Hamiltonian
 $\hat{H}_{\alpha,0}=\hat{H}$
with the $\mathcal{PT}$ regularized Calogero Hamiltonian  $\hat{H}_{\alpha,m}$, 
\be\label{AmIntert}
\hat{\A}_{\alpha,m}^-\hat{H}_{\alpha,0}=\hat{H}_{\alpha,m}\hat{\A}_{\alpha,m}^-\,,\qquad
\hat{\A}_{\alpha,m}^+\hat{H}_{\alpha,m}=\hat{H}_{\alpha,0}\hat{\A}_{\alpha,m}^+\,.
\ee
 The $\mathcal{PT}$-symmetric 
 system $\hat{H}_{\alpha,m}$ is regular on a real line, and due 
 to its relation to the  free particle via the intertwining relations
 (\ref{AmIntert}), it turns out to be perfectly invisible system
 with  a unique $L^2(\R)$ integrable state $\Xi_{0,m}$
 of zero energy at the very edge
 of the doubly degenerate continuous part of the spectrum. 
 The integral $\hat{\mathcal{P}}_{\alpha,m}$ separates 
 the states of the same energy in the doubly degenerate continuous part of the spectrum 
 as well as detects the unique bound state of the system 
 $\hat{H}_{\alpha,m}$ by annihilating it \cite{JuanMP1,JuanMP2}.

The commutation relations between the symmetry generators 
(\ref{LacNoviIN}) and $\hat{H}_{\alpha,m}$, $\hat{D}_{\alpha}$ and $\hat{K}_{\alpha}$ produce an
 extended  non-linear
algebra, that includes, particularly,   the  Lie algebraic relations 
\begin{eqnarray}
&
[\hat{H}_{\alpha,m}\hat{\mathcal{P}}_{\alpha,m}]=0\,,\qquad
[\hat{H}_{\alpha,m},\hat{\mathcal{X}}_{\alpha,m}]=-i\hat{\mathcal{P}}_{\alpha,m}\,,
 &\\&
[\hat{D}_\alpha,\hat{\mathcal{P}}_{\alpha,m}]=\frac{i}{2}(2m+1)\hat{\mathcal{P}}_{\alpha,m}\,,\qquad
[\hat{D}_\alpha,\hat{\mathcal{X}}_{\alpha,m}]=\frac{i}{2}(2m-1)\hat{\mathcal{X}}_{\alpha,m}\,.
&
\end{eqnarray}
The action of operators (\ref{LacNoviIN})  on Jordan states (\ref{JordanNonH}) with $\nu=m$  yields
\begin{eqnarray}
&
\hat{\mathcal{P}}_{\alpha,m}\Xi_{n,m}^{\alpha}\propto \Omega_{n-2m-1,m}^{\alpha}\,,\qquad
\hat{\mathcal{X}}_{\alpha,m}\Xi_{n,m}^{\alpha}\propto\Omega_{n-2m,m}^{\alpha}\,,&\\&
\hat{\mathcal{P}}_{\alpha,m}\Omega_{n,m}^{\alpha}\propto \Xi_{n,m}^{\alpha}\,,\qquad
\hat{\mathcal{X}}_{\alpha,m}\Omega_{n,m}^{\alpha}\propto\Xi_{n,m}^{\alpha}\,, &\\&
\ker \hat{\mathcal{P}}_{\alpha,m}=\text{span}\{\Xi_{0,m}^{\alpha},\ldots,\Xi_{2m,m}^{\alpha}\}\,,\qquad
\ker \hat{\mathcal{X}}_{\alpha,m}=\text{span}\{\Xi_{0,m}^{\alpha},\ldots,\Xi_{2m-1,m}^{\alpha}\}\,.
&
\end{eqnarray}

To relate a non-Hermitian  $\mathcal{PT}$-symmetric asymptomatically  free system
 like $\hat{H}_{\alpha,m}$ with its confined version, 
we introduce an extended CBT operator  
 \begin{eqnarray}
\hat{\mathfrak{S}}_\alpha=
\hat{I}_{\alpha}\hat{\mathfrak{S}}_0\hat{I}_{-\alpha}\,,\qquad
\hat{\mathfrak{S}}_\alpha^{-1}=\hat{I}_{\alpha}\hat{\mathfrak{S}}_0^{-1}\hat{I}_{-\alpha}\,,
\label{CBTM}
\end{eqnarray}
which in this case yields 
\begin{eqnarray}
&\label{GenConBrNONH}
\hat{\mathfrak{S}}_\alpha(\hat{H}_{\alpha,m})\hat{\mathfrak{S}}_\alpha^{-1}=-\omega \hat{\mathcal{J}}_{-,\alpha}\,,\quad
\hat{\mathfrak{S}}_\alpha(i\hat{D}_\alpha)\hat{\mathfrak{S}}^{-1}_\alpha= \hat{\mathcal{J}}_{0,\alpha}\,,\quad
\hat{\mathfrak{S}}_\alpha(\hat{K}_\alpha)\hat{\mathfrak{S}}_\alpha^{-1}=\frac{1}{\omega}
\hat{\mathcal{J}}_{+,\alpha}\,,&
\\&\label{nonH-JJJ}
\hat{\mathcal{J}}_{0,\alpha}=\frac{1}{2\omega}(\hat{H}_\alpha+\omega^2\hat{K}_\alpha)\,,\qquad
\hat{\mathcal{J}}_{\pm,\alpha}=-\frac{1}{2\omega}(\hat{H}_\alpha-\omega^2\hat{K}_\alpha\pm 2i\omega\hat{D}_\alpha)\,. &
\end{eqnarray}
Here we have  $\hat{\mathcal{J}}_{0,\alpha}=\hat{I}_\alpha\hat{\mathcal{J}}_{0,0}\hat{I}_{-\alpha}$
and $\hat{\mathcal{J}}_{\pm, \alpha}=\hat{I}_\alpha\hat{\mathcal{J}}_{\pm,0}\hat{I}_{-\alpha}$, and this implies that 
$\hat{\mathcal{J}}_{+, \alpha}$ is not the Hermitian conjugation (with respect to the usual inner product on $\R$) of
$\hat{\mathcal{J}}_{-, \alpha}$. 
Operators  $(\hat{\mathcal{J}}_{0,0},\hat{\mathcal{J}}_{\pm,0})$ correspond to the  AFF model generators 
(\ref{AFFsl2Rgen}), which are singular for $x\in \R$. 

On the other hand,
\begin{eqnarray}
\label{RegAFF}
&\hat{H}_{\alpha,\nu}^{AFF}=
\hat{I}_\alpha (\hat{H}_{0,\nu}^{AFF})
\hat{I}_{-\alpha} 
=-\frac{1}{2}\frac{d^2}{dx^2}+\frac{\nu(\nu+1)}{2(x+i\alpha)^2}+\frac{\omega^2}{2}(x+i\alpha)^2\,,&
\end{eqnarray}
is the non-Hermitian Hamiltonian of the $\mathcal{PT}$-regularized AFF model
with arbitrary value of the parameter $\nu>-1/2$. 
Note that for  $\nu=0$ we 
recover the $\mathcal{PT}$-symmetric one-dimensional harmonic oscillator with $x$ displaced for imaginary 
constant $i\alpha$  
being the simplest case of non-Hermitian systems 
introduced in \cite{Bender}.
The corresponding   eigenstates of (\ref{RegAFF}) are 
\begin{eqnarray}
&
\psi_{n,\nu}^{\alpha}\propto \hat{\mathfrak{S}}_{\alpha}\Omega_{n,\nu}^{\alpha}\,,\qquad 
\phi_{n,\nu}^{\alpha}\propto \hat{\mathfrak{S}}_{\alpha}\Xi_{n,\nu}^{\alpha}\,,
&
\end{eqnarray}
whose explicit form is 
 \begin{eqnarray}&
\psi_{n,\nu}^{\alpha}= (x+i\alpha)^{\nu+1}e^{-\frac{\omega (x^2-\alpha^2)}{2}+
i\omega \alpha x} L_{n}^{(\nu+\frac{1}{2})}(\omega (x+i\alpha)^2 )
\,,\qquad 
\phi_{n,\nu}^{\alpha}=\psi_{n,-\nu-1}^{\alpha}\,.\label{LaguerreKLein4}&
\end{eqnarray}

\begin{figure}[hbt!]
\begin{subfigure}[l]{0.4\linewidth}
\includegraphics[scale=0.6]{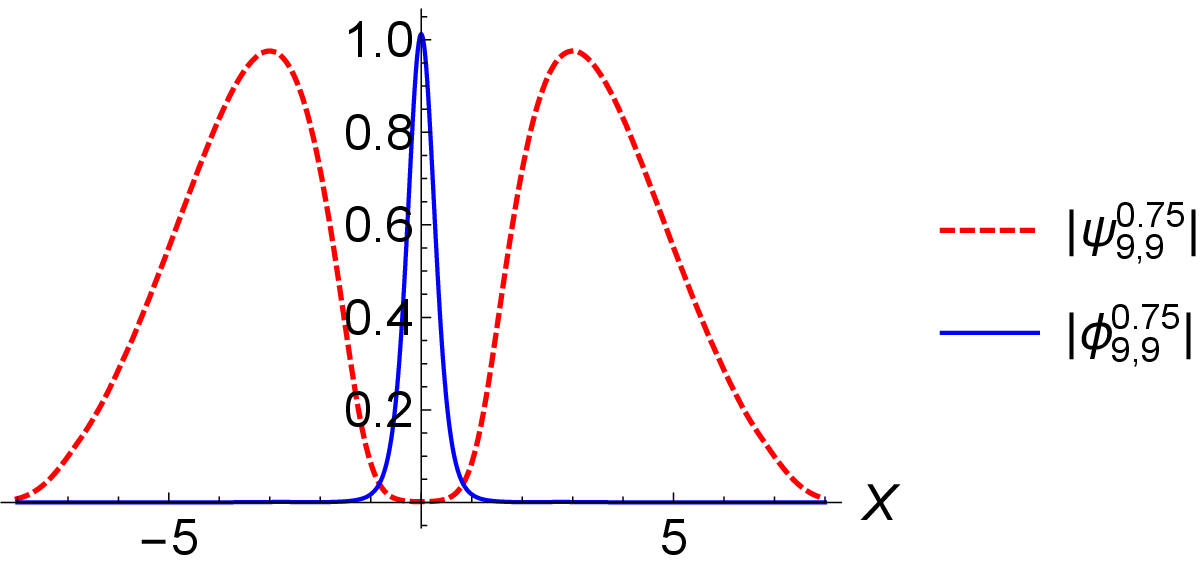}
\caption{\small{$n=m=9\,,\,\alpha=0.75$
}
}
\label{fig1aSec3}
\end{subfigure}
\hskip2cm
\begin{subfigure}[l]{0.4\linewidth}
\includegraphics[scale=0.6]{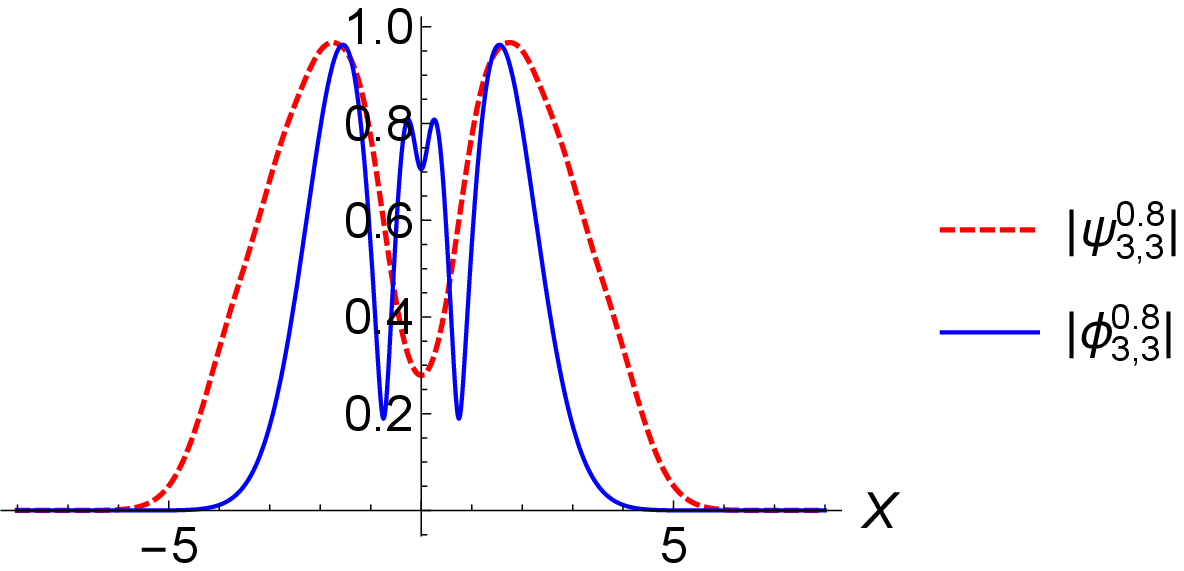}
\caption{\small{$n=m=3$, $\,\,\alpha=0.8$ 
}}
\label{fig1bSec3}
\end{subfigure}

\begin{subfigure}[c]{0.32\linewidth}
\includegraphics[scale=0.4]{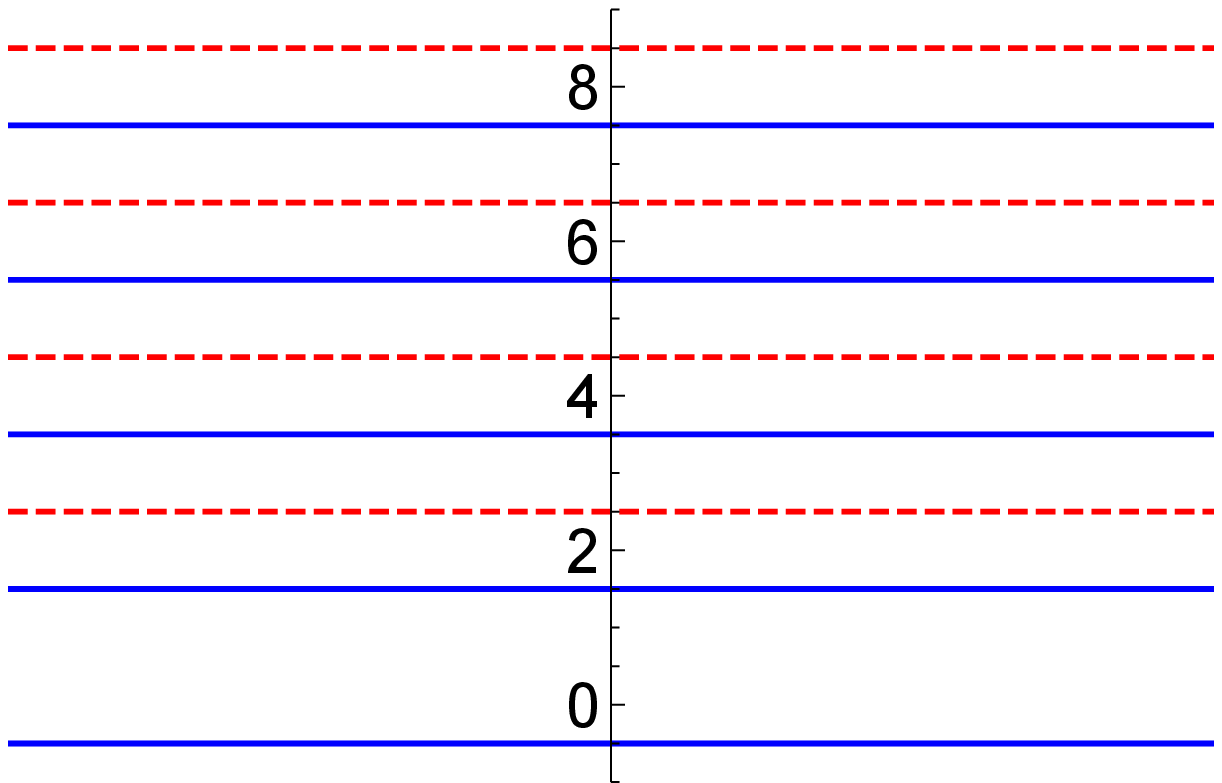}
\caption{\small{$m=1$}}
\label{fig1cSec3}
\end{subfigure}
\begin{subfigure}[c]{0.32\linewidth}
\includegraphics[scale=0.4]{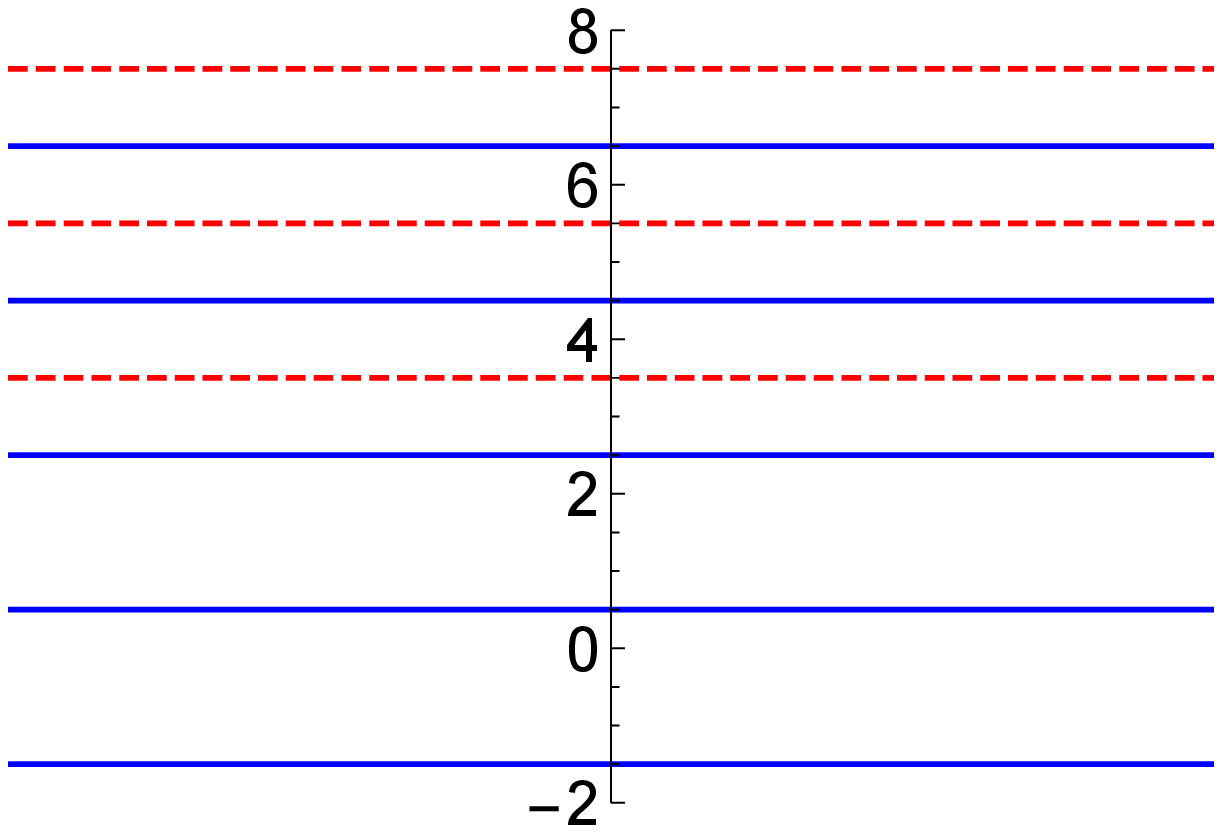}
\caption{\small{$m=2$}}
\label{fig1dSec3}
\end{subfigure}
\begin{subfigure}[c]{0.32\linewidth}
\includegraphics[scale=0.4]{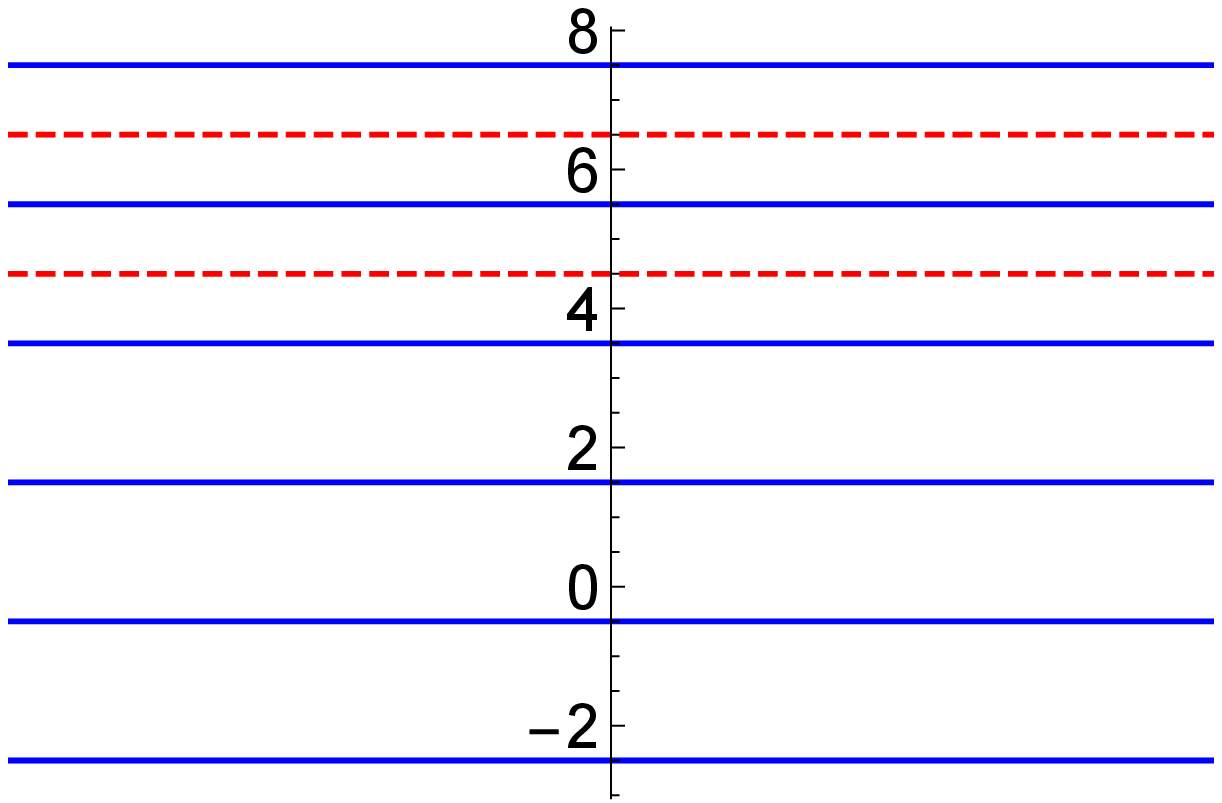}
\caption{\small{$m=3$}}
\label{fig1eSec3}
\end{subfigure}
\caption{\small{
In Figs. \ref{fig1aSec3} and \ref{fig1bSec3},  the modules 
of the states  $\phi_{n,m}^{\alpha}$ and $\psi_{n,m}^{\alpha}$ (normalized by their respective maxima)
with some  values  of $n$, $m$ and 
$\alpha$
are compared.  On Figs. \ref{fig1cSec3}, \ref{fig1dSec3} and \ref{fig1eSec3}, 
the energy levels  $\mathcal{E}_{n,m}$ (continuos blue lines) and 
$E_{n,m}$ (dashed red lines) are shown for different values of the parameter $m$.}    
}
\end{figure}

These functions satisfy the eigenvalue equations 
\begin{eqnarray}
&\label{SchrEqPT1}
\hat{H}_{\alpha,\nu}^{AFF}\psi_{n,\nu}^{\alpha}=E_{n,\nu}\psi_{n,\nu}^{\alpha}\,,\qquad
E_{n,\nu}=\omega(2n+\nu+\frac{3}{2})\,, 
&\\&
\hat{H}_{\alpha,\nu}^{AFF}\phi_{n,\nu}^{\alpha}=\mathcal{E}_{n,\nu}\phi_{n,\nu}^{\alpha}\,,\qquad
\mathcal{E}_{n,\nu}=\omega(2n-\nu+\frac{1}{2})\,,&
\end{eqnarray}
and  
they are $L^2(\R)$ normalizable,  
being of the form 
of a regular  polynomial times a Gaussian term, see Figs. \ref{fig1aSec3} and \ref{fig1bSec3}.  
The notable here is that we have two different towers of states, where 
the distance between two consecutive energy levels in each tower is given by  
$\Delta E=E_{n,\nu}-E_{n-1,\nu}=\mathcal{E}_{n,\nu}-\mathcal{E}_{n-1,\nu}=2\omega$. 
On the other hand,   
$\delta E= E_{n,\nu}-\mathcal{E}_{n,\nu}=\omega(2\nu+1)$. 
The last relation means that 
when $\nu=\ell-\frac{1}{2}$, $\delta E=\ell \Delta E$,  $\ell=1,2,\ldots$,
one could conclude that 
there emerges  a double degeneracy in the spectrum 
because of  the relation 
$E_{s-\ell,\ell-\frac{1}{2}}=\mathcal{E}_{s,\ell-\frac{1}{2}}$ with $s\geq \ell$.
However, due to the Laguerre polynomial identity 
\begin{eqnarray}
&\frac{(-\eta)^{s}}{s!}L_{\ell}^{(s-\ell)}(\eta)=\frac{(-\eta)^{\ell}}{\ell!}L_{s}^{(\ell-s)}(\eta)\,,&
\end{eqnarray}
one can deduce that $\psi_{s-\ell,\ell-\frac{1}{2}}\propto \phi_{s,\ell-\frac{1}{2}}$,  and 
so, such a double degeneracy does not really exist. 
For a similar  phenomenon observed earlier  in the 
Darboux transformations of the AFF model 
see ref. \cite{InzPly3}.
On the other hand, when $\nu=m$ one has    
$\delta E=\omega(2m+1)$. In this case  the levels of the  tower $E_{n,m}$ appear 
in the middle between the levels corresponding 
to $\mathcal{E}_{n,m}$, and the resulting spectrum is divided in two parts.
 One 
part corresponds to a semi-infinite equidistant part with energy  levels
separated by  $\omega$.  In another,  finite  part,  equidistant  separation between energy levels  is 
$\Delta E=2\omega$, see Figs. 
\ref{fig1cSec3}-\ref{fig1eSec3}. 

The action of the operators $\hat{\mathcal{J}}_{\alpha,\pm}$ is obtained via the application of $\hat{\mathfrak{S}}_\alpha$,
that yields 
\be
\hat{\mathcal{J}}_{\alpha,\pm}\psi_{n,\nu}^{\alpha}\propto \psi_{n\pm 1,\nu}^{\alpha}\,,\qquad
\hat{\mathcal{J}}_{\alpha,\pm}\phi_{n,\nu}^{\alpha}\propto \phi_{n\pm 1,\nu}^{\alpha}\,,\qquad 
\ker{\hat{\mathcal{J}}_{\alpha,-}}=\text{span}\{\psi_{0,\nu}^{\alpha},\phi_{0,\nu}^{\alpha}\}\,. 
\ee
This tells us that the states associated to each tower of energy levels can be produced by the $\mathfrak{sl}(2,\R)$
generators starting from any fixed state, and  also shows
 that there is no way to relate the states from the two towers 
 when $\nu$ is not integer.
In the integer case $\nu=m$ we have the operators 
\begin{eqnarray}
&
\hat{\mathcal{A}}_{\alpha,m}=\hat{\mathfrak{S}}_{\alpha}(\hat{\mathcal{P}}_{\alpha,m})\hat{\mathfrak{S}}_{\alpha}^{-1}
\,,\qquad 
\hat{\mathcal{B}}_{\alpha,m}=
\hat{\mathfrak{S}}_{\alpha}(\hat{\mathcal{X}}_{\alpha,m})\hat{\mathfrak{S}}_{\alpha}^{-1}\,,
&\\&
[\hat{H}_{\alpha,m}^{AFF},\hat{\mathcal{A}}_{\alpha,m}]=-\omega(2m+1)\hat{\mathcal{A}}_{\alpha,m}\,,\qquad
[\hat{H}_{\alpha,m}^{AFF},\hat{\mathcal{B}}_{\alpha,m}]=-\omega(2m-1)\hat{\mathcal{B}}_{\alpha,m}\,.
&
\end{eqnarray}
With the help of  the CBT, one learns that  the operator 
$\hat{\mathcal{A}}_{\alpha,m}$ annihilates the states $\phi_{j,m}$ with $j=0,1,\ldots,m$,
while the  operator 
$\hat{\mathcal{B}}_{\alpha,m}$ annihilates the states $\phi_{l,m}$ with $l=0,1,\ldots,m-1$.
Among  these states we have all the 
 eigenfunctions corresponding to the part separated from  the infinite equidistant part of the spectrum. 
 One finds also that they effectively 
relate the states of one tower with the states of another, 
\begin{eqnarray}
&
\hat{\mathcal{A}}_{\alpha,m}\phi_{n,m}^{\alpha}\propto \psi_{n-2m-1,m}^{\alpha}\,,\qquad
\hat{\mathcal{B}}_{\alpha,m}\phi_{n,m}^{\alpha}\propto\psi_{n-2m,m}^{\alpha}\,,&\\&
\hat{\mathcal{A}}_{\alpha,m}\psi_{n,m}^{\alpha}\propto \phi_{n,m}^{\alpha}\,,\qquad
\hat{\mathcal{B}}_{\alpha,m}\psi_{n,m}^{\alpha}\propto\phi_{n,m}^{\alpha}\,. 
&
\end{eqnarray}

Finally, via application of $\hat{\mathfrak{S}}_\alpha$ to  eigenstates of $\hat{H}_{\alpha,\nu}$
we get the overcomplete set of coherent states by allowing $\kappa$ to take  complex values,
\begin{eqnarray}
&\hat{\mathfrak{S}}_\alpha\psi_{\kappa,\nu}^{\alpha}(\frac{1}{\sqrt{2}}x)=
2^{\frac{1}{4}}e^{-\frac{1}{2}(x+i\alpha)^2+\frac{1}{4}\kappa^2}
\sqrt{x+i\alpha}\,{J}_{\nu+1/2}(\kappa (x+i\alpha)):=
\Psi_\nu^\alpha(x,\kappa)\,,&\\
&\hat{\mathfrak{S}}_\alpha\phi_{\kappa,\nu}^{\alpha}(\frac{1}{\sqrt{2}}x)=
2^{\frac{1}{4}}e^{-\frac{1}{2}(x+i\alpha)^2+\frac{1}{4}\kappa^2}
\sqrt{x+i\alpha}\,{Y}_{\nu+1/2}(\kappa (x+i\alpha)):=\Phi_\nu^{\alpha}(x,\kappa)\,.&
\end{eqnarray}

Considering the issue of the inner product, one  could try to take it in the form 
$\bra{\chi_1}\hat{I}_{-2\alpha}\ket{\chi_2}$, and then we obtain 
$\bra{\psi_{n,\nu}^{\alpha}}\hat{I}_{-2\alpha}\ket{\psi_{n,\nu}^{\alpha}}\propto
\bra{\psi_{n,\nu}^{0}}\ket{\psi_{n,\nu}^{0}}$. 
However the quantity $\bra{\phi_{n,\nu}^{\alpha}}\hat{I}_{-2\alpha}\ket{\phi_{n,\nu}^{\alpha}}$
diverges when $\nu\not=0$, since functions $\phi_{n,\nu}^{0}$ are singular in the origin, 
see Eq. (\ref{LaguerreKLein4}).
Note, however, here that in the case of $\mathcal{PT}$-symmetric harmonic oscillator 
($\nu=0$) we do not have these problems since 
the  states 
$\phi_{n,0}^{0}$ can be written in terms of  even Hermite polynomials, 
which do not have singularities in the real line.
So, the interesting open problem
 is to find a Hermitian system  $\hat{\mathcal{H}}_\nu$, defined on 
the entire real line,  with the same spectrum of 
$\hat{H}_{\alpha,\nu}$,  and an operator $\hat{\mathcal{O}}_\alpha$ such that 
  $\hat{\mathcal{O}}_\alpha\hat{\mathcal{H}}_{\nu}\hat{\mathcal{O}}_\alpha^{-1}=\hat{H}_{\alpha,\nu}$. 
  One can expect that such a  Hamiltonian will be   a non-local operator
  of the nature similar to that  considered in \cite{hidsusypara},
  see also \cite{AssisFring2}.
 
\subsection{A $\mathcal{PT}$-symmetric reinterpretation of CBT}
\label{SecReinPT}
According to  \cite{PTmetric1,PTmetric2}, the $\mathcal{PT}$ symmetric  Hamiltonian operator
\be
\label{Sw-H}
\mathscr{H}_{\alpha,\beta,\gamma}=\alpha \hat{a}^{+}\hat{a}^{-}+\beta (\hat{a}^-)^{2}+\gamma (\hat{a}^+)^{2} \,,\qquad 
x\in \R\,,
\ee
where $\hat{a}^+$ and $\hat{a}^-$ are the Hermitian conjugate raising and lowering 
ladder operators (\ref{d-lader1}), 
is characterized by  a purely  real spectrum if the real parameters $\alpha$, $\beta$ and $\gamma$ satisfy  
the relation $\alpha^2-4\beta\gamma\geq 0$. In the particular case in which 
$\alpha=0$, $\beta=\omega$ and $\gamma=-\omega$,  this operator takes the form 
\be\label{Sw-H-p}
\hat{\mathscr{H}}_{0,\omega,-\omega}=2i\omega\hat{D} \,,
\ee
where $\hat{D}$ corresponds to the dilatation operator appearing in (\ref{1dfreeparticlegen}).
The eigenstates of this Hamiltonian are the functions $\Omega_n$ 
presented  in Eq. (\ref{JordanOsci}), and the eigenvalue problem  corresponds 
to the third equation in  (\ref{HosinOm}). From Eq.  (\ref{JordanTransformed}) 
one deduces that the CBT generator $\hat{\mathfrak{S}}$ works
 as the operator that relates the Hamiltonian 
(\ref{Sw-H-p}) with the harmonic oscillator system described by 
the  Hermitian Hamiltonian. 
Indeed, from equations (\ref{GenConBr}) we obtain the  
$\mathcal{PT}$-symmetric conjugation 
\be
\label{PT-Conjugation}
 \hat{\mathscr{H}}_{0,\omega,-\omega}\hat{\Theta}=  
 \hat{\mathscr{H}}_{0,\omega,-\omega}^\dagger\hat{\Theta}\,,\qquad
 \hat{\Theta}=(\hat{\mathfrak{S}})^{2}\,,
\ee
implying that the $\mathcal{PT}$-symmetric normalization of the eigenstates $\Omega_{n}$ is just 
equivalent to  normalization of eigenstates of the quantum harmonic oscillator
under the usual inner product in $\R$. 
In the same vein, we note that 
at the classical level, the time evolution produced by the Hamiltonian
$\mathscr{H}_{0,\omega,-\omega}=2i\omega D$ in the variables $x$ and 
$p$ is governed by the equations  
$
\dot{x}=\{x,\mathscr{H}_{0,\omega,-\omega}\}=i\omega x$ and  
$\dot{p}=\{p,\mathscr{H}_{0,\omega,-\omega}\}=-i\omega p$,
which resemble  the equations of motion of 
  the classical analog $a^+$ and $a^-$ of the first order ladder operators 
of the harmonic oscillator system. Off course, both systems are related to each 
other by the classical version of the conformal bridge transformation reviewed in  Section
\ref{SecClassicalCBT}.

The  model (\ref{Sw-H}) can be generalised up to a
concrete realization of the $\mathfrak{so}(2,1)\cong\mathfrak{sl}(2,\R)$
generators, by changing the harmonic oscillator 
operators by 
\begin{eqnarray}
&
\hat{\mathcal{H}}_{\alpha,\beta,\gamma}=\alpha \hat{\mathcal{J}}_0+
\beta \hat{\mathcal{J}}_{-}+\gamma \hat{\mathcal{J}}_{+} 
=\frac{\alpha-\beta-\gamma}{2\omega}\hat{H}+
\frac{\omega(\alpha+\beta+\gamma)}{2}\hat{K}+i(\beta-\gamma)\hat{D}\,.&
\label{Sw-H-g}
\end{eqnarray}
Relation (\ref{PT-Conjugation}) holds since it is true by the conformal algebraic arguments. 
This also means that 
the $\mathcal{PT}$-symmetric normalization of the considered physical states of the system, 
which are also the rank 
$[n/2]+1$ Jordan states of zero energy of $\hat{H}$, correspond to the normalization of the eigenstates of the system 
given by the Hamiltonian $\hat{\mathcal{J}}_0$. 
In this way, if we select the realization (\ref{HermitianCal}), where the generators  are defined 
on $\R^+$, the physical eigenstates of (\ref{Sw-H-g}) 
with $\alpha=0$,  $\beta=-\gamma$ correspond to  (\ref{Omega1}),  
and the corresponding eigenvalue equation is given  by 
Eq. (\ref{actionHKAFF2}).

\section{CBT for higher-dimensional Euclidean systems}
\label{SecCBT2d}

Let us start with  realization  of the 
 operators $ \hat{H}$, $ \hat{D} $ and $ \hat{K} $ in higher-dimensional systems. 
 In the simplest case of  a  free particle in $\R^d$, the generators of its  
 $ \mathfrak{so}(2,1)$ conformal symmetry are given by
 (in this section we restore  the dimensional constants):
\begin{eqnarray}\label{HDKd}
&
\hat{H}=\sum_{i=1}^{d}\hat{H}_i\,,\qquad
\hat{D}=\sum_{i=1}^{d}\hat{D}_i\,,\qquad
\hat{K}=\sum_{i=1}^{d}\hat{K}_i\,,\label{TotalSo(2,1)}
&\\&
\hat{H}_i=\frac{-\hbar^2}{2m}\frac{\partial^2}{\partial x_i^{2}}\,,\qquad
\hat{D}_i=-i\frac{\hbar}{2}\left(x_i\frac{\partial}{\partial x_i}+\frac{1}{2}\right)\,,\qquad
\hat{K}_i=\frac{m}{2}x_i^2\,.\label{PartialSo(2,1)}
\end{eqnarray} 
These symmetry generators are complemented by the Heisenberg algebra generators
\begin{eqnarray}
\label{FreePHe}&
\hat{p}_j=-i\hbar\frac{\partial}{\partial x_j}\,,\qquad
\hat{\xi}_{j}= m x_j\,,\qquad
[\hat{\xi}_j,\hat{\xi}_k]=[\hat{p}_j,\hat{p}_k]=0\,,\qquad
[\hat{\xi}_j,\hat{p}_k]=i\hbar m \delta_{jk}\,,&
\end{eqnarray}  
and the angular momentum tensor 
$
\hat{M}_{ij}=\frac{1}{m}(\hat{\xi}_i\hat{p}_j-\hat{\xi}_j\hat{p}_i)\,.
$
Together, all these  generators produce a 
$d$-dimensional Schr\"odinger symmetry  of a free particle \cite{Nied1}. 

The commutation relations in (\ref{FreePHe}) imply that different conformal 
bridge transformations can be applied for each spatial direction, and each 
of them works in the same way as in the one-dimensional case
considered in  Sec. \ref{CBTExp1}. 

First, let us consider  the  isotropic CBT  produced  by the operator
\be
\label{MultCBT}
\hat{\mathfrak{S}}=\Pi_{i=1}^{d}\hat{\mathfrak{S}}_i\,,\qquad
\hat{\mathfrak{S}}_i=e^{-\frac{\omega}{\hbar} \hat{K}_i} 
e^{\frac{\hat{H}_i}{2\hbar \omega}}
e^{\frac{i}{\hbar}\ln(2)\hat{D}_i}\,,\qquad
[\hat{\mathfrak{S}}_i,\hat{\mathfrak{S}}_j]=0\,.
\ee
This  operator generates a composed  CBT with  equal  frequencies in each direction. 
For this reason, it commutes with the angular momentum tensor, and so, 
 is rotationally invariant. 
 Applying the similarity transformation given by (\ref{MultCBT}) to the free particle,
we obtain, in accordance with Eqs. (\ref{GenConBr}) and (\ref{JJJ}), 
 the $d$-dimensional isotropic  harmonic oscillator, with    the $\mathfrak{sl}(2,\R)$ 
 generators  $\hat{\mathcal{J}}_\pm$ to be quadratic radial ladder operators.
 Also one gets 
\begin{eqnarray}\label{pxapm}
&
\hat{\mathfrak{S}}(\hat{p}_j)\hat{\mathfrak{S}}^{-1}=
-i\sqrt{m\hbar\omega}\,\hat{a}_j^-\,,\qquad 
\hat{\mathfrak{S}}(\hat{\xi}_{j})\hat{\mathfrak{S}}^{-1}=
\sqrt{\frac{m\hbar}{\omega}}\,\hat{a}_j^+\,,
&
\end{eqnarray}
where  $\hat{a}_i^+$
are the first order ladder operators for each direction, 
\begin{eqnarray}
&\label{d-lader-espacial}
\hat{a}_i^\pm=\sqrt{\frac{m\omega}{2\hbar}}\left(x_i \mp
 \frac{\hbar}{m\omega}\frac{\partial}{\partial x_i}\right)\,,\qquad
[\hat{a}_i^\pm,\hat{a}_j^\pm]=0\,,\qquad [\hat{a}_i^-,\hat{a}_j^+]=\delta_{ij}\,.
&
\end{eqnarray}

The second option is the anisotropic CBT  composed from generators 
with different values of frequencies $\omega_i>0$,
\be
\label{AniMultCBT}
\hat{\mathfrak{S}}_{\omega_1,\ldots\omega_d}=\Pi_{i=1}^{d}\hat{\mathfrak{S}}_{\omega_i}\,,\qquad
\hat{\mathfrak{S}}_{\omega_i}=e^{-\frac{\omega_i}{\hbar} \hat{K}_i} 
e^{\frac{\hat{H}_i}{2\hbar \omega_i}}
e^{\frac{i}{\hbar}\ln(2)\hat{D}_i}\,,\qquad
[\hat{\mathfrak{S}}_{\omega_i},\hat{\mathfrak{S}}_{\omega_j}]=0\,.
\ee
Via the similarity transformation, this operator and its inverse transform the linear combination 
$
2i\hat{\mathcal{D}}^{\epsilon_1,\ldots,\epsilon_d}_{\omega_1,\ldots,\omega_d}=
2i\sum_{i=1}^{d}\omega_i\epsilon_{i}\hat{D}_i
$
into the $d$-dimensional anisotropic oscillator 
Hamiltonian $
\hat{H}_{\omega_1,\ldots,\omega_d}^{\epsilon_1,\ldots,\epsilon_d}=
\sum_{i}^{d} \epsilon_{i} \hat{H}_{\text{os}}^{\omega_i}$ with 
$\hat{H}_{\text{os}}^{\omega_i}=
\hbar\omega_i(\hat{a}_{\omega_i}^{+}\hat{a}_{\omega_i}^{-}+\frac{1}{2})
$, and $
\hat{a}_{\omega_i}^\pm=\sqrt{\frac{m\omega_i}{2\hbar}}
\left(x_i \mp \frac{\hbar}{m\omega_i}\frac{\partial}{\partial x_i}\right),
$
where each $\epsilon_i$ can be chosen as $1$ or $-1$.
On the other hand, up to multiplicative constants, one also gets
$\hat{\mathfrak{S}}_{\omega_1,\ldots\omega_d}:(\hat{\xi}_i,\hat{p}_i)
\rightarrow (\hat{a}_{\omega_i}^+,\hat{a}_{\omega_i}^-)$.

Systems in $d$  Euclidean dimensions,   such as the free particle 
and isotropic or anisotropic harmonic oscillators,  
can have hidden symmetries 
generated by higher order  integrals of  motion \cite{Cariglia}. 
 Since these integrals are always written in terms of $ \hat{x} $ and $ \hat{p} $ 
 (or in terms of (\ref{d-lader-espacial})),  the 
 CBT maps the symmetry generators of one system into those of another. 
 In practical terms, application of  the CBT to
  the operators that commute with $ 2i \hat{D} $ ($ 2i \hat{\mathcal{D}}$) produces 
 the Hamiltonian symmetries of the isotropic  (anisotropic) case.
This scheme allows a 
reinterpretation from the perspective of  $ \mathcal{PT} $ symmetry, 
in the spirit of Sec. \ref{SecReinPT},
since the operators $ 2i \hat{D} $ and $ 2i \hat{\mathcal{D}} $ are
 a generalization of the Hamiltonian (\ref{Sw-H-p})  
 to $ d $-dimensions, in the isotropic and anisotropic case,
 respectively.

To see some concrete  applications of the higher dimensional 
CBT in detail, we will consider only the case $ d = 2 $. 
In particular, we focus our attention on the 
exotic rotationally invariant harmonic oscillator (ERIHO)  system \cite{InzPly8}, 
which corresponds to the 
planar isotropic harmonic oscillator extended by a Zeeman type  term. 
This model is  
generated by 
the isotropic CBT by generalizing the already considered constructions. 
It represents   a one parametric family of systems revealing  different phases, 
two of which correspond to  the Landau problem. Additionally,
in \cite{InzPly8} it also was shown that in spite of the  explicit rotationally 
invariant nature, the model  is unitary 
equivalent to the planar anisotropic harmonic  oscillator (AHO)
via the application of a  certain  $\mathfrak{su}(2)$ rotation
accompanied by an anisotropic $\mathfrak{so}(1,1)$ Bogolyubov  
transformation \cite{Bogol}.

\subsection{The ERIHO system:  classical case}
\label{SecRIAHO}
Starting from the two-dimensional free particles system, 
let us consider the following complex combination of 
its symmetry generators 
\begin{eqnarray}
&\label{AnotherSw}
2iD_{0}+gp_\varphi=x_j\Delta_{jk}p_k \qquad g \in \R\,, &\\&
\label{DeltaTe}
\Delta_{jk}=i\delta_{jk}+g\epsilon_{jk}\,,\qquad 
\Delta_{jk}\Delta_{jl}=(g^2-1)\delta_{jl}\,,\qquad 
\det \Delta=g^2-1\,.
\end{eqnarray}

The generator $ p_\varphi = M_ {12}=\epsilon_{ij}x_ip_j $ of $\mathfrak{so}(2)$ 
rotations  is invariant
under the classical isotropic CBT. As a result,  (\ref{AnotherSw}), 
multiplied by $\omega$,  is transformed into
the classical Hamiltonian of the ERIHO system, 
\begin{eqnarray}
\label{gHamil}
&H_{g}=H_{\text{osc}}+ g\omega  p_\varphi\,,\qquad
H_{\text{osc}}=\frac{1}{2m}p_ip_i+\frac{1}{2}m\omega^2 x_ix_i\,.&
\end{eqnarray}
System (\ref{gHamil})   admits the following three different
 physical interpretations \cite{InzPly8}.
 
First,  $H_g$  corresponds to the Hamiltonian of a planar
particle in a non-inertial frame  rotating with angular velocity 
$\Omega=g\omega$ and  subjected to the action of 
the  isotropic harmonic trap $U=\frac{1}{2}k x_ix_i$.
The cases $k>m\Omega^2$, $k=m\Omega^2$ and 
$0<k<m\Omega^2$ correspond, respectively, 
 to the phases $0<g^2<1$, $g^2=1$ and $g^2>1$ of the system
  (\ref{gHamil}), while the inertial case $\Omega=0$, $k=m\omega^2$
  corresponds to the phase of  the  isotropic  oscillator 
  of (\ref{gHamil}) with $g=0$.
 
Second, in the cases $g=+1$ and $g=-1$,
(\ref{gHamil}) takes the form of the Hamiltonian of Landau problem 
in symmetric gauge with different orientation (sign) of the magnetic field $B$
and $\omega=g\omega_B$, $\omega_B=qB/2mc$, where $q$ is 
the charge of a particle.
Then the phases with $0\leq g^2<1$ and $g^2>1$ of (\ref{gHamil})
correspond to the extended Landau problem in the presence of  the 
additional harmonic potential term $\frac{1}{2}m\Lambda x_ix_i$
with $\Lambda>0$ and $0>\Lambda>-\omega_B^2$, respectively,
where $\omega=\sqrt{\Lambda+\omega_B^2}$, and $g=\omega_B/\omega$.
The repulsive critical, $\Lambda=-m\omega_B^2$, and supercritical,
$\Lambda<-m\omega^2$, cases of the extended Landau problem 
have no analogs in the system (\ref{gHamil}).

Finally, in terms of the classical analogues of the 
circular ladder operators, 
\begin{eqnarray}
\label{gen1}
&
b_1^-= \frac{1}{\sqrt{2}}(a_1^--ia_2^-)\,, \quad b_1^+=(b_1^-)^*\,,
\quad
b_2^-= \frac{1}{\sqrt{2}}(a_1^-+ia_2^-)\,,\quad b_2^+=(b_2^-)^*\,,
&\\&
a_i^\pm=\sqrt{\frac{m\omega}{2}}\left(\,x_i \mp \frac{i}{m\omega}\,p_i\right)\,,
\label{gen1+}&
\end{eqnarray}
Hamiltonian (\ref{gHamil}) takes the form 
\begin{eqnarray}
\label{gHamilbb+}
H_g= \omega\left(\ell_1 b_1^+b_1^-+\ell_2b_2^+b_2^- \right)\,,\qquad
\ell_1=1+g\,,\qquad \ell_2=1-g\,.
\end{eqnarray}
It looks like the anisotropic harmonic oscillator Hamiltonian, but  
system  (\ref{gHamil}) is manifestly rotational invariant.

In correspondence with relations (\ref{DeltaTe}) and the comments on different interpretations,
 it is expected that the system (\ref{gHamil}) 
  should have essentially different physical
properties and symmetries 
 in the cases 
$g^2<1$ and $g^2>1$, as well as when $g=\pm 1$. Indeed,
the system corresponds to  the planar  isotropic harmonic oscillator when $g=0$,
meanwhile, as it was already mentioned,
the model at $g=\pm 1$ represents  the Landau problem in the symmetric gauge. 
In the case $|g|<1$, the Hamiltonian  (\ref{gHamil}) formally looks like the  Euclidean AHO  
with different frequencies 
 $\omega_1\neq  \omega_2$,
$\omega_i=\ell_i\omega$, contrary to the case  
of  $|g|>1$, when  (\ref{gHamil}) has 
 instead the form of a Hamiltonian of the Minkowskian  AHO
 with frequencies  of two different signs.
This last family of systems 
resembles  the  Pais-Uhlenbeck  oscillator, which  recently attracted  a considerable  attention
in relation to the $\mathcal{PT}$-symmetry, see  Refs. \cite{BenMan,Smilga,MostPU}.
Finally, in the limit $g\rightarrow \infty$, one has 
\be
\label{MinwoskianAHO}
g^{-1}H_g\rightarrow\omega p_\varphi=\omega (b^+_1b^-_1- b^+_2b^-_2)=b_i^+\eta_{ij}b_j^-\,,\qquad
\eta=\text{diag}(1,-1),
\ee
which can be interpreted as  the isotropic 
Minkowskian 
oscillator.

From the point of view of the $\mathcal{PT}$ symmetry, 
the generator (\ref{AnotherSw}) is a generalization of the classical 
analogue of the $\mathcal {PT}$ invariant Hamiltonian of the form 
(\ref{Sw-H-p}), extended now  by the angular momentum taken 
with arbitrary coupling constant. Here,  the  
isotropic CBT provides us the transformation  that connects this system with its real 
(Hermitian in the quantum case) counterpart $ H_g $. 

By solving the equations of motion for 
$b_j^\pm$, $\dot{b}^\pm_j=\pm i \omega\ell_j b^\pm_j$, $j=1,2$,
and using the relation $\sqrt{m\omega}(x_1+ix_2)=b^+_1+b^-_2$, 
we get the trajectories of the system, 
\be\label{traj}
z(t)=x_1(t)+ix_2(t)=R_1e^{i\gamma_1}e^{i\omega \ell_1 t}+
R_2e^{-i\gamma_2}e^{-i\omega \ell_2 t}\,,
\ee
where  $R_i\geq 0$ and $\gamma_i\in \R$ are the  integration constants. 
 The energy and angular momentum  of the system are given by
 $E_g=m\omega^2(\ell_1R^2_1+\ell_2R_2^2)$, 
  $p_\varphi=m\omega(R_1^2-R_2^2)$.
Notice that for   $g^2<1$ 
the exponents in (\ref{traj}) evolve in opposite directions, while 
 in the case of $g^2>1$ they change  in the same direction that depends 
 on the sign of $g$.
On the other hand,  at $g=+1$ ($g=-1$),  one gets $\omega_2=0$  ($\omega_1=0$),
and the orbit is a circumference of radius $R_1$ ($R_2$) centered at 
$(X_1,X_2)$ with
$Z=X_1+iX_2=R_2e^{-i\gamma_2}$ ($Z=R_1e^{i\gamma_1}$).
 
\begin{figure}[hbt!]
\begin{center}
\begin{subfigure}[c]{0.28\linewidth}
\includegraphics[scale=0.3]{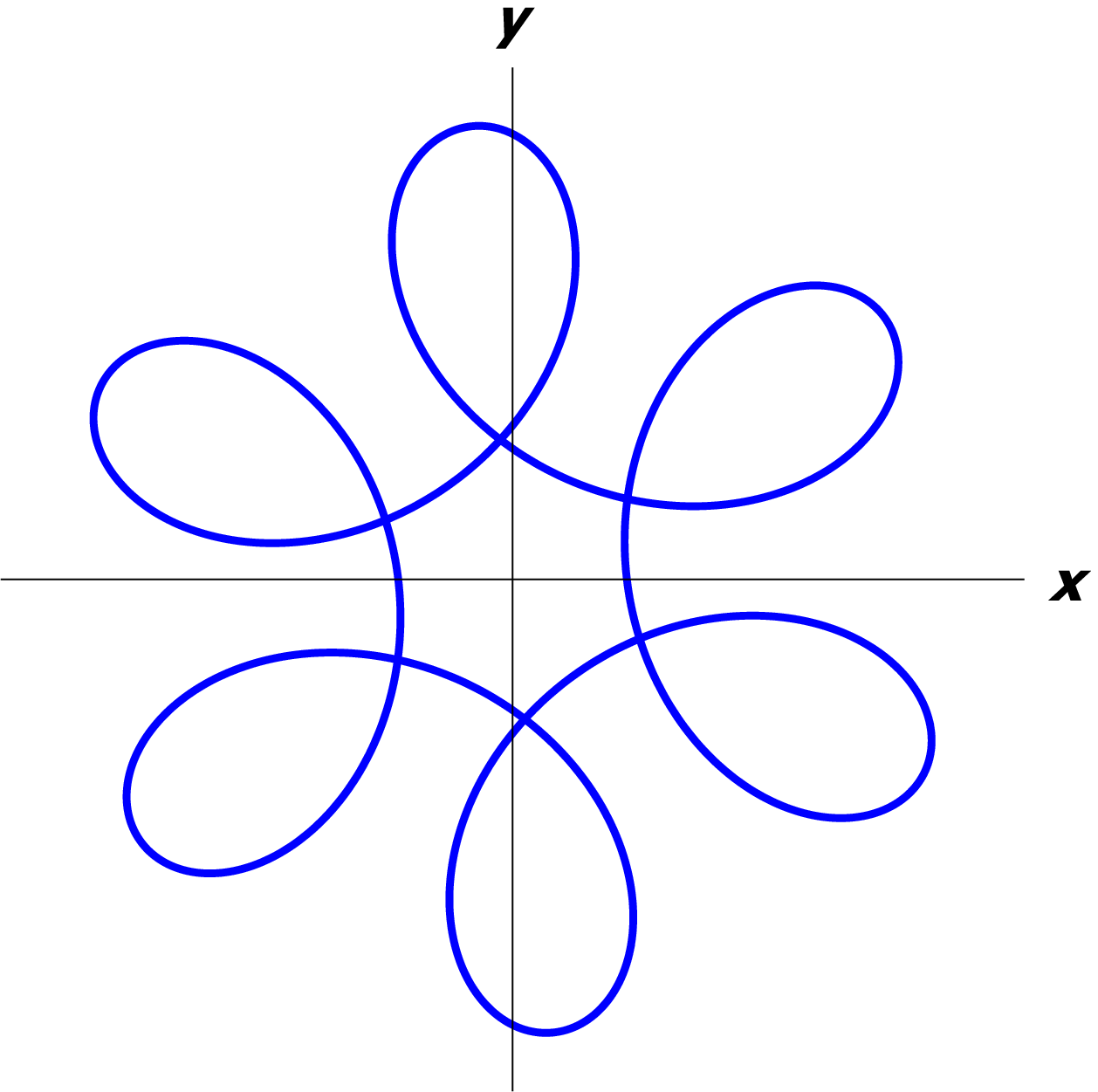}
\caption{\small{$g=2/3$, $R_1<R_2$}}
\end{subfigure}
\begin{subfigure}[c]{0.28\linewidth}
\includegraphics[scale=0.3]{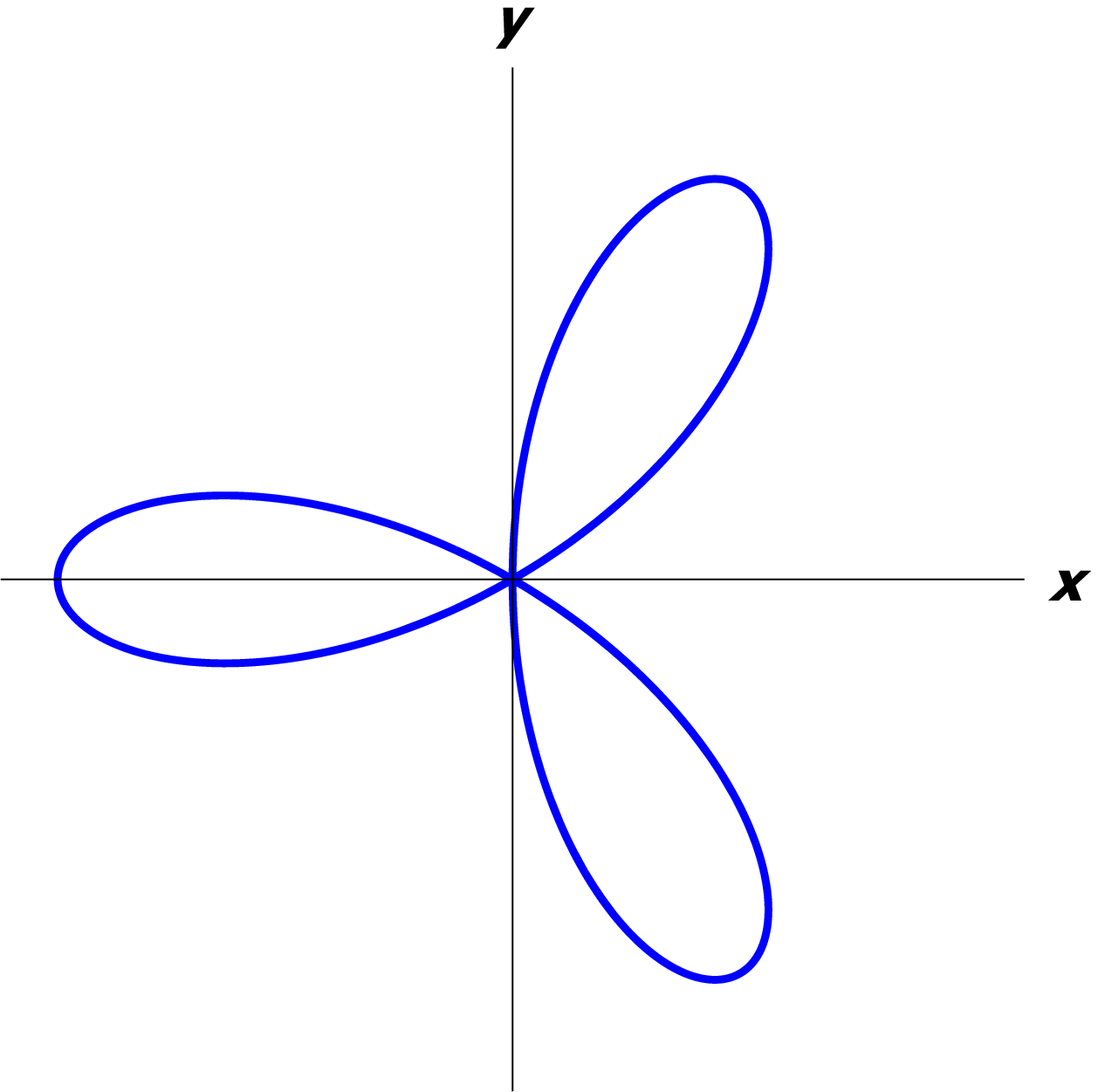}
\caption{\small{$g=1/3$, $R_1=R_2$}}
\end{subfigure}
\begin{subfigure}[c]{0.28\linewidth}
\includegraphics[scale=0.3]{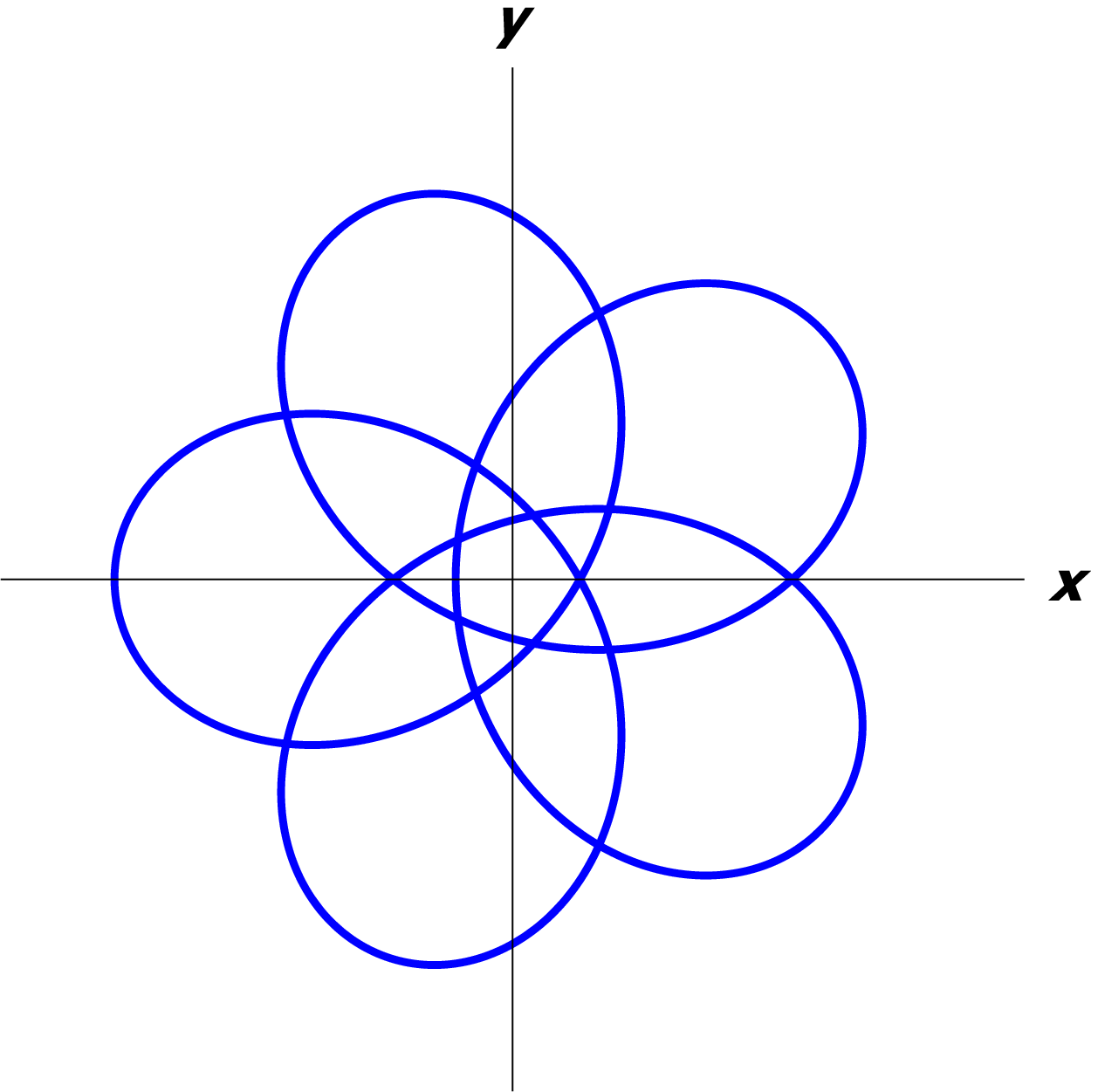}
\caption{\small{$g=3/5$, $R_1>R_2$}}
\end{subfigure}
\begin{subfigure}[c]{0.28\linewidth}
\includegraphics[scale=0.3]{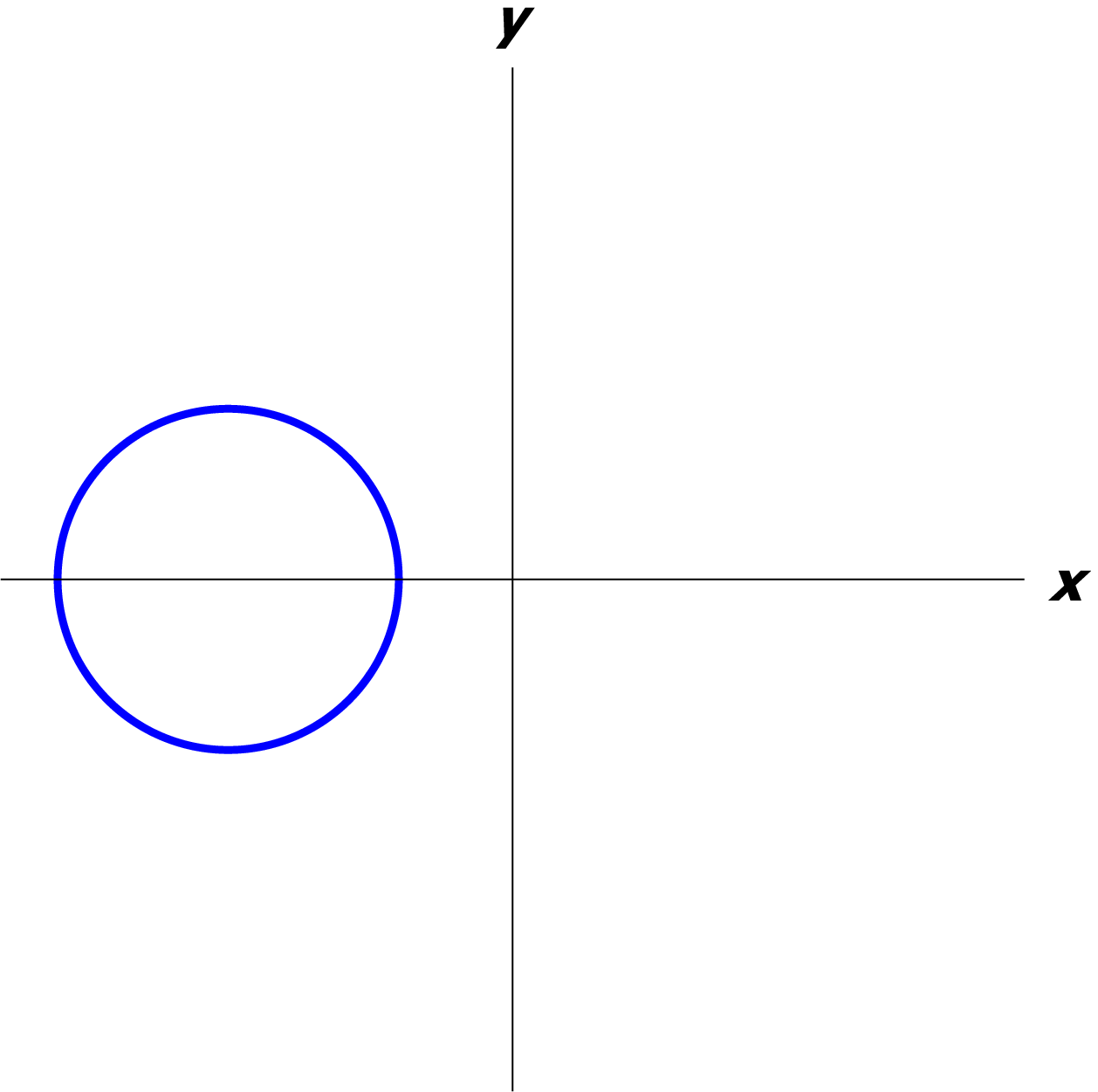}
\caption{\small{$g=1$, $R_1<R_2$}}
\end{subfigure}
\begin{subfigure}[c]{0.28\linewidth}
\includegraphics[scale=0.3]{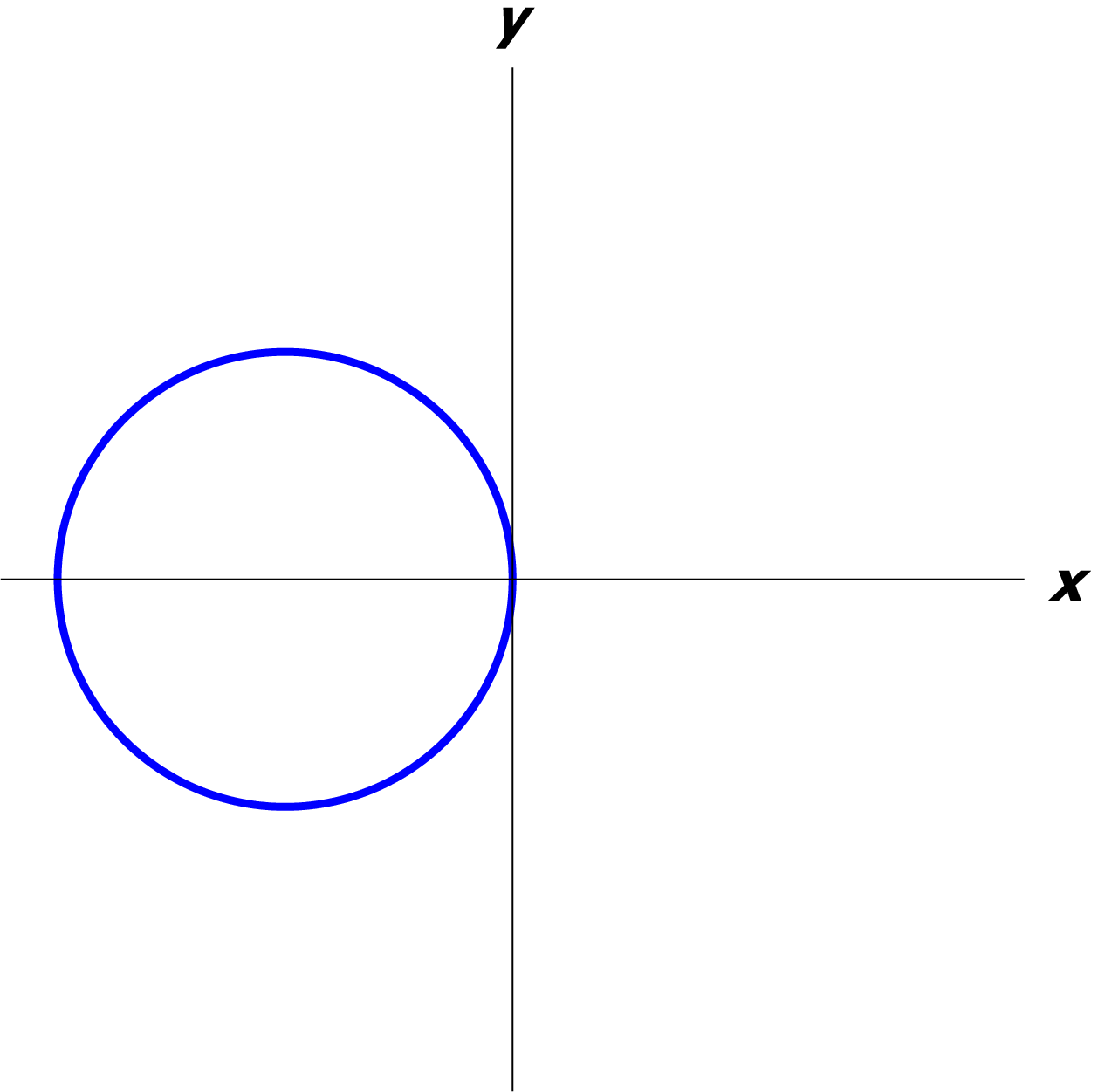}
\caption{\small{$g=1$, $R_1=R_2$}}
\end{subfigure}
\begin{subfigure}[c]{0.28\linewidth}
\includegraphics[scale=0.3]{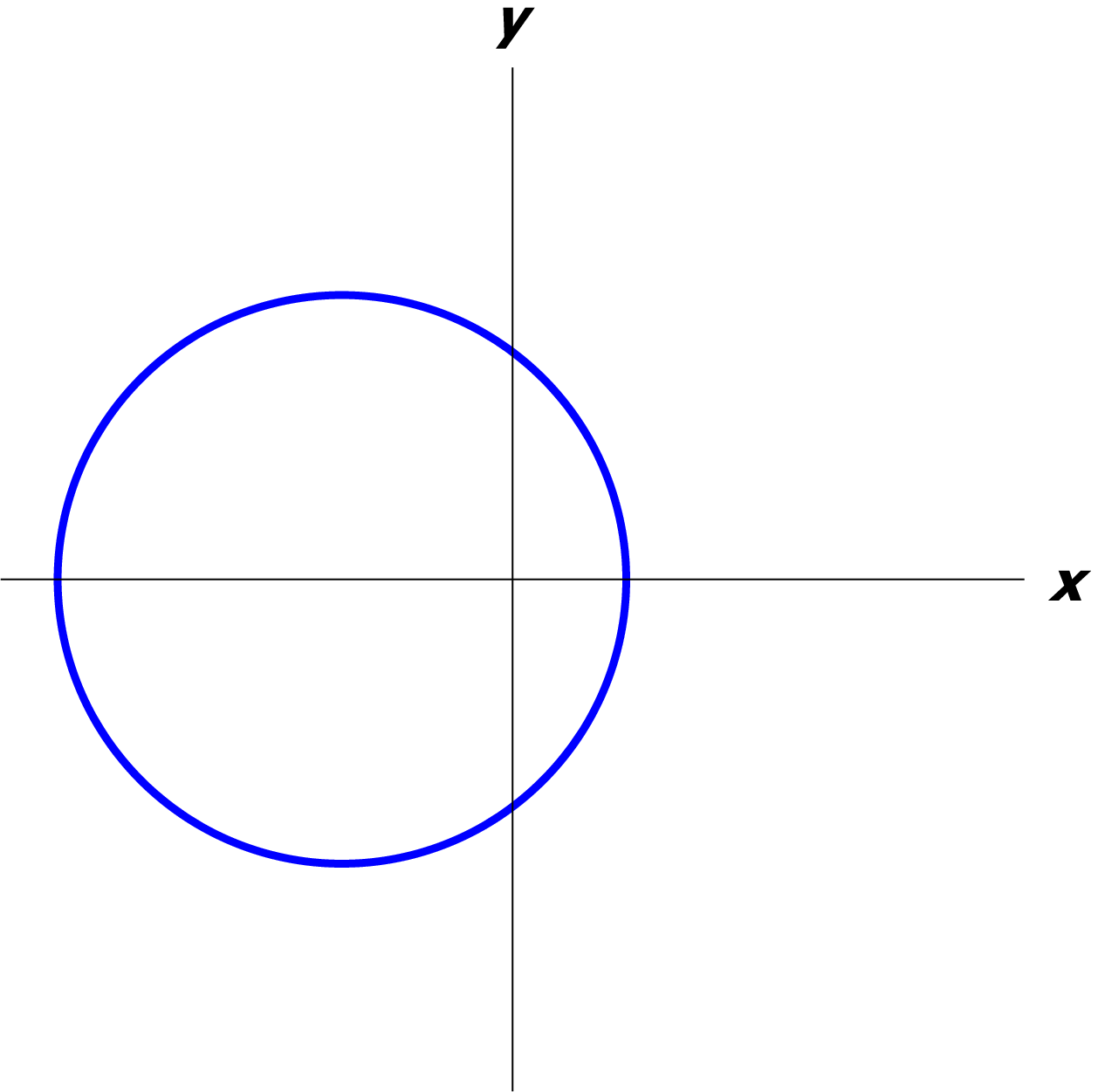}
\caption{\small{$g=1$, $R_1>R_2$}}
\label{Fig(f)}
\end{subfigure}
\begin{subfigure}[c]{0.28\linewidth}
\includegraphics[scale=0.3]{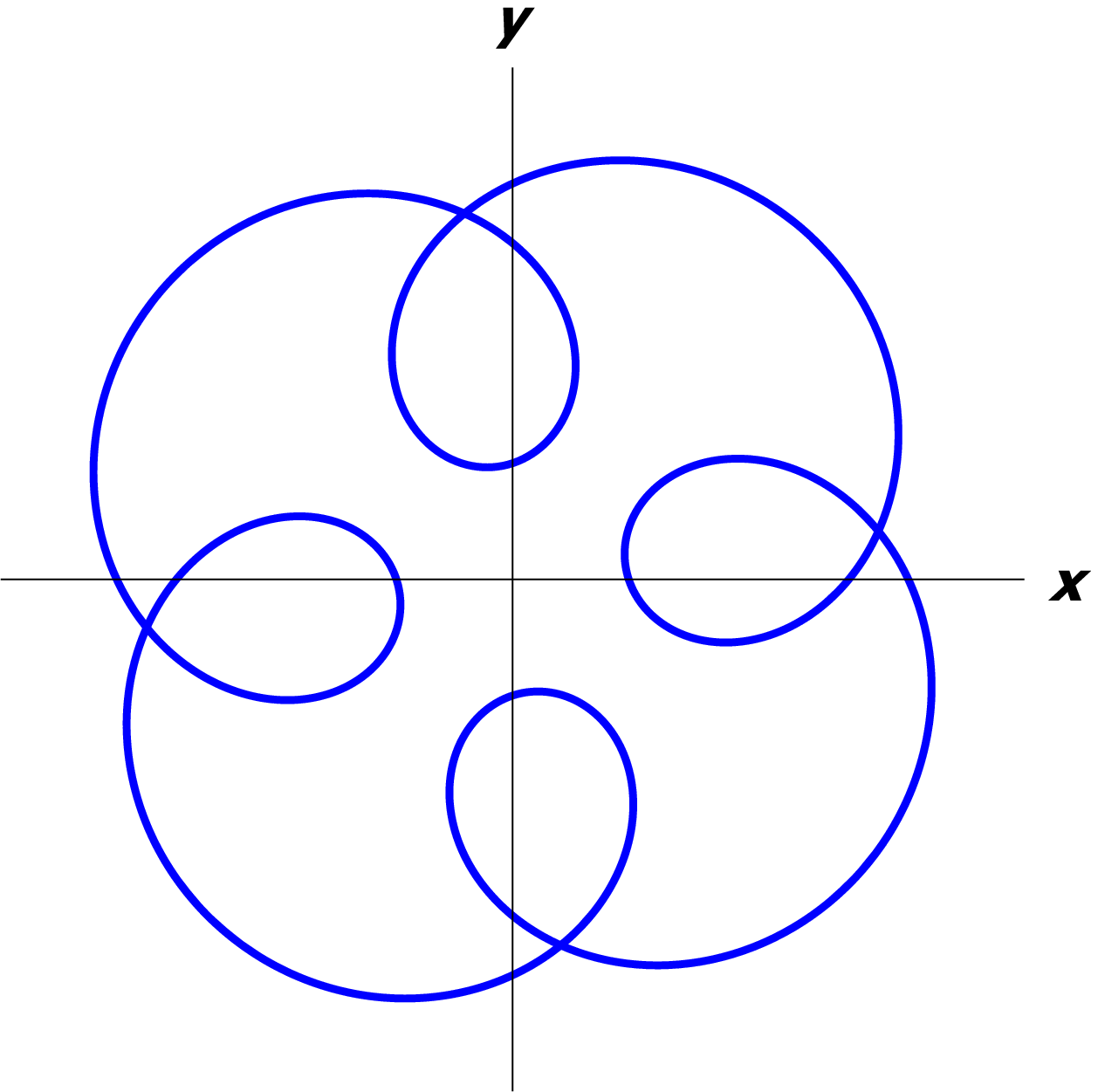}
\caption{\small{$g=3/2$, $R_1<R_2$}}
\end{subfigure}
\begin{subfigure}[c]{0.28\linewidth}
\includegraphics[scale=0.3]{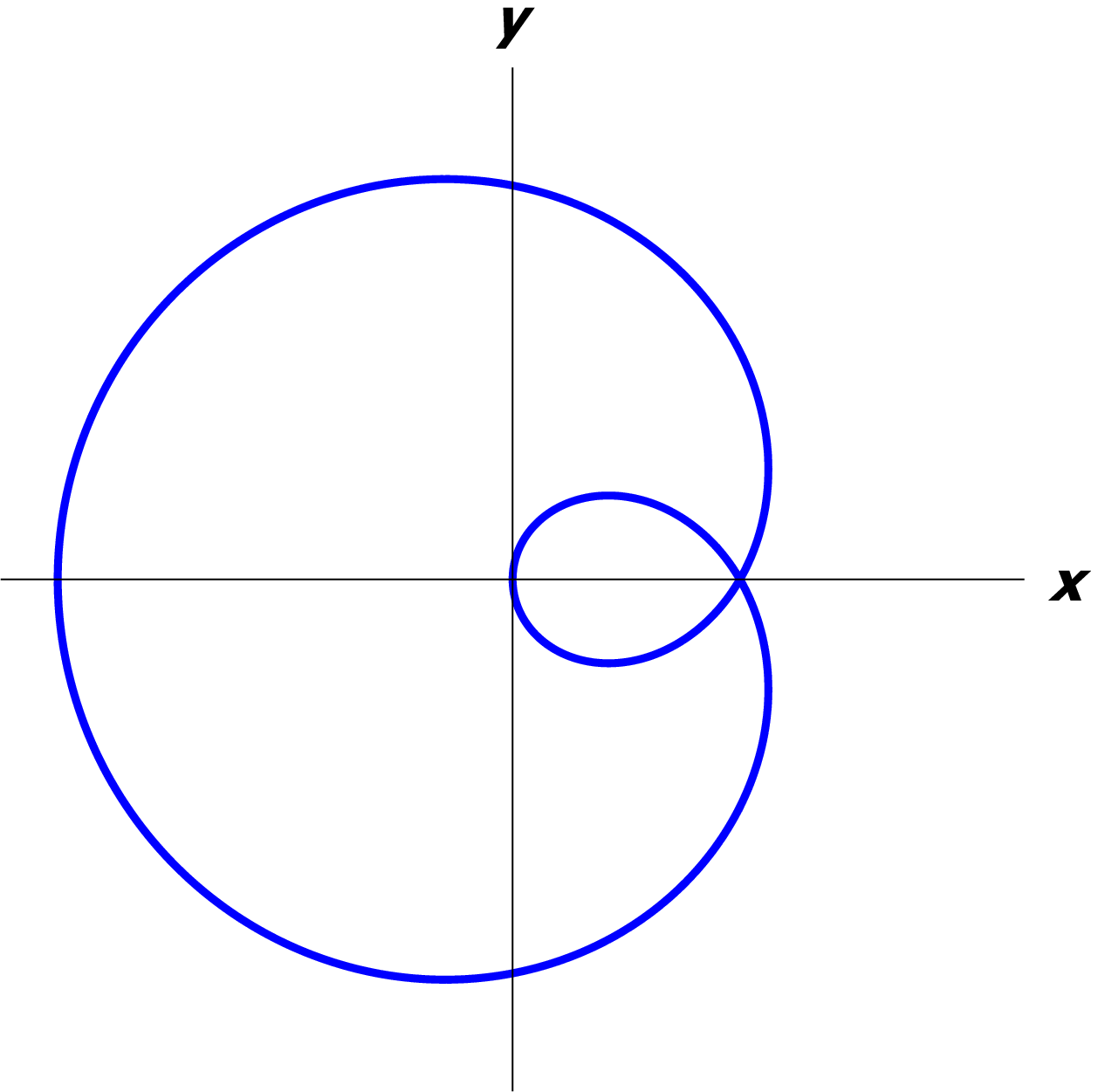}
\caption{\small{$g=3$, $R_1=R_2$}}
\end{subfigure}
\begin{subfigure}[c]{0.28\linewidth}
\includegraphics[scale=0.3]{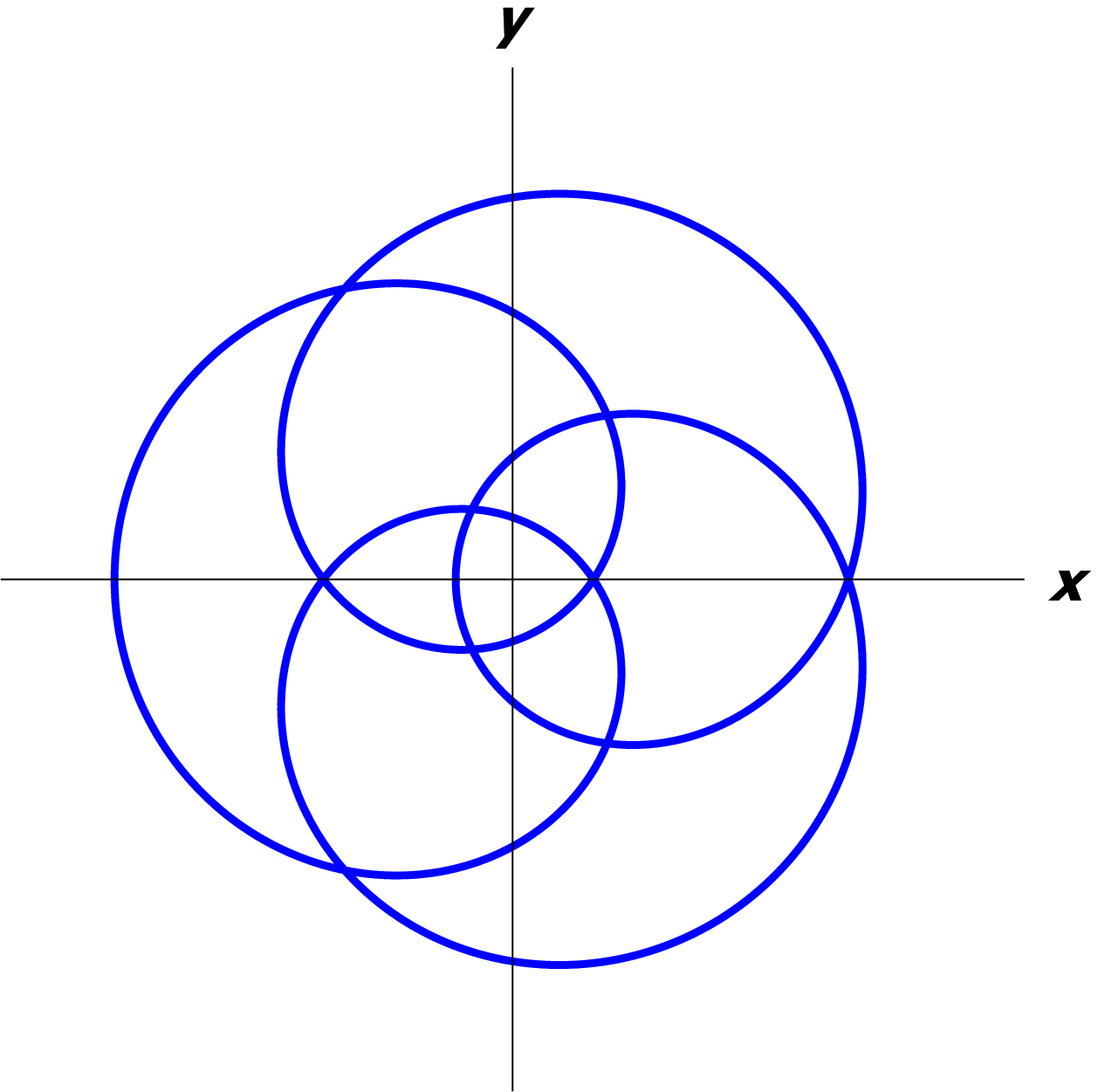}
\caption{\small{$g=5/3$, $R_1>R_2$}}
\end{subfigure}
\end{center}
\caption{\small{Trajectories for some rational values of $g$.
In cases b), e) and h),  $p_\varphi=0$ and trajectories pass through
 the origin. For $R_1\neq R_2$, $\text{sign}\,(p_\varphi)=\text{sign}\,(R_1-R_2)$.
}}
\label{Fig1}
\end{figure}

\begin{figure}[hbt!]
\begin{center}
\begin{subfigure}[c]{0.28\linewidth}
\includegraphics[scale=0.3]{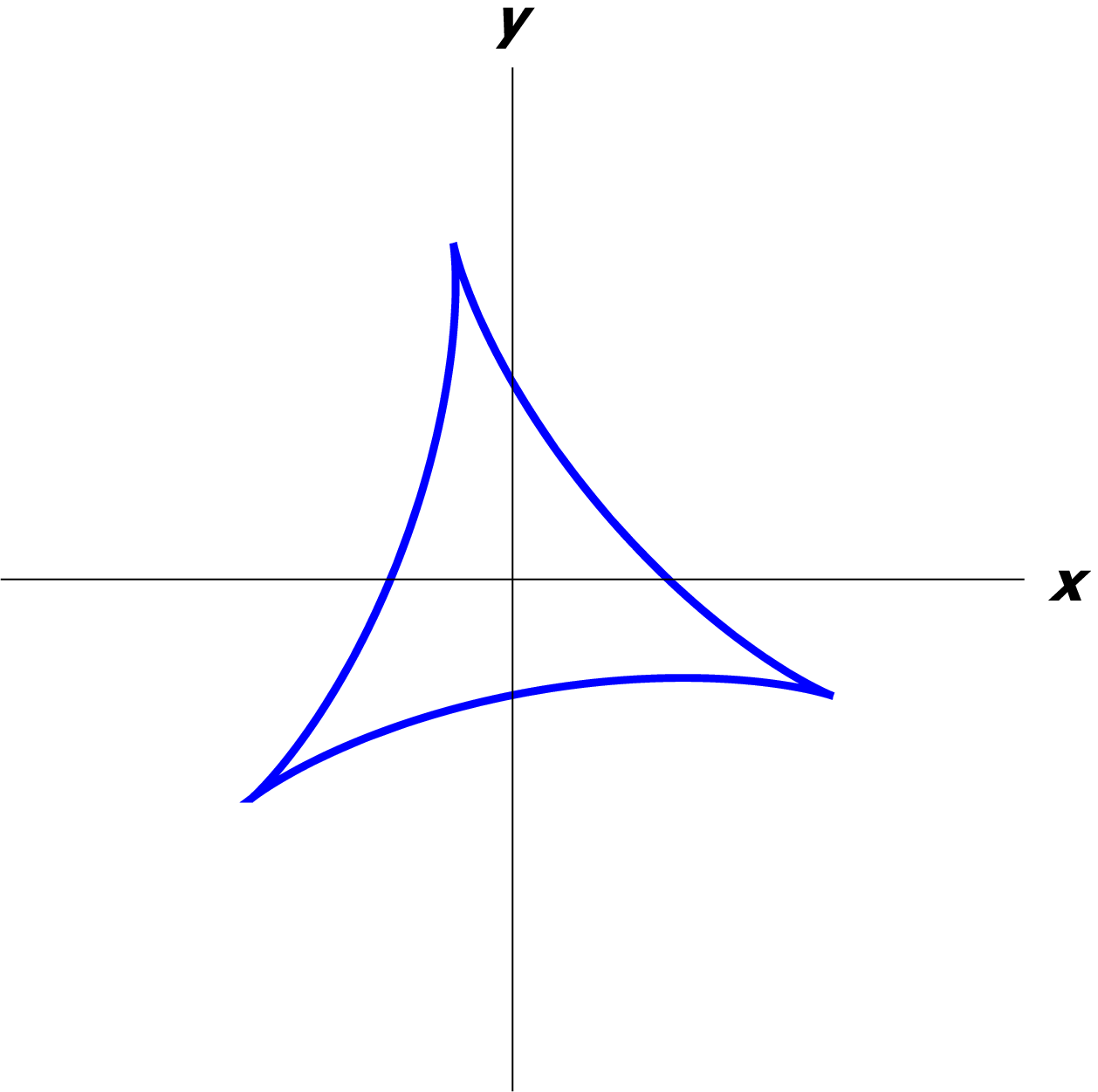}
\caption{\small{$g=1/3$, $R_2/R_1=2$}}
\end{subfigure}
\begin{subfigure}[c]{0.28\linewidth}
\includegraphics[scale=0.3]{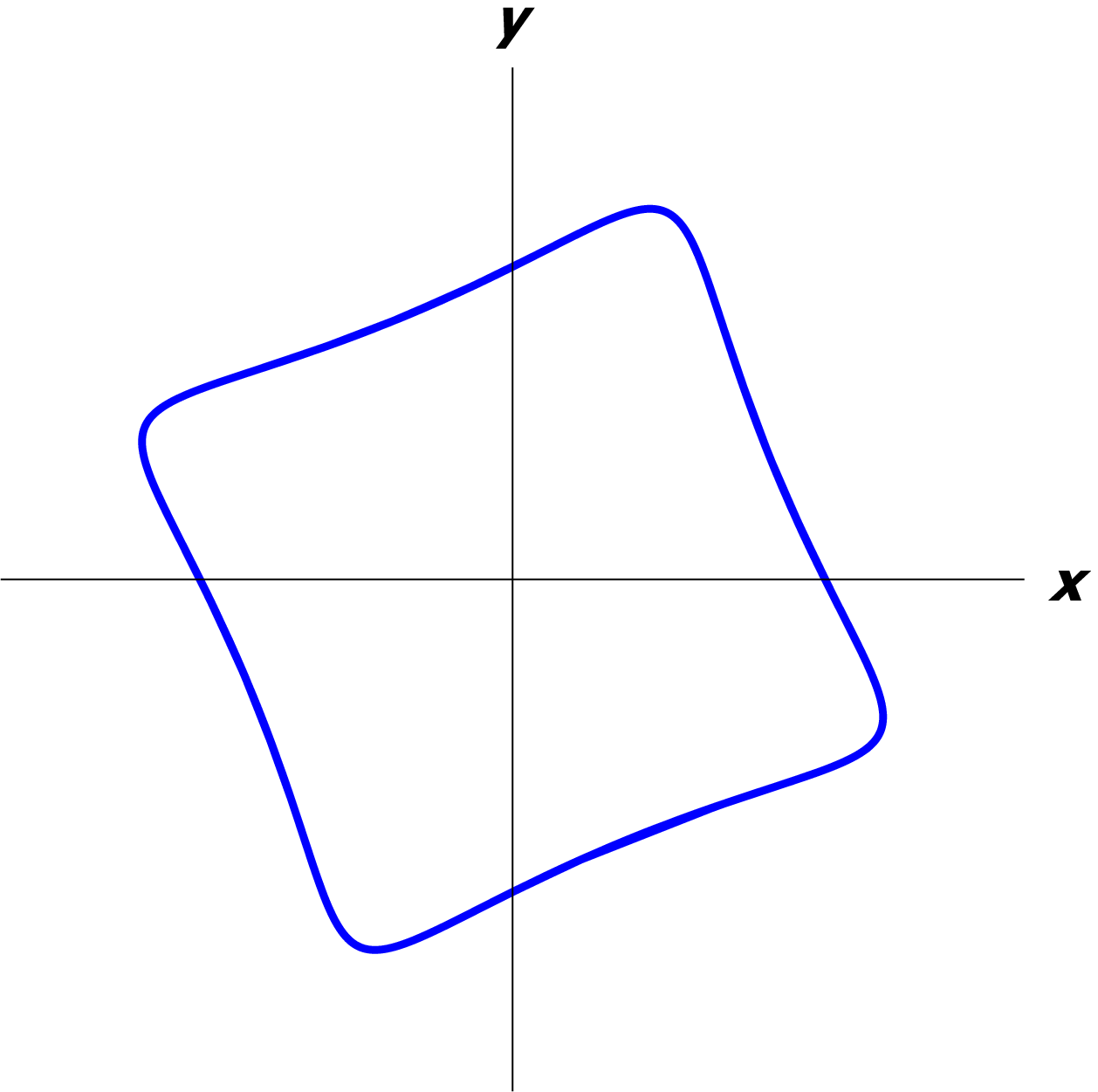}
\caption{\small{$g=1/2$, $R_2/R_1=6$}}
\end{subfigure}
\begin{subfigure}[c]{0.28\linewidth}
\includegraphics[scale=0.3]{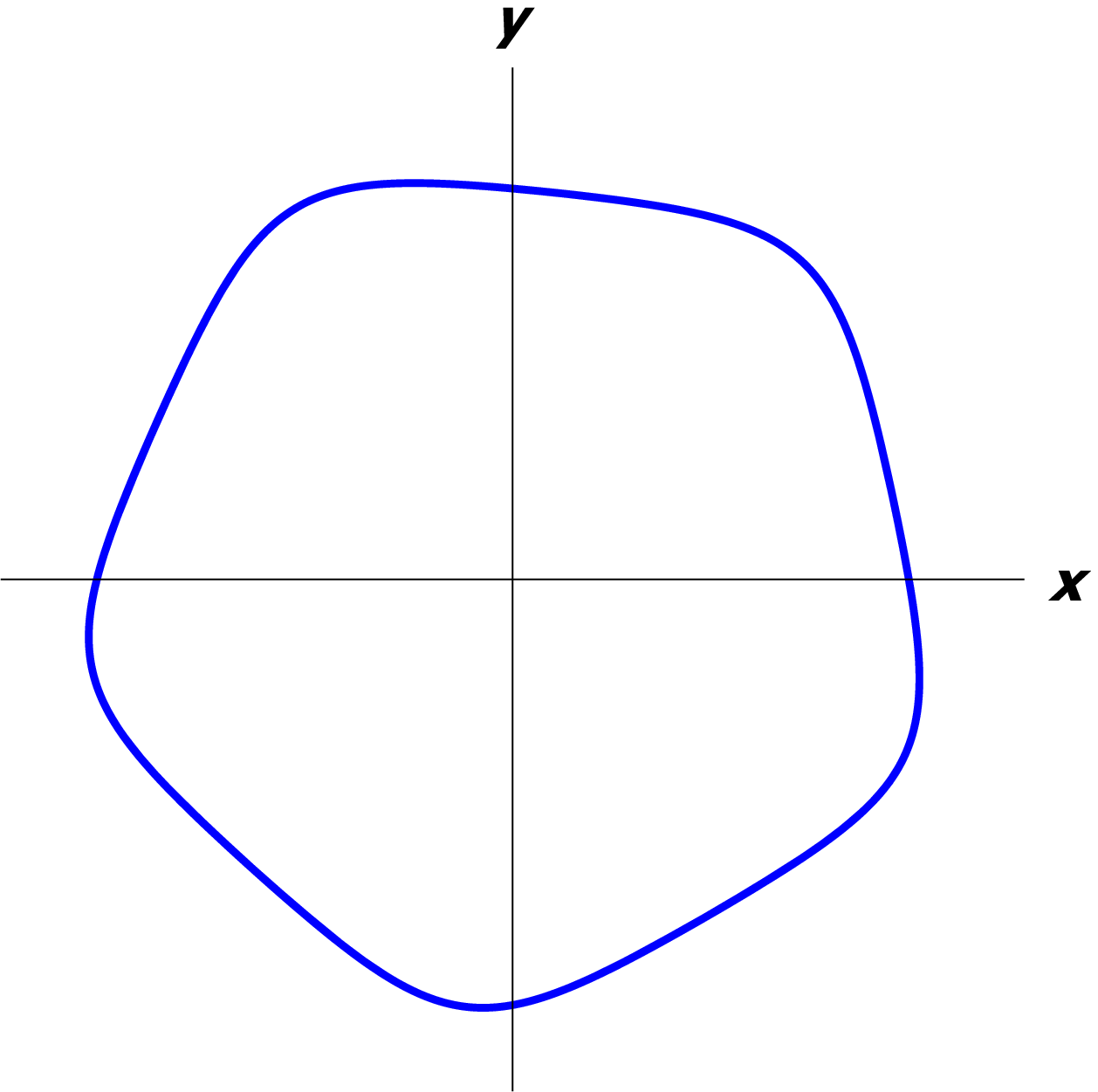}
\caption{\small{$g=3/5$, $R_2/R_1=20$}}
\end{subfigure}
\begin{subfigure}[c]{0.28\linewidth}
\includegraphics[scale=0.3]{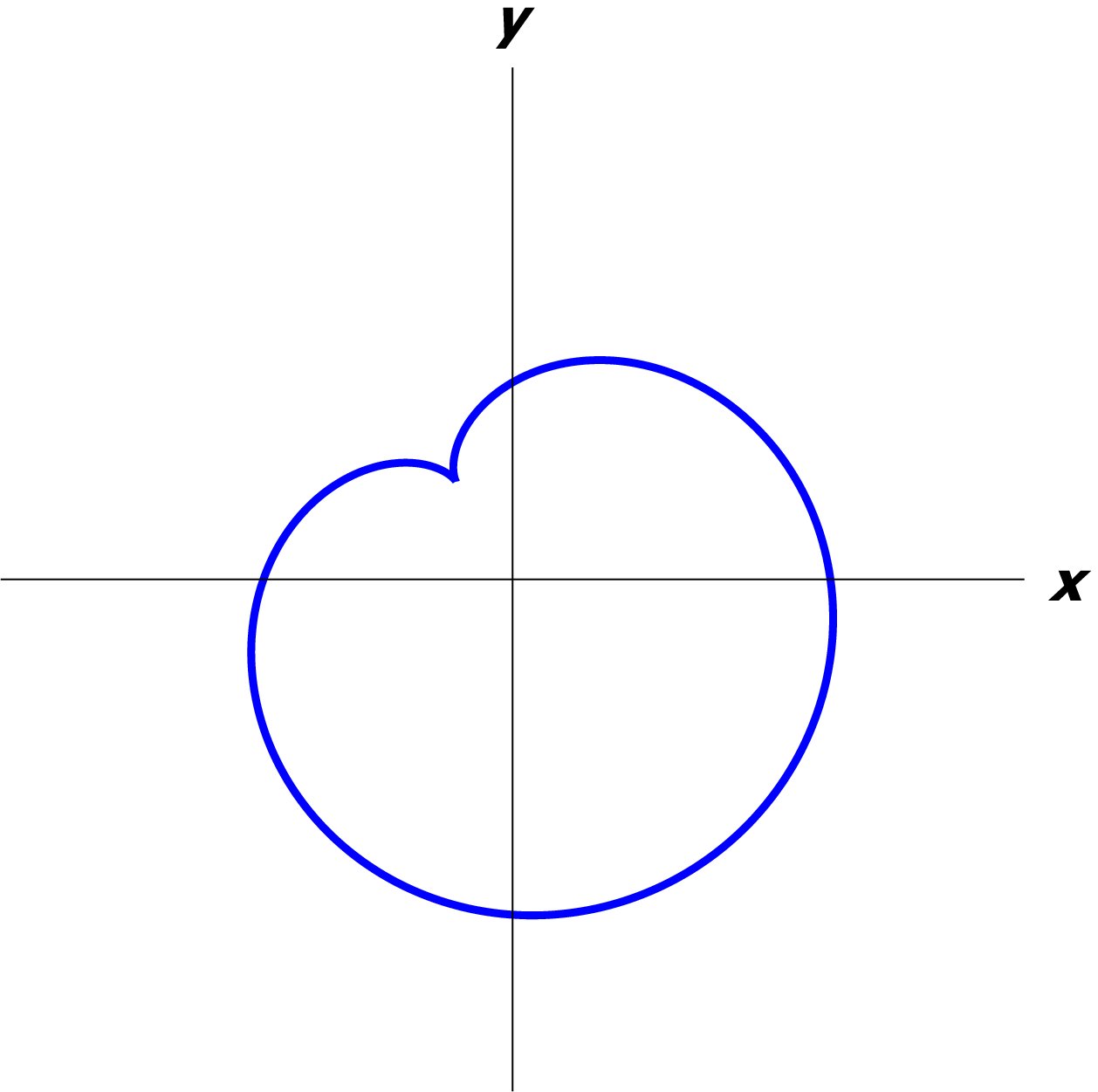}
\caption{\small{$g=3$, $R_2/R_1=2$}}
\end{subfigure}
\begin{subfigure}[c]{0.28\linewidth}
\includegraphics[scale=0.3]{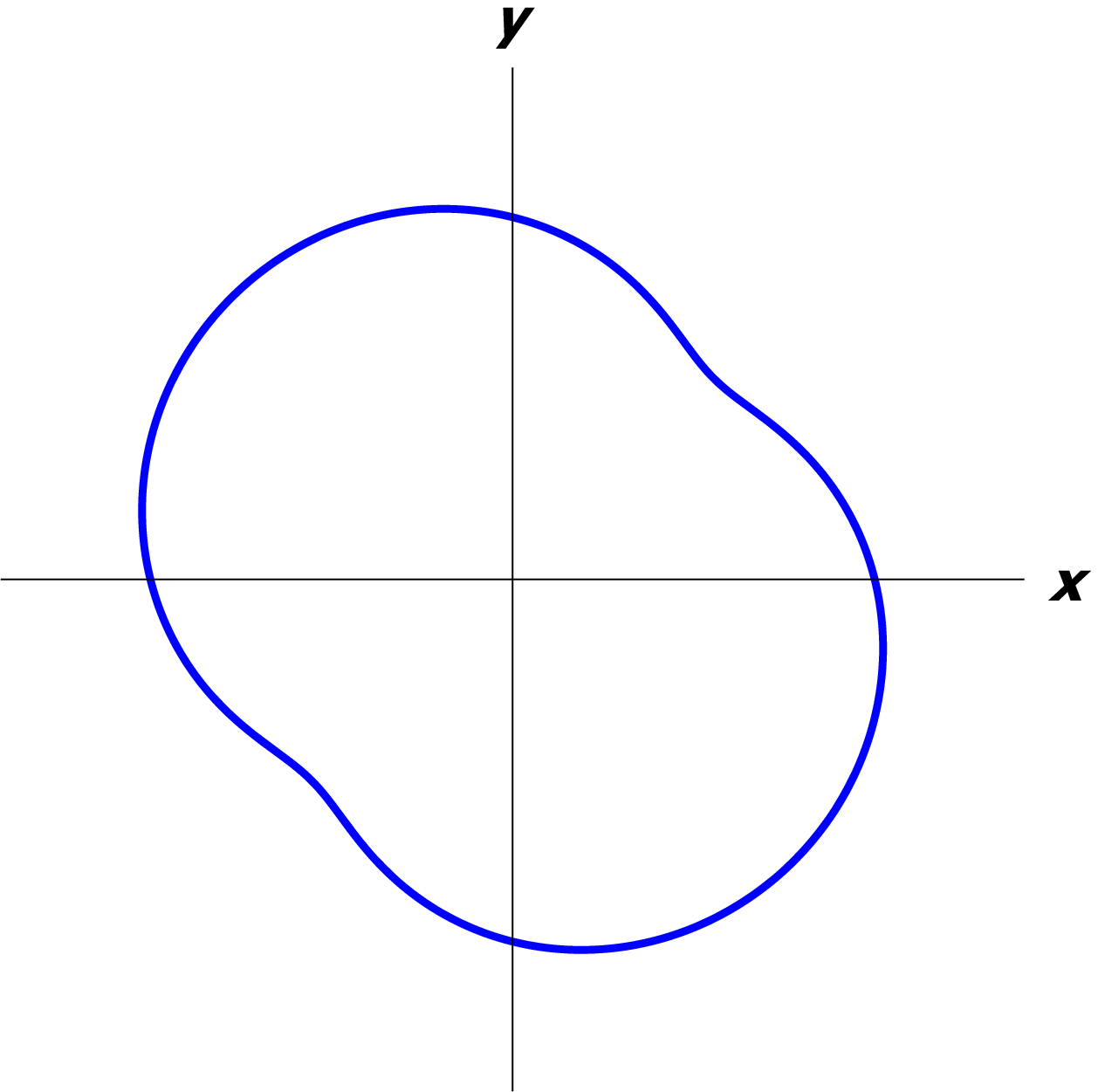}
\caption{\small{$g=2$, $R_2/R_1=6$}}
\end{subfigure}
\begin{subfigure}[c]{0.28\linewidth}
\includegraphics[scale=0.3]{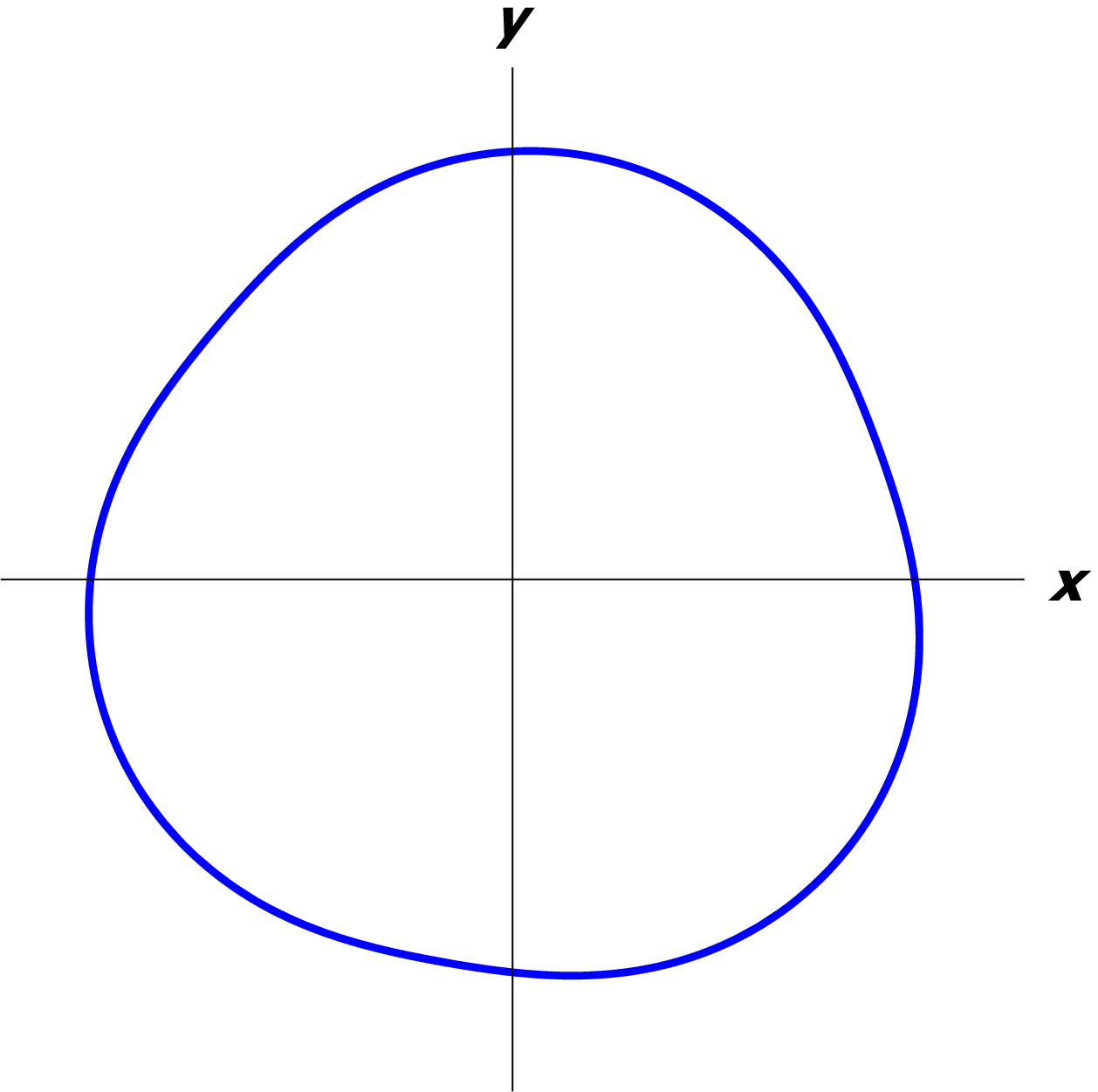}
\caption{\small{$g=5/3$, $R_2/R_1=20$}}
\end{subfigure}
\end{center}
\caption{\small{Trajectories for some rational values of $g$ 
and $R_1/R_2$.  The ``dual" figures  (a)  and (d), see below,  correspond to a general 
 case
$R_1|\ell_1|=R_2|\ell_2|$ of the trajectories with cusps,
in which velocity turns into zero. 
}}
\label{Fig2}
\end{figure}

In general case,  the trajectory is closed for arbitrary choice of
the integration constants 
iff the condition 
$ \ell_1 / \ell_2 = q_2 / q_1 $ 
with 
$q_1, q_2 \in \Z $ 
is  fulfilled.  
This implies rational values for the parameter  
$g=(q_2-q_1)/(q_1+q_2)$. Some  trajectories for rational values of $g$ are shown in 
 Figs. \ref{Fig1} and \ref{Fig2}. In  the case of Minkowskian isotropic oscillator (\ref{MinwoskianAHO}),
 the trajectories are obtained  by applying the transformation $\omega\rightarrow \omega/|g|$, 
 and taking the  limit
 $|g|\rightarrow \infty$ in (\ref{traj}). As a result one gets a circle 
 centered in the origin of the coordinate 
 system, $z(t)=e^{i\epsilon \omega t}(R_1e^{i\gamma_1}+R_2e^{i\gamma_2})$,
 where $\epsilon=\pm 1$ for $g\rightarrow \pm \infty$ \cite{InzPly8}. 
 
 The closed character of the trajectories for rational values of the parameter 
 $g$
  indicates that some additional true integrals  of motion  have to appear 
 in the corresponding systems. To obtain them, let us employ the classical CBT. 
 
 In the case in which we select $g$ as the irreducible rational number 
 \be
 \label{gmenos}
 g_{<}^{s_1,s_2}=(s_2-s_1)/(s_1+s_2)\,,\qquad s_1,s_2=1,2,\ldots\,,\qquad | g_{<}^{s_1,s_2}|< 1\,,
 \ee
 it is easy to see that the phase space functions of the classical free particle
\begin{eqnarray}
&
S_{s_1,s_2}^+=(\xi_+)^{s_1}(p_+)^{s_2}\,,\qquad
\hat{S}_{s_1,s_2}^-=(p_-)^{s_1}(\xi_-)^{s_2}\,,&
\end{eqnarray}
where  $p_\pm=p_1\pm ip_2$ and $
\xi_\pm=\xi_1 \pm i\xi_2$,
Poisson commute with (\ref{AnotherSw}). After the application of 
the classical CBT we get (up to certain constant
multiplicative factors)
\begin{eqnarray}
&\label{higherorderA}
\mathcal{L}_{s_1,s_2}^{+}=(b_1^+)^{s_1} (b_2^-)^{s_2}\,,
\qquad \mathcal{L}_{s_1,s_2}^{-}=(\mathcal{L}_{s_1,s_2}^{+})^*\,.
&
\end{eqnarray}
These new generators are the true integrals of motion for the system (\ref{gHamil}).
Together with   $p_\varphi$ and $H_g$, they generate 
a non-linear deformation of 
the $\mathfrak{u}(2)\cong \mathfrak{su}(2)\oplus \mathfrak{u}(1)$ algebra \cite{InzPly8},
which in the case of  $s_1=s_2=1$, $g=0$, 
reduces to  the $\mathfrak{u}(2)\cong \mathfrak{su}(2)\oplus \mathfrak{u}(1)$
Lie algebraic symmetry of the isotropic oscillator.  
As $\{\mathcal{L}_{s_1,s_2}^{+},\mathcal{L}_{s_1,s_2}^{-}\}$ is a polynomial
of $H_g$ and $p_\varphi$ (of order $s_1+s_2$), 
effectively (\ref{higherorderA}) provides us with only one new 
integral independent from $H_g$ and $p_\varphi$.

If instead of (\ref{gmenos}) 
we chose  now the irreducible fraction 
\be
 g_{>}^{s_1,s_2}=(s_2+s_1)/(s_2-s_1)\,,\qquad
| g_{>}^{s_1,s_2}|> 1\,,\label{gmas}
\ee
 one can show that
the  polynomials  of the free particle Heisenberg generators 
\begin{equation}
\Xi_{s_1,s_2}^+=(\xi_+)^{s_1}(\xi_-)^{s_2}\,,\qquad
\Xi_{s_1,s_2}^-=(p_+)^{s_1}(p_-)^{s_2}\,
\end{equation}
Poisson commute with (\ref{AnotherSw}), and after the application of 
the  conformal bridge transformation, we obtain 
(up to certain multiplicative constants)
the true integrals of motion for our system,
\begin{eqnarray}
&
\label{higherorderB}
\mathcal{J}_{s_1,s_2}^+=(b_1^+)^{s_1} (b_2^+)^{s_2}\,,\qquad
 \mathcal{J}_{s_1,s_2}^{-}=(\mathcal{J}_{s_1,s_2}^{+})^*\,.
&
\end{eqnarray}
As in the previous case, these integrals generate a non-linear algebra, which corresponds here 
to a deformation of the
 $\mathfrak{gl}(2,\R)=\mathfrak{sl}(2,\R) \oplus\mathfrak{u}(1)$ algebra \cite{InzPly8}.  

The symmetries of the phases $g=1$  and $g=-1$ of the Landau problem 
as well as of the isotropic Minkowskian oscillator, $|g|=\infty$,
can also be reproduced   by the described CBT, see ref.   \cite{InzPly8}.

We do not discuss here the complete 
algebraic structure generated by computing  the Poisson brackets 
between the  true integrals and the rest of dynamical symmetries of the system. 
Nevertheless we notice,  
that in the case $g=g_<^{s_1,s_2}$ ($g=g_>^{s_1,s_2}$) the integrals 
$\mathcal{J}_{s_1,s_2}^\pm$ ($\mathcal{L}_{s_1,s_2}^\pm$) are dynamical,  
and can be generated via the Poisson brackets between 
the integrals $\mathcal{L}_{s_1,s_2}^{\pm}$ ($\mathcal{J}_{s_1,s_2}^\pm$) and the quantities 
 $\mathcal{L}_{1,1}^{\pm}:=\mathcal{L}_\pm$ ($\mathcal{J}_{1,1}^\pm:=\mathcal{J}_{\pm}$),
which also are dynamical integrals of the system. 
In this process we generate a large but still finite number (depending on the values of 
the integer parameters $s_1$ and $s_2$)
of dynamical
integrals, which together with four true integrals 
generate a finite non-linear algebra of the $ W $ type. 
From this point of view 
we also have a kind of transmutation 
of symmetries for the 
``dual" pairs of 
the systems with 
$g=g_<^{s_1,s_2}$
 and $g=g_>^{s_1,s_2}=1/g_<^{s_1,s_2}$,
 where the non-linearly 
 deformed  $\mathfrak{u}(2)$ 
 and     $\mathfrak{gl}(2,\R)$
 subalgebras generated by the sets 
 ($H_g,\mathcal{L}_{2},\mathcal{L}_{s_1,s_2}^{\pm}$) 
 and 
  ($H_g,\mathcal{J}_{0},\mathcal{J}_{s_1,s_2}^{\pm}$) 
change their role in the sense 
of the true and dynamical sub-symmetries. For more details see
ref.  \cite{InzPly8}.

\subsection{Quantum  case of the ERIHO system}
\label{SecRIAHOq}

At the quantum level we have 
\begin{eqnarray}
&
\label{Hg}
\hat{H}_g=\hat{H}_{\text{osc}}+\omega\hat{p}_\varphi=
\hat{\mathfrak{S}}\omega(2i\hat{D}+g\hat{p}_{\varphi})\hat{\mathfrak{S}}^{-1}=
 \hbar\omega(\ell_1\hat{b}_1^+\hat{b}_1^-  + \ell_2\hat{b}_2^+\hat{b}_2^- +1)\,,&\\&
 \ell_1=1+g\,,\qquad \ell_2=1-g\,,
&
\\&
\label{Qb+-}
\hat{b}_1^-= \frac{1}{\sqrt{2}}(\hat{a}_1^--i\hat{a}_2^-)\,, \quad \hat{b}_1^+=(\hat{b}_1^-)^\dagger\,,
\quad
\hat{b}_2^-= \frac{1}{\sqrt{2}}(\hat{a}_1^-+i\hat{a}_2^-)\,,\quad \hat{b}_2^+=(\hat{b}_2^-)^\dagger\,.&
\end{eqnarray} 
Notice here that the operator $2i\hat{D}+g\hat{p}_{\varphi}$, to which we apply
the CBT,  is $\mathcal{PT}$ symmetric if  $\mathcal{P}$ is  identified as a 
spatial reflection operator in two dimensions, 
$\mathcal{P}x_1=-x_1\mathcal{P}$, 
$\mathcal{P}x_2=x_2\mathcal{P}$. 
To obtain the eigenstates and the spectrum of this system, analogously to the procedure
described in Sec. \ref{SecAFF1d}, 
we first have to solve  the eigenvalue equation 
\begin{eqnarray}\label{Dpeig}
&
(2i\hat{D}+g\hat{p}_\varphi)\phi_{\lambda}=\hbar\left( (1+g)z\frac{\partial}{\partial z}+
(1-g)z^{*}\frac{\partial}{\partial z^{*}}
 \right)\phi_{\lambda}=\lambda\phi_{\lambda}\,,\qquad z=x_1+ix_2\,.
&
\end{eqnarray}
The well defined in $\R^2$ solutions of this equation correspond to 
$ \phi_{n_1,n_2}=z^{n_1}(z^*)^{n_2}\,,$
 where $n_1$ and $n_2$ are non-negative integers. These are the Jordan states of the 
two-dimensional free particle that satisfy the equations $\hat{H}\phi_{n_1,n_2}=
-\frac{2\hbar}{m}n_1n_2 \phi_{n_1-1,n_2-1}$, 
 $2i\hat{D}\phi_{n_1,n_2}=\hbar(n_1+n_2+1)\phi_{n_1,n_2}$,
$\hat{p}_\varphi\phi_{n_1,n_2}=\hbar(n_1-n_2)\phi_{n_1,n_2}$, 
which imply that $(2i\hat{D}+g\hat{p}_\varphi)\phi_{n_1,n_2}=\hbar(n_1\ell_1+n_2\ell_2+1)\phi_{n_1,n_2}$. 
 The isotropic two-dimensional CBT produces (up to multiplicative constants) a map
 \be
 \hat{\mathfrak{S}}:\, (\hat{\xi}_+,\hat{\xi}_-,\hat{p}_+,\hat{p}_-)\quad
 \rightarrow\quad
 (\hat{b}^+_{1},\hat{b}^+_{2},\hat{b}^-_{2},\hat{b}^-_{1})\,,
 \ee
 where $\hat{p}_\pm=\hat{p}_1\pm i\hat{p}_2$ and  
$\hat{\xi}_\pm=\hat{\xi}_1\pm i\hat{\xi}_2$,
as well the map (\ref{GenConBr}) with $\hat{H}$, $\hat{D}$ and $\hat{K}$
given by (\ref{HDKd}) with $d=2$, while  $\hat{p}_\varphi$ 
is left invariant.  
 By computing the 
action of the generators $ (\hat{H},\hat{D},\hat{K},\hat{p}_\varphi,\hat{\xi}_+,\hat{\xi}_-,
\hat{p}_+,\hat{p}_-)$ on the states $\phi_{n_1,n_2}$, 
and with the subsequent application of the 
CBT generator $\hat{\mathfrak{S}}$ from the left,  one obtains the equations 
\begin{eqnarray}
&
\hat{b}_1^\pm\Psi_{n_1,n_2}=\sqrt{n_1+\beta_\pm}\Psi_{n_1\pm 1,n_2}\,,\qquad
\hat{b}_2^\pm \Psi_{n_1,n_2}=\sqrt{n_2+\beta_{\pm}}\Psi_{n_1,n_2\pm 1}\,,\label{specgen1} &\\&
\hat{\mathcal{J}}_\pm \Psi_{n_1,n_2}\sqrt{(n_1+\beta_\pm)(n_2+\beta_\pm)}\Psi_{n_1\pm1,n_2\pm1}\,,&\\&
\hat{H}_g \Psi_{n_1,n_2}=E_{n_1,n_2}\Psi_{n_1,n_2}\,,\qquad
 \hat{p}_\varphi \Psi_{n_1,n_2}=\hbar(n_1-n_2)\Psi_{n_1,n_2}\,,\label{Sch-Eq-RI} &\\&
 E_{n_1,n_2}=\hbar\omega (\ell_1 n_1+\ell_2 n_2+1)\,, 
 \qquad \beta_\pm=\frac{1\pm1}{2}\,.\label{Momen-Eq-RI}
&
\end{eqnarray}
Here the physical eigenstates $\Psi_{n_1,n_2}(x_1,x_2)$ are given by
 \begin{eqnarray}
&
 \hat{\mathfrak{S}}\phi_{n_1,n_2}=\mathcal{N}_{n_1,n_2}\Psi_{n_1,n_2}\,,\qquad
 \mathcal{N}_{n_1,n_2}=\left(\frac{2\hbar}{m\omega}\right)^{\frac{n_1+n_2}{2}} \sqrt{n_1!n_2! \pi}\,,&\\
\label{Psinn}
&
\Psi_{n_1,n_2}= 
\sqrt{\frac{m\omega}{\hbar \pi n_1!n_2!}} 
\,H_{n_1,n_2}\left(\sqrt{\frac{m\omega}{\hbar}}x_1,\sqrt{\frac{m\omega}{\hbar}}x_2\right)
e^{-\frac{m\omega}{2\hbar}(x_1^2+x_2^2)}\,,&
\end{eqnarray}
where  the functions 
are the generalized Hermite polynomials of two indexes \cite{Hermite}. 

From equations (\ref{specgen1}) one deduces that the operators 
$\hat{b}^\pm_i$ are the spectrum generating ladder operators of the system for arbitrary values of 
$g$. Eqs. 
(\ref{Sch-Eq-RI}) and (\ref{Momen-Eq-RI}) yield  the energy spectrum of the system and 
 the angular momentum value of each stationary  state. 
In dependence on the value of $g$, the spectrum has the following properties. 
It is  degenerate iff $g$ is a rational number, that  we assume from now on. 
 The spectrum is positive,   has a finite  degeneracy,  and the ground state is not degenerate when $|g|<1$. 
In the case  $|g|>1$,   it is not bounded from below,  and has infinite degeneracy in each energy level. 
Finally,  we have the spectrum of the Landau problem when $|g|=1$, see \cite{InzPlyWipf1}. 

In the case in which $g$ is equal to (\ref{gmenos}) one gets that the integrals 
\begin{eqnarray}
\hat{\mathcal{L}}_{s_1,s_2}^+=(\hat{b}_1^+)^{s_1}(\hat{b}_2^-)^{s_2}\,,\qquad
\hat{\mathcal{L}}_{s_1,s_2}^-=(\hat{\mathcal{L}}_{s_1,s_2}^-)^\dagger \,,
\label{OperatorsA}
\end{eqnarray}
which are the direct quantum analogs of $\mathcal{L}_{s_1,s_2}^\pm$, act as follows, 
\begin{eqnarray}
&
\hat{\mathcal{L}}_{s_1,s_2}^\pm\Psi_{n_1,n_2}=\sqrt{\frac{\Gamma(n_1+\beta_\pm s_1+1)\Gamma(n_2+\beta_\mp s_2+1)}
{\Gamma(n_1-\beta_\mp s_1+1)\Gamma(n_2-\beta_\pm s_2+1)} }\Psi_{n_1\pm s_1,n_2 \mp s_2}\,.&
\label{Lintdeg+}
\end{eqnarray} 
Besides, when $g$ corresponds to the case  (\ref{gmas}), the action of the  quantum analogs of the integrals 
$\hat{\mathcal{J}}_{s_1,s_2}^\pm$, 
 \be
\hat{ \mathcal{J}}_{s_1,s_2}^+=(\hat{b}_1^+)^{s_1}(\hat{b}_2^+)^{s_2}\,,\qquad
 \hat{\mathcal{J}}_{s_1,s_2}^-=(\hat{\mathcal{J}}_{s_1,s_2}^-)^\dagger\,,
 \ee
yields
 \begin{eqnarray}
&
\hat{\mathcal{J}}_{s_1,s_2}^\pm\Psi_{n_1,n_2}=
\sqrt{\frac{\Gamma(n_1+\beta_\pm s_1+1)\Gamma(n_2+\beta_\pm s_2+1)}{\Gamma(n_1-\beta_\mp s_1+1)
\Gamma(n_2-\beta_\mp s_2+1)}}\Psi_{n_1\pm s_1,n_2 \pm s_2}\,.\label{Jintdeg-}
&
\end{eqnarray}

All the normalizable eigenfunctions  with the same energy can be obtained by repeated  application 
of these operators to some fixed state $\Psi_{n_1,n_2}$.
When considering the case (\ref{gmenos}), 
 the action of 
the integrals $\hat{\mathcal{L}}^\pm_{s_1,s_2}$ with  both upper signs  produces a finite list of states.
This happens due to obligatorily appearance  of the poles in the Gamma function in a denominator of some coefficients. 
In contrast, when $g$ is given by Eq. (\ref{gmas}), equations (\ref{Jintdeg-}) imply that the repeated action of 
 $\hat{\mathcal{J}}^-_{s_1,s_2}$ at some step  annihilates a state, but the repeated 
 application of 
  $\hat{\mathcal{J}}^+_{s_1,s_2}$ will never produces zero. 
  The described properties of the quantum integrals reflect 
   the properties of the spectrum in dependence on
  the corresponding rational value of  $g$.

As in the previous section, we can construct the coherent states of the system. 
The way to obtain them is to apply the CBT operator to  the eigenstates 
of the free particle Hamiltonian. For this, we consider the plane wave
$e^{\frac{1}{\sqrt{2}}(\alpha_1 z+ \alpha_2 z^*)}$,
which, in dependence on the  values of the parameters 
$\alpha_1,\alpha_2\in \C$ can be a physical or  non-physical,  formal eigenstate of $\hat{H}$. 
The resulting $L^2(\R^2)$ integrable functions are eigenstates of operators 
$\hat{b}_i^-$ with eigenvalues
$\sqrt{\frac{\hbar}{m\omega }}\alpha_i$, $i=1,2$, and they hold their shape under the  
time translations and rotations
\cite{InzPly8}.

\section{CBT in cosmic strings  and   Dirac monopole backgrounds}
\label{SecCBT3d}

Here we discuss applications of CBT  with  non-trivial realizations of conformal 
generators in more than one dimension. In the first subsection we consider 
the relationship between the free particle and the harmonic oscillator 
on a cosmic string background \cite{InzPly7}. In the second subsection we comment on the
three-dimensional example in   the  
Dirac monopole background \cite{InzPlyWipf2}. 
This second example corresponds to  a direct 
generalization of the relationship between the one-dimensional 
 Calogero type  model and the AFF conformal mechanics studied in 
Sec. \ref{SecCalAFF}  to the case of three-dimensional spaces.

\subsection{CBT in a cosmic string background}
The metric corresponding to the $(2+1)$ cosmic string space-time  is given by \cite{SokSta,Vilenkin}
\begin{eqnarray}
&
dS^2=-c^2 dt^2+ds^2\,,\quad
ds^2=\left(1-\frac{8\mu G}{c^2}\ln(\frac{r}{r_0})\right)(dr^2+r^2d\varphi^2)\,,
&
\end{eqnarray}
where $G$ is Newton constant, $c$ is the speed of light, 
$\mu$ is the linear mass density of the cosmic string and 
$r_0$ corresponds to the  cosmic string  radius. By introducing the new coordinate
\begin{eqnarray}
&
r'^2=\left(1-\frac{8\mu G}{c^2}\ln(\frac{r}{r_0})\right)r^2\,,\quad 
\alpha ^2 dr'^2= \left(1-\frac{8 \mu G}{c^2}\ln(\frac{r}{r_0})\right)dr^2\,,\quad\alpha=\frac{1}{1-\frac{4\mu G}{c^2}}>0\,,&
\end{eqnarray}
one gets  (renaming $r'\rightarrow r$)
\be
\label{Metric}
ds^2= \alpha^2 dr^2 + r^2 d\varphi^2\,.
\ee
When $\alpha>1$, which implies $\mu>0$, metric (\ref{Metric}) 
is obtained from the three-dimensional Euclidean metric reduced 
to the conic surface $z=\lambda_E r$. On the other hand,   
when $0\leq\alpha<1$, that means $\mu<0$, metric (\ref{Metric}) is obtained 
by reducing a $(2+1)$ dimensional Minkowski space metric $ds^2=-c^2d\tau^2 +dr^2+r^2d\varphi^2$ 
to the surface 
$c\tau=\lambda r $, $0<\lambda<1$. Such metric also appears in condensed matter 
systems  \cite{Visser,KatVol,Cramer,Volovik,Manton}.

The non-relativistic action of a free particle in this space is   
$I=\int L dt$, 
$L=\frac{m}{2}g_{ij}\frac{dx_i}{dt}\frac{dx_j}{dt}=\frac{m}{2}\left(\alpha^2 \dot{r}^2+r^2\dot{\varphi}^2\right)$,
and the classical Hamiltonian corresponds to 
 \begin{eqnarray}&
H^{(\alpha)}=\frac{1}{2m}\left(\frac{p_r^2}{\alpha^2}+\frac{p_{\varphi}^2}{r^2}\right) \,.&\end{eqnarray}

The formal analogs of the momenta integrals and the Galilean boosts generators  are given by 
 \begin{eqnarray}
&\label{Pi+-}
\Pi_\pm=\Pi_1 \pm i \Pi_2= \left(\frac{p_r}{\alpha}\pm i \frac{p_\varphi}{r}\right)e^{\pm i\frac{\varphi}{\alpha}}\,,&\\&
\label{X+-}
\Xi_\pm=\Xi_1 \pm i\Xi_2=
\left[
\alpha m r-t \left(\frac{p_r}{\alpha}\pm i \frac{p_\varphi}{r}\right)\right]e^{\pm i\frac{\varphi}{\alpha}}\,.
&
\end{eqnarray} 
These  are well defined phase space functions only when $ \alpha^{-1} $ 
is an integer, while in the general case they are multi-valued. 
Despite this obstacle, 
we can use these formal conserved quantities  to construct the  well defined integrals for the system. 
In the general case of 
$\alpha$,
we have the $\mathfrak{sl}(2,\R)\oplus \mathfrak{u}(1)$ generators 
which are the Hamiltonian $H^{(\alpha)}$,
 the dilatations generator $D$, the generator of the special 
conformal transformations $K$, and the generator of rotations $J_0$, 
\begin{eqnarray}\label{HDkuku}
&
H^{(\alpha)}=\frac{1}{2m}\Pi_+\Pi_-\,,\qquad 
D=\frac{1}{4m}(\Xi_+\Pi_-+ \Pi_+\Xi_+)\,,\label{ConeHK}
&\\&
K=\frac{1}{2m}\Xi_+\Xi_-\,, \qquad
J_0= \frac{i}{4m}(\Xi_+\Pi_- - \Pi_+\Xi_+)=\frac{\alpha}{2} p_\varphi\,.\label{ConeDPhi}
& 
\end{eqnarray}
For the case of rational values of $\alpha=q/k$, with $q,k=1,2,\ldots,$ one can construct
\begin{eqnarray}
&\label{O}
\mathcal{O}_{\mu,\nu}^\pm=(\Xi_\pm)^\mu (\Pi_\pm)^\nu\,,
\qquad 
\mu=0,1,\ldots,q,\qquad  \nu=q-\mu\,, &\\&
\mathcal{S}_{\mu',\nu'}^\pm=(\Xi_\pm)^{\mu'} (\Pi_\pm)^{\nu'}\,,
\label{S}
\qquad 
\mu'=0,1,\ldots,2q, \qquad \nu'=2q-\mu' \,.
&
\end{eqnarray}
Here, the generators $ \mathcal{O}_{\mu,\nu}^\pm$ ($\mathcal{S}_{\mu',\nu'}^\pm$) have
 the angular dependence $e^{\pm ik\varphi}$ ($e^{\pm i2k\varphi}$), and 
therefore, 
they are well defined phase space functions. 
The finite sets of generators (\ref{O}) and (\ref{S}) are obtained 
by taking repeated Poisson brackets between $K$ (or $H^{(\alpha)}$) with 
$\mathcal{O}_{0,q}^\pm$ ($\mathcal{O}_{q,0}^\pm$) 
and $\mathcal{S}_{0,2q}^\pm$ ($\mathcal{S}_{2q,0}^\pm$) respectively. On the other hand, the brackets 
$ \{\mathcal{O}_{\mu,\nu}^+,\mathcal{O}_{\lambda,\sigma}^-\} $  and 
$ \{\mathcal{S}_{\mu',\nu'}^+,\mathcal{S}_{\lambda',\sigma'}^-\} $
are polynomial  functions of
$ m $, $ D $, $ J_0 $, and $ H^{(\alpha)}$ only. These properties 
imply that  the sets $\mathcal{U}_1=\{H^{(\alpha)}$, $K$, $D$, $J_0$, $\mathcal{O}_{\mu,\nu}^\pm\}$ and 
  $\mathcal{U}_2=\{H^{(\alpha)}$, $K$, $D$, $J_0$, $\mathcal{S}_{\mu',\nu'}^\pm\}$ 
  generate independent non-linear subalgebras.
  The complete symmetry algebra  of the system corresponds to 
  $\mathcal{U}_1\cup\mathcal{U}_2$ and also
  one can show that $\mathcal{U}_1$ is an ideal subalgebra  \cite{InzPly7}. 
   For subsequent  application of the conformal bridge transformation, 
 it is useful to write down explicitly the brackets 
 \begin{eqnarray}
 &\label{commutingDOS}
 \{D,\mathcal{O}_{\mu,\nu}^\pm\}=\frac{\nu-\mu}{2}\mathcal{O}_{\mu,\nu}^\pm\,,\qquad
 \{D,\mathcal{S}_{\mu',\nu'}^\pm\}=\frac{\nu'-\mu'}{2}\mathcal{S}_{\mu',\nu'}^\pm\,.
 &
 \end{eqnarray}
 From  them  one sees that  in the case $q=2n$, the integrals that Poisson commute with 
$D$  correspond to $(\mathcal{O}_{n,n}^\pm,\mathcal{S}_{2n,2n}^\pm=(\mathcal{O}_{n,n}^\pm)^2)$,
while  in the case $q=2n+1$, only the integral $\mathcal{S}_{2n+1,2n+1}^\pm$  are 
dilatation invariant.

\begin{figure}[hbt!]
\begin{center}
\hskip1.5cm
\begin{subfigure}[c]{0.28\linewidth}
\includegraphics[scale=0.25]{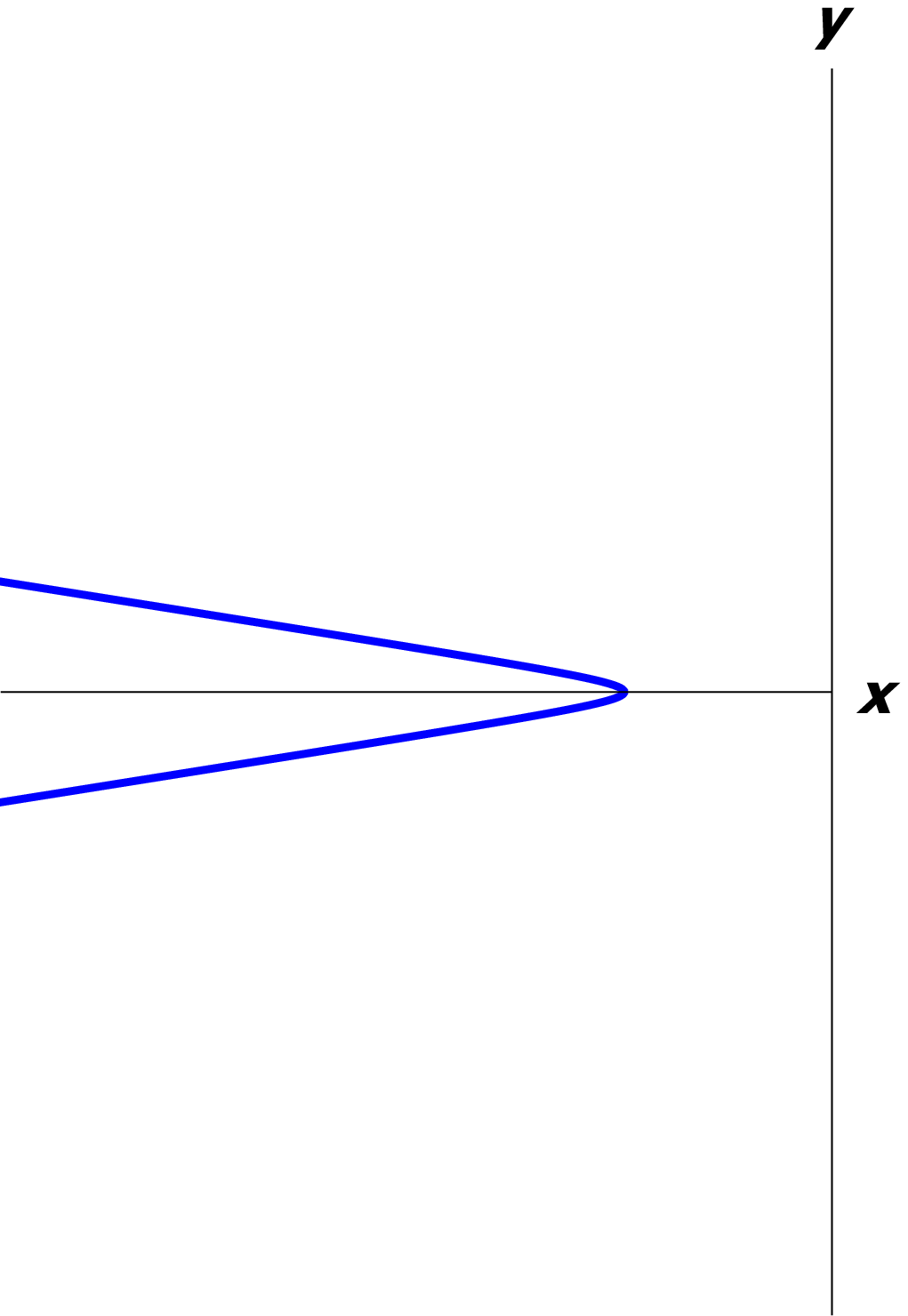}
\caption{\small{$\alpha=1/10$}}
\end{subfigure}
\begin{subfigure}[c]{0.28\linewidth}
\includegraphics[scale=0.25]{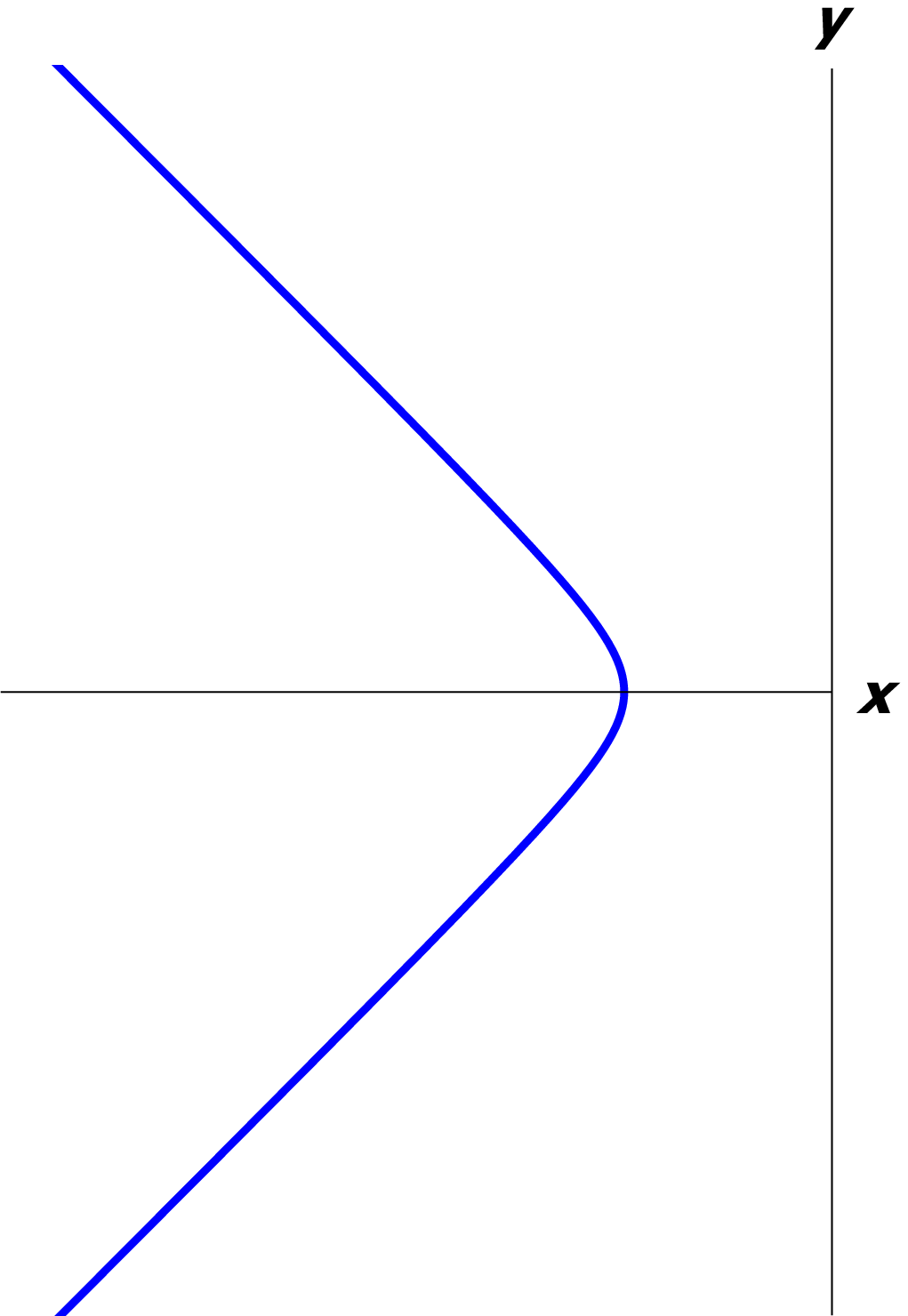}
\caption{\small{$\alpha=1/2$}}
\end{subfigure}
\begin{subfigure}[c]{0.28\linewidth}
\includegraphics[scale=0.25]{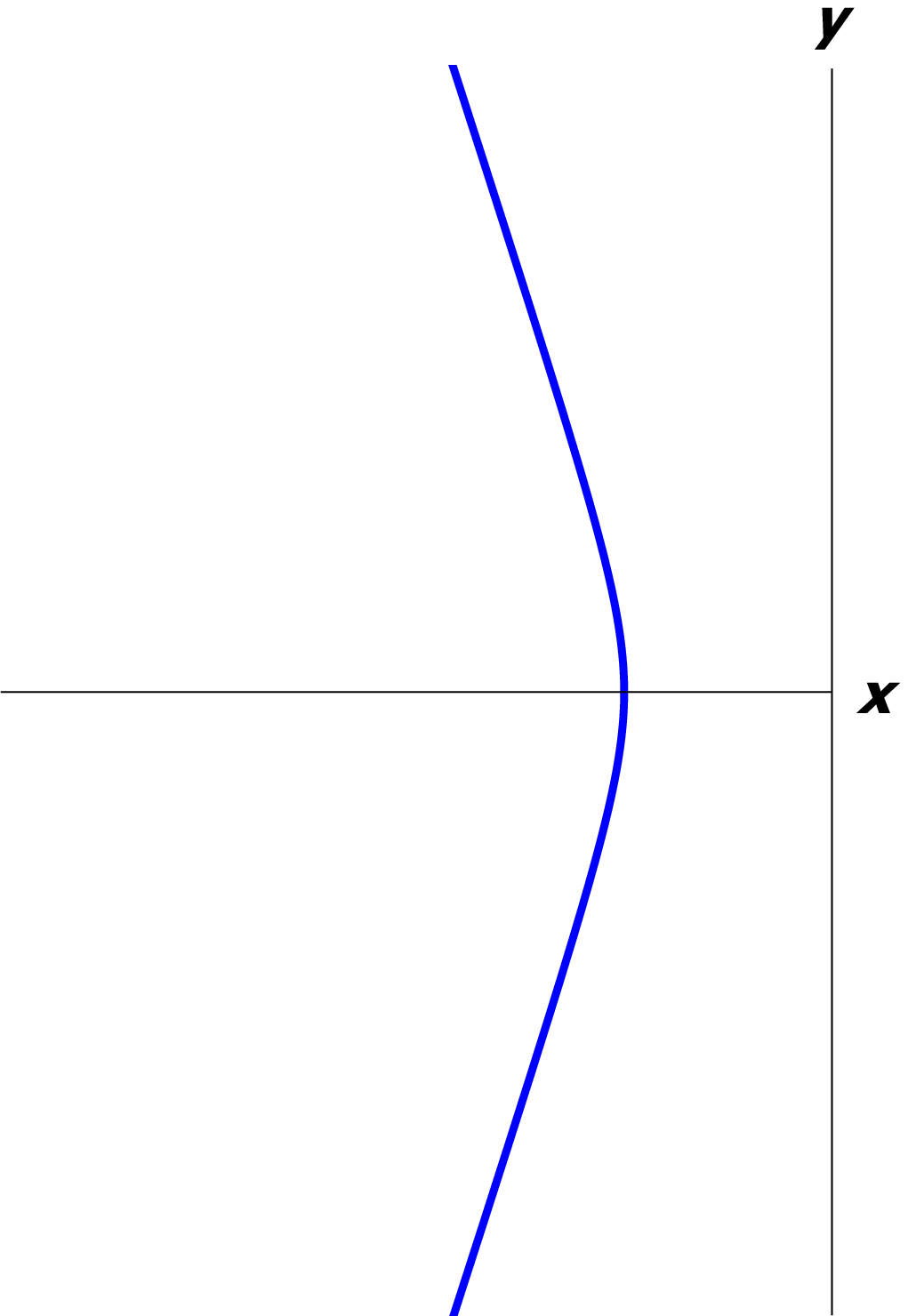}
\caption{\small{$\alpha=4/5$}}
\end{subfigure}
\vskip0.5cm
\hskip0.5cm
\begin{subfigure}[c]{0.25\linewidth}
\includegraphics[scale=0.25]{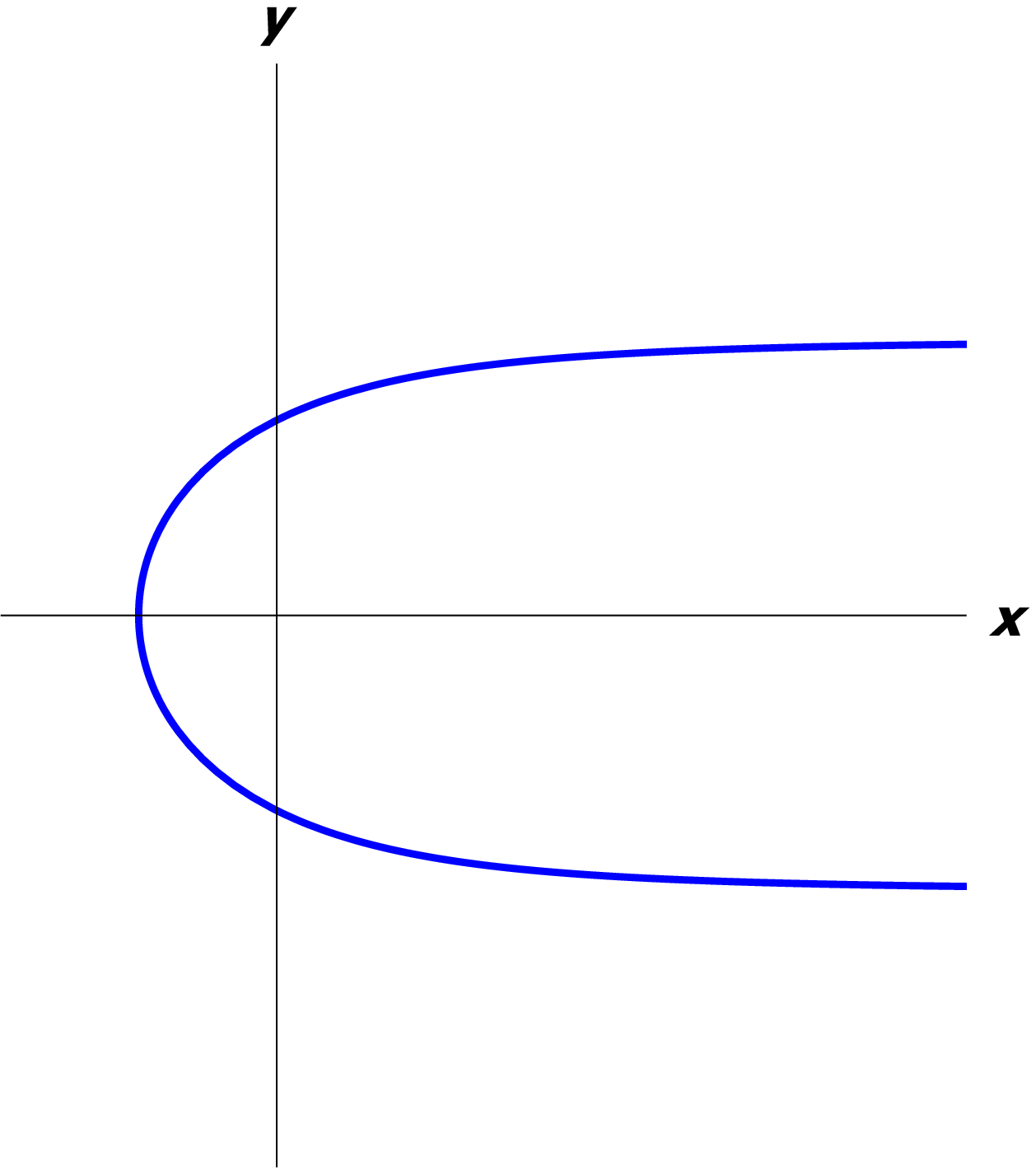}
\caption{\small{$\alpha=2$}}
\end{subfigure}
\begin{subfigure}[c]{0.2\linewidth}
\includegraphics[scale=0.25]{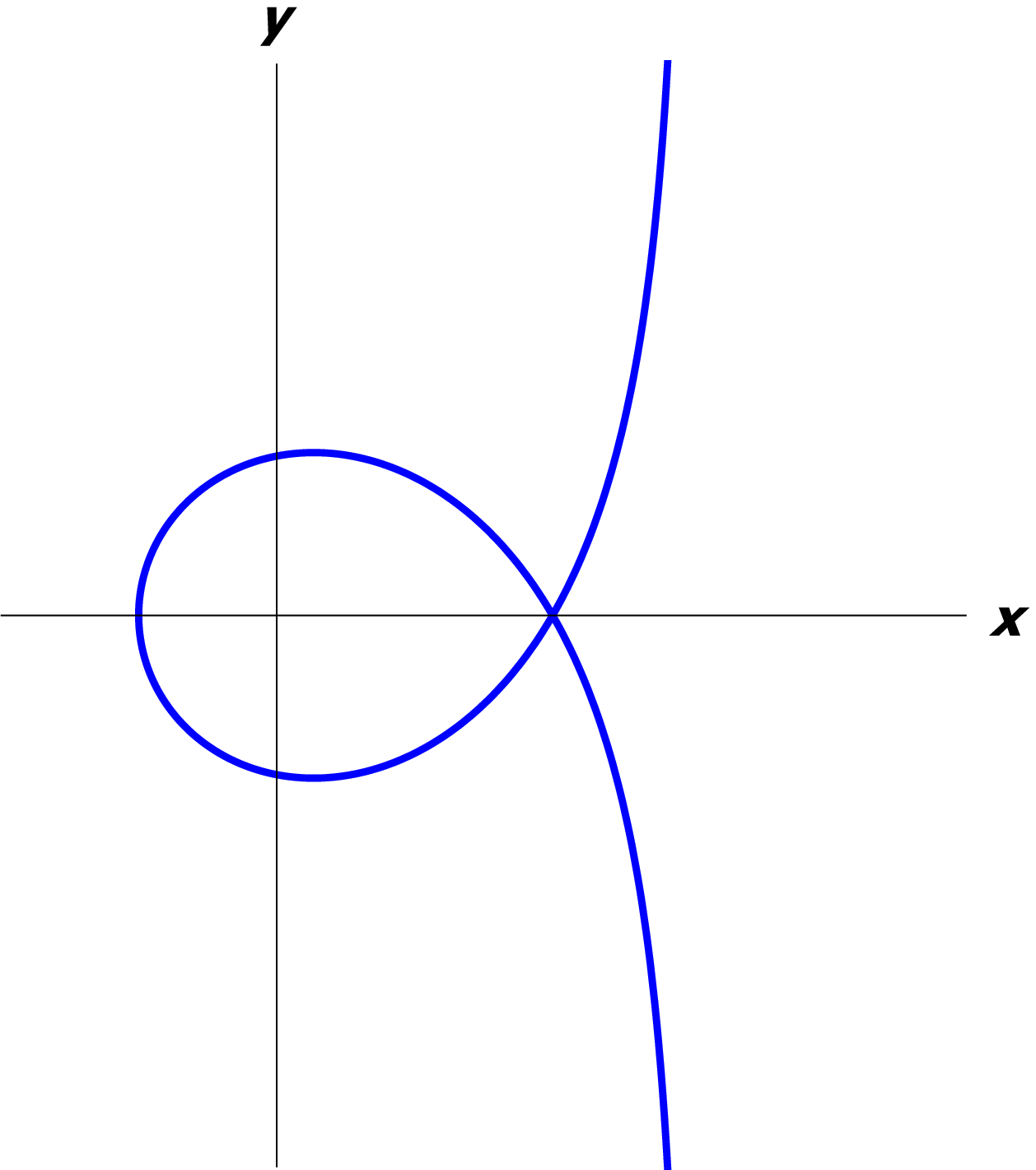}
\caption{\small{$\alpha=3$}}
\end{subfigure}
\begin{subfigure}[c]{0.2\linewidth}
\includegraphics[scale=0.25]{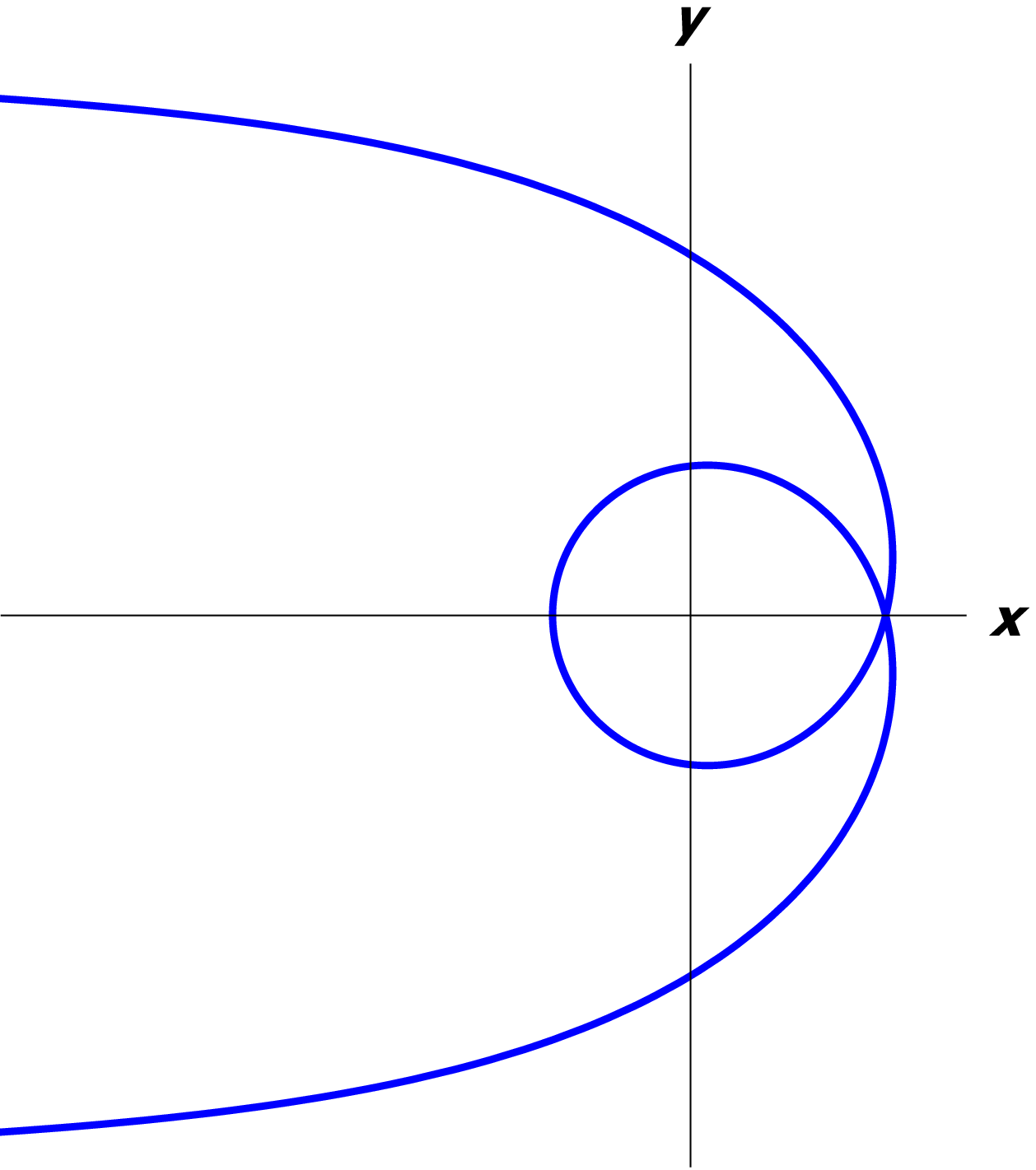}
\caption{\small{$\alpha=4$}}
\end{subfigure}
\begin{subfigure}[c]{0.25\linewidth}
\includegraphics[scale=0.25]{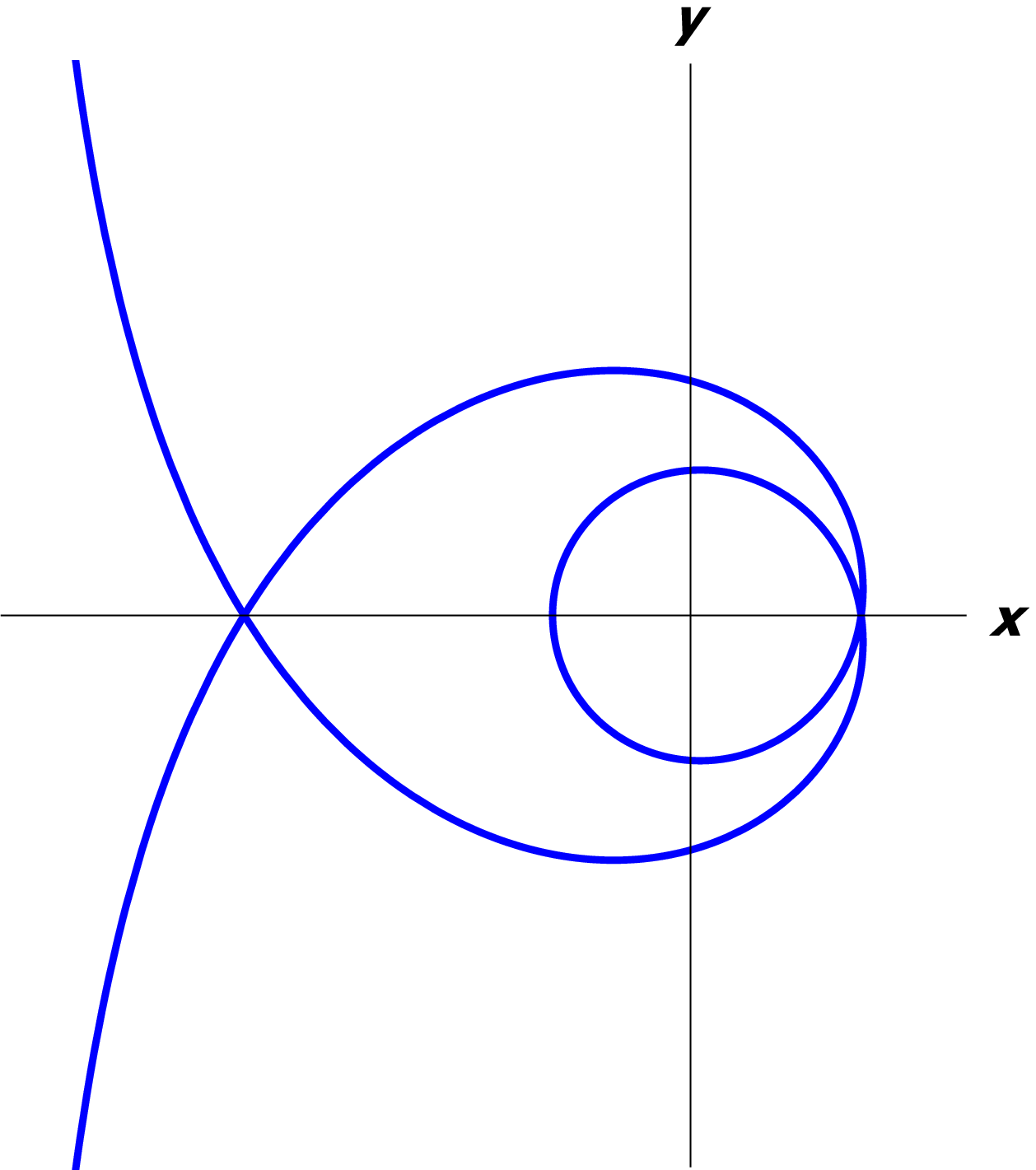}
\caption{\small{$\alpha=5$}}
\end{subfigure}
\end{center}
\caption{\small{Some examples of the geodesic motion in the conical geometry 
in coordinates $x=r\cos\varphi$, $y=r\sin\varphi$.
From the first three figures one sees that for $0<\alpha<1$,
the dynamics resembles that of the repulsive Kepler-Coulomb problem.
When $\alpha>1$ and is even, $\alpha=2n$, the particle experiences a backscattering.
When $\alpha$ is odd, $\alpha=2n+1$, 
the particle approaches the initial direction asymptotically after $n$ times 
circling the vertex of the cone.}
 }
\label{figure1Sec4}
\end{figure}

These properties associated with the parameter $\alpha$ can be predicted by analyzing 
the classical trajectories 
\begin{eqnarray}
&r(\varphi)=\frac{r_*}{\cos\left((\varphi-\varphi_*)/\alpha\right)}\,,\qquad
r_*=\frac{p_\varphi}{\sqrt{2mH^{(\alpha)}}}\,,
\qquad -\frac{\pi}{2}\alpha\leq \varphi-\varphi_* \leq \frac{\pi}{2}\alpha\,,&
\end{eqnarray}
from where we learn that the scattering angle is $\varphi_{\text{scat}}=\alpha\pi$. 
Some examples of the trajectories are shown  on Fig \ref{figure1Sec4}. 
 
Though in a free case special values of  the parameter $\alpha$ 
associated with existence of additional non-trivial integrals of motion reveal themselves
in dynamics only in rational values of the scattering angle in comparison with a flat case
where $\varphi_{\text{scat}}=\pi$, they will explicitly be detected  in the dynamics after
applying the conformal bridge transformation. 

After quantization, the Hamiltonian operator, its eigenstates and its spectrum are given by 
\begin{eqnarray}
\label{free-Hamiltonian}&
\hat{H}^{(\alpha)}=-\frac{\hbar^2}{2m}\frac{1}{\sqrt{g}}
\frac{\partial}{\partial x^i}\sqrt{g}g^{ij}\frac{\partial}{\partial x^j}
=-\frac{\hbar^2}{2m}
\left(\frac{1}{\alpha^2r}\frac{\partial}{\partial r}\left(r\frac{\partial}{\partial r}\right) +
\frac{1}{r^2}\frac{\partial^2}{\partial\varphi}\right)\,,&\\
\label{FreeParticleE}&
\psi_{\kappa,l}^\pm (r,\varphi)=\sqrt{\frac{\kappa}{2\pi \alpha}}J_{\alpha l}(\kappa r)e^{\pm il 
 \varphi}\,,\qquad E=\frac{\hbar^2\kappa^2}{2m\alpha^2}\,, \quad \kappa\geq 0\,, \quad l=0,1,\ldots\,.
 \end{eqnarray} 
The eigenfunctions satisfy $\braket{\psi_{\kappa,l}^\pm}{\psi_{\kappa',l'}^\mp }=\delta_{ll'}\delta(\kappa-\kappa')$,
where 
$
\braket{\Psi_1}{\Psi_2}=
\int_V \Psi_1^*\Psi_2  \sqrt{g}  dV=
\int_0^{\infty}\alpha r dr\int_{0}^{2\pi}d\varphi \Psi_1^*\Psi_2.
$
 The quantum versions 
of the formal integrals $\Pi_\pm$  are given by 
\begin{eqnarray}&
\hat{\Pi}_{\pm}=
e^{\pm \frac{i\varphi}{2\alpha}}\left(
\frac{1}{\alpha} \hat{p}_r \pm \frac{i}{r} \hat{p}_\varphi
\right)e^{\pm \frac{i\varphi}{2\alpha}}
=
-i\hbar \frac{1}{\alpha}e^{\pm i\frac{\varphi}{\alpha}}\left(
\frac{\partial}{\partial r}
\pm i\frac{\alpha}{ r}\frac{\partial}{\partial\varphi} 
\right)\,,&
\end{eqnarray}
and from the exponential factors one deduces that the action of these operators on eigenstates 
produce non-physical solutions in the general case. Explicitly we have 
\begin{eqnarray}
\label{FormalAction1}&
\hat{\Pi}_\pm\psi_{\kappa,l}^\pm (r,\varphi)= i 
\frac{\hbar\kappa}{\alpha}\sqrt{\frac{\kappa}{2\pi \alpha}}J_{\alpha l+1}(\kappa r)e^{\pm i(l+\frac{1}{\alpha })\varphi}\,,&\\&
\label{FormalAction2}
\hat{\Pi}_\pm\psi_{\kappa,l}^\mp (r,\varphi)=- i \frac{\hbar\kappa}{\alpha}
\sqrt{\frac{\kappa}{2\pi \alpha}}J_{\alpha l-1}(\kappa r)e^{\pm i(l-\frac{1}{\alpha})\varphi}\,.
&
\end{eqnarray} 

Quantum analogs of the generators (\ref{ConeHK}) and (\ref{ConeDPhi}) can be constructed 
straightforward
for arbitrary values of $\alpha$. But this is not the case for the integrals corresponding 
to rational values of this geometrical parameter. In fact,
with the help of expressions (\ref{FormalAction1}) and (\ref{FormalAction2}) one can show that 
the well defined symmetry operators  that are  the quantum analogs of the integrals 
(\ref{O}), (\ref{S}) can only be constructed for the special case of integer values of $\alpha=q$.
This reveals a  kind of the quantum anomaly in the system,  since this is the only case 
in which the action of the corresponding operators do not produce functions outside the Hilbert 
space constructed  from  the eigenstates (\ref{FreeParticleE}), see \cite{InzPly7} for more details.

\begin{figure}[hbt!]
\begin{center}
\hskip0.5cm
\begin{subfigure}[c]{0.27\linewidth}
\includegraphics[scale=0.25]{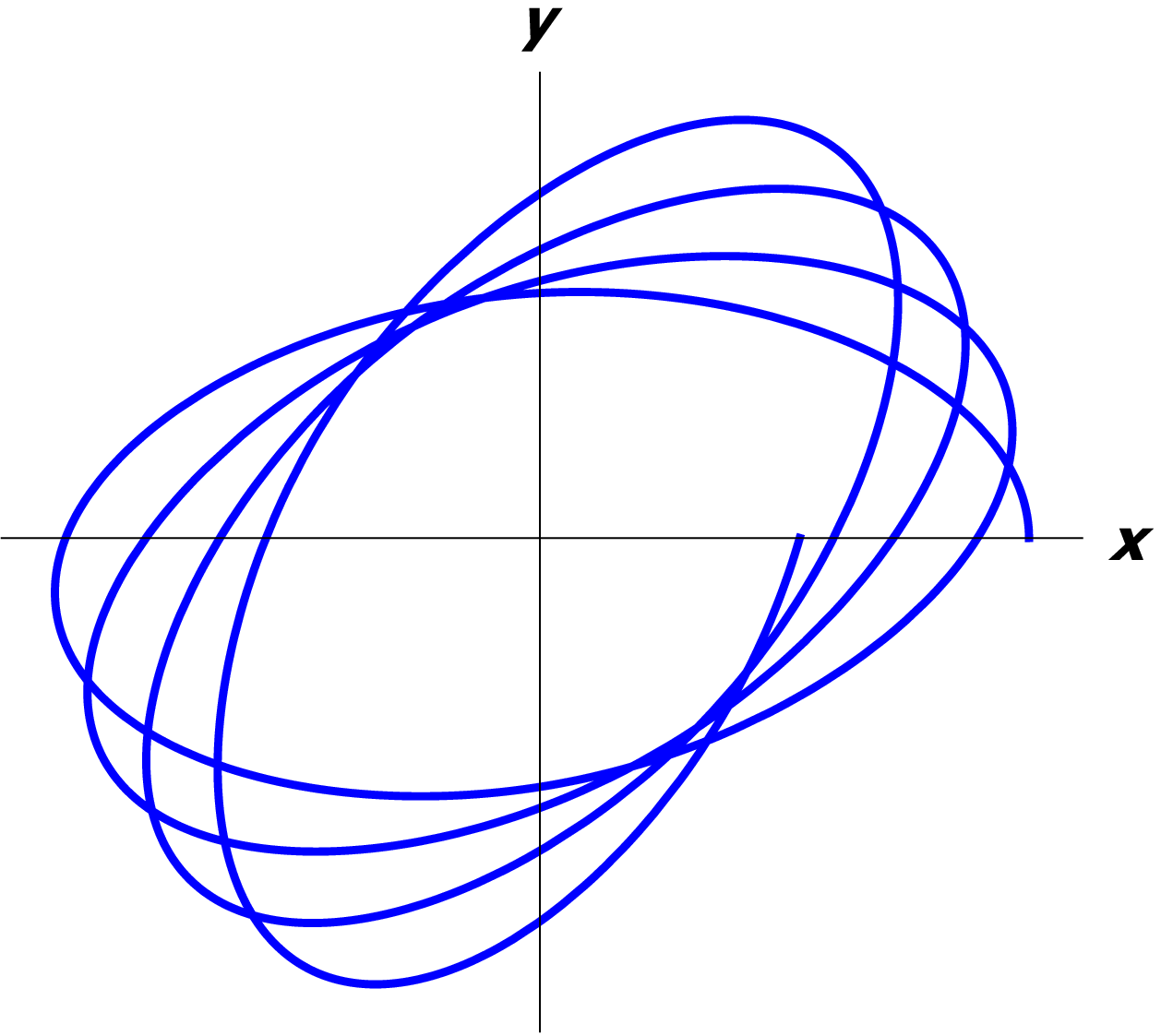}
\caption{\small{$\alpha=\sqrt{1.1}$}}
\end{subfigure}
\begin{subfigure}[c]{0.25\linewidth}
\includegraphics[scale=0.25]{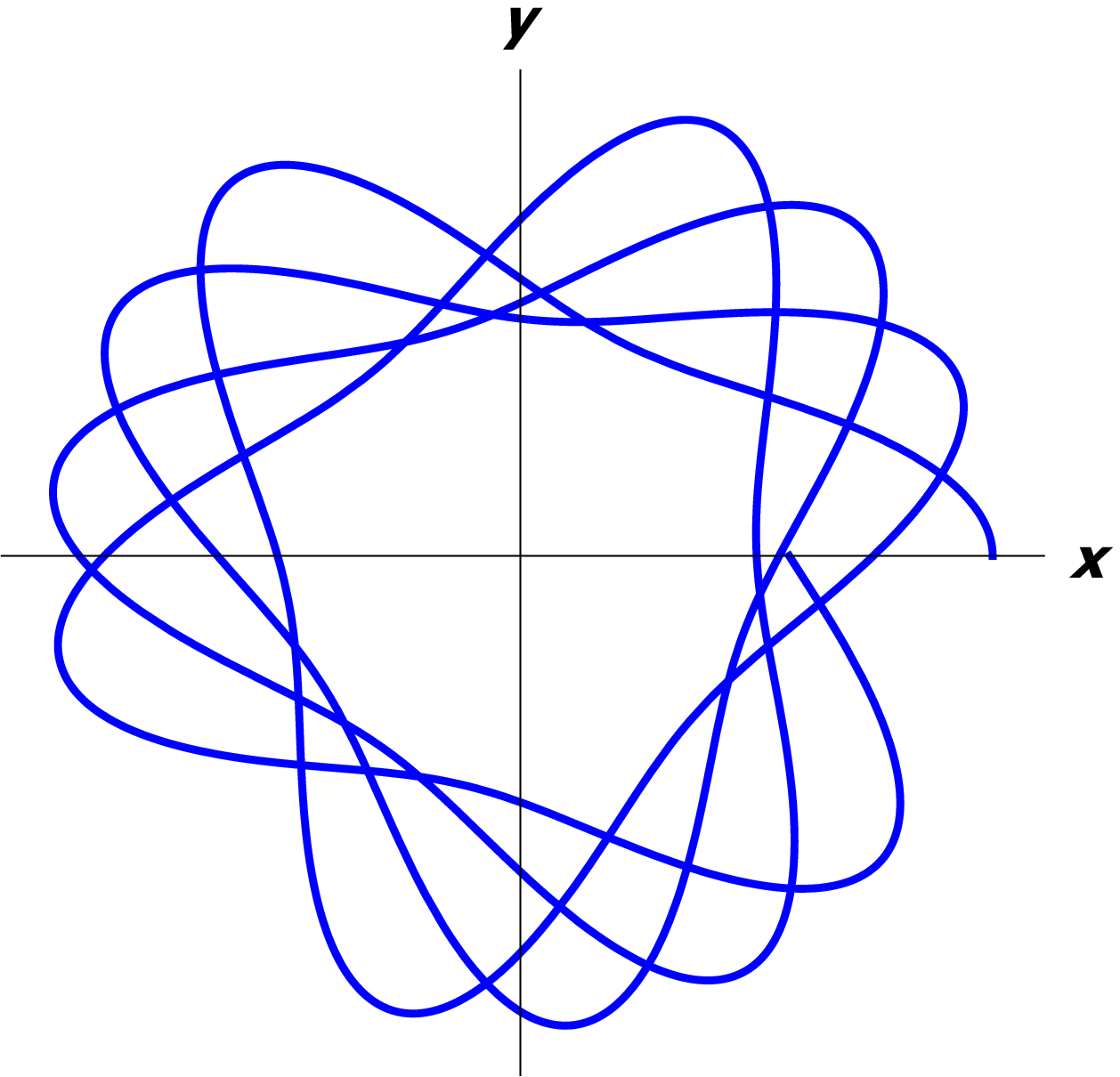}
\caption{\small{$\alpha=1/\sqrt{2}$}}
\end{subfigure}

\hskip0.5cm
\begin{subfigure}[c]{0.35\linewidth}
\includegraphics[scale=0.25]{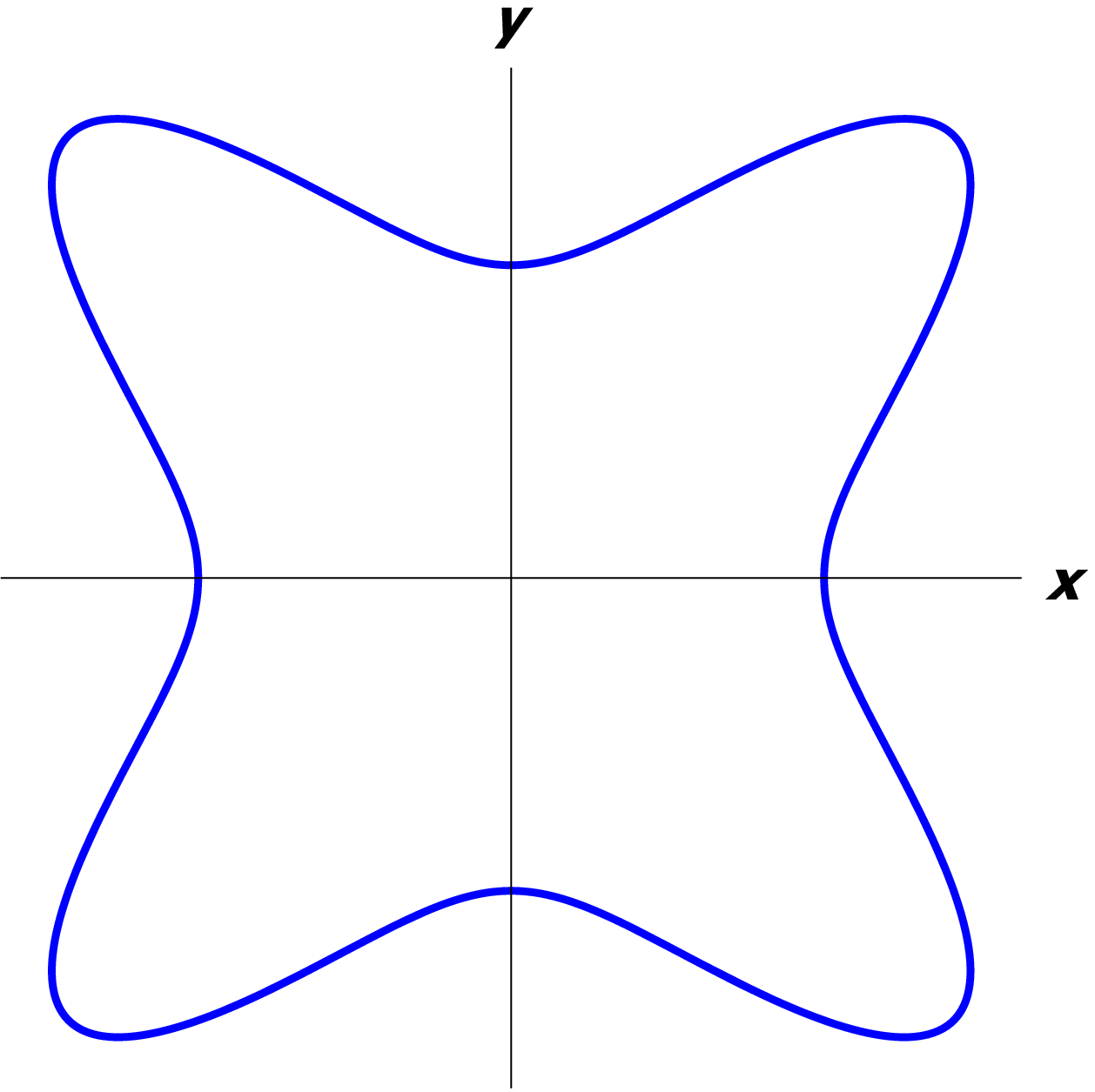}
\caption{\small{$\alpha=1/2$}}
\end{subfigure}
\begin{subfigure}[c]{0.26\linewidth}
\includegraphics[scale=0.25]{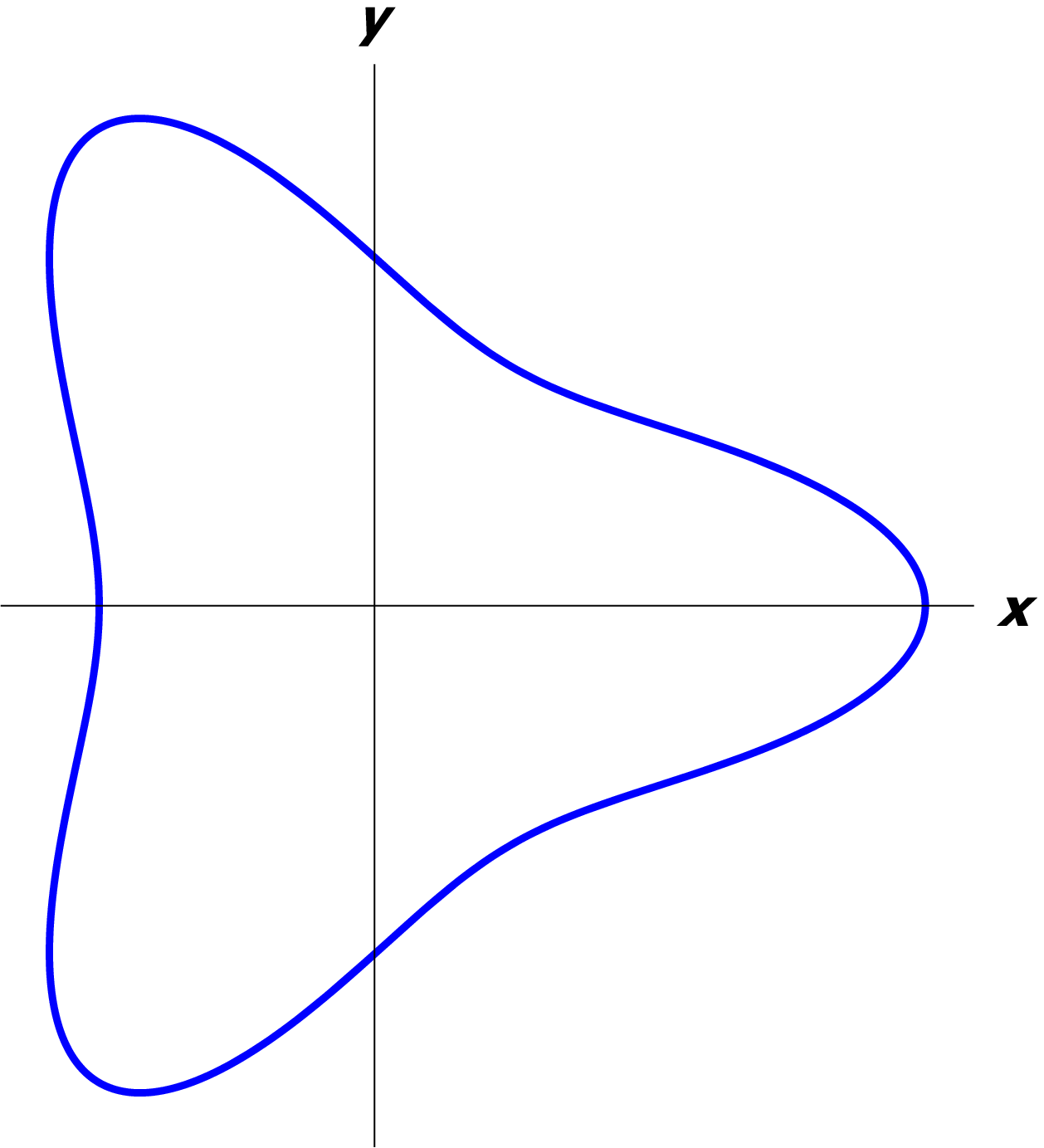}
\caption{\small{$\alpha=2/3$}}
\end{subfigure}
\begin{subfigure}[c]{0.25\linewidth}
\includegraphics[scale=0.25]{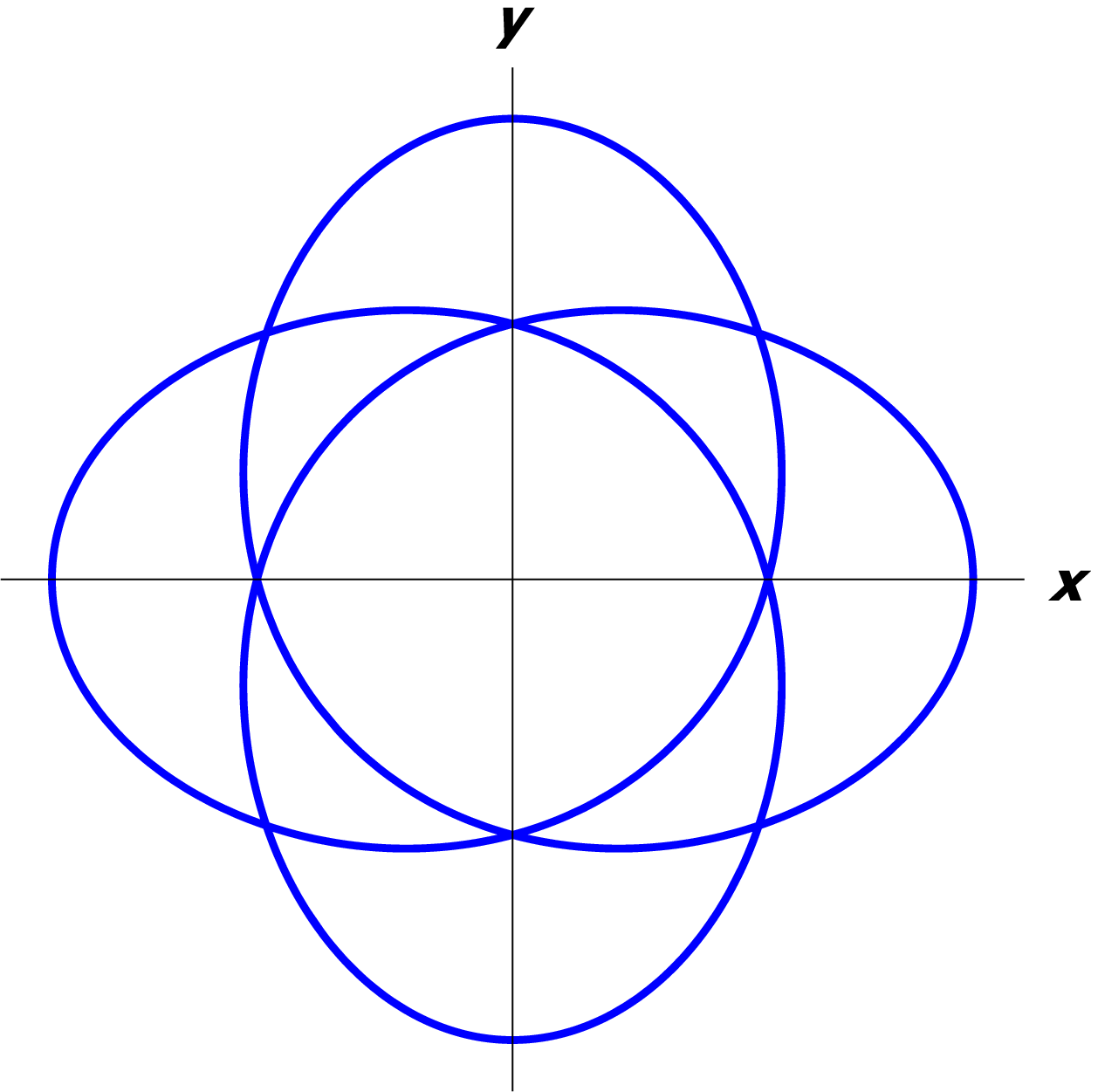}
\caption{\small{$\alpha=3/2$}}
\end{subfigure}

\hskip0.5cm
\begin{subfigure}[c]{0.27\linewidth}
\includegraphics[scale=0.25]{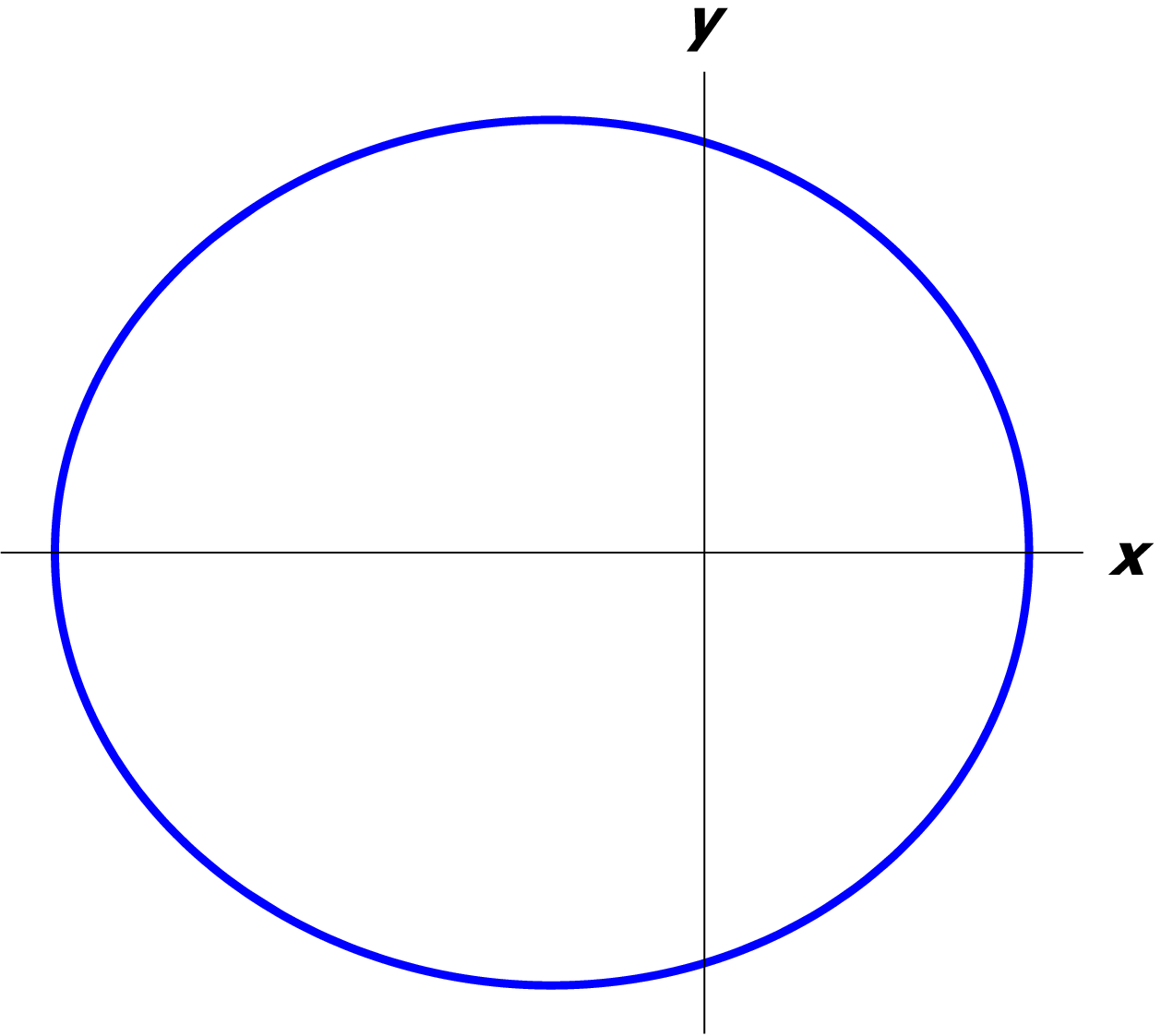}
\caption{\small{$\alpha=2$}}
\end{subfigure}
\begin{subfigure}[c]{0.26\linewidth}
\includegraphics[scale=0.25]{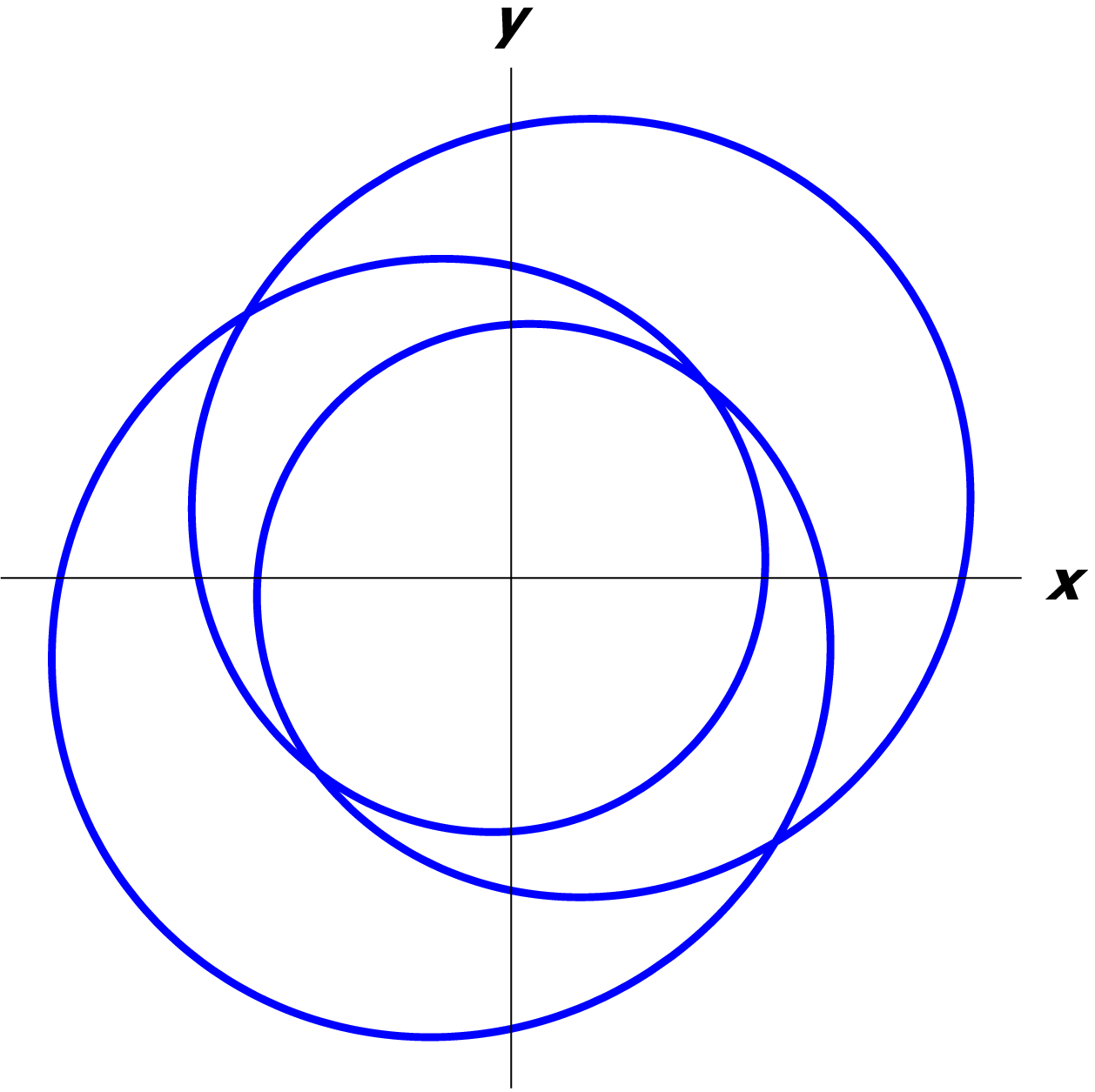}
\caption{\small{$\alpha=3$}}
\end{subfigure}
\begin{subfigure}[c]{0.25\linewidth}
\includegraphics[scale=0.25]{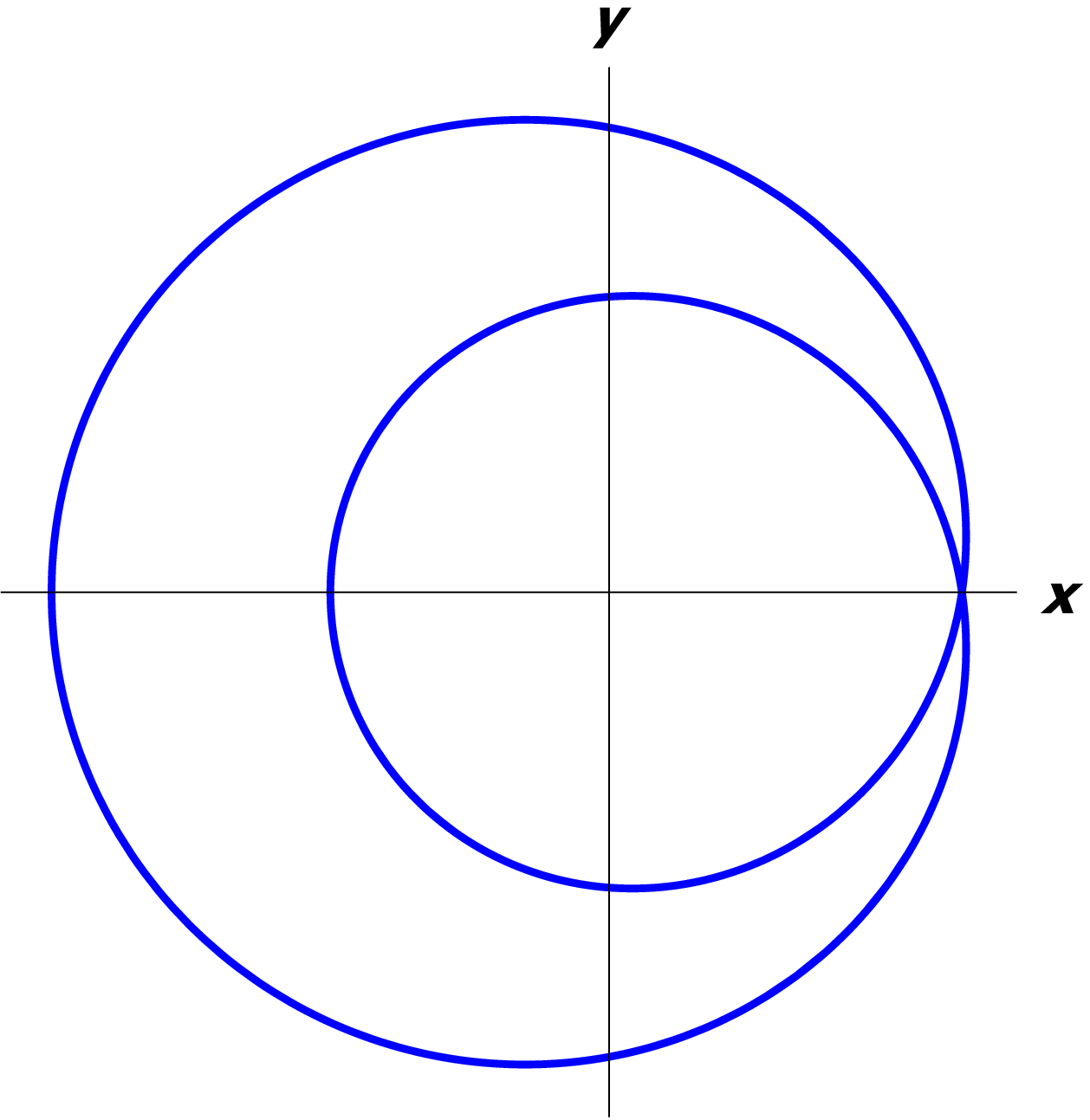}
\caption{\small{$\alpha=4$}}
\end{subfigure}
\end{center}
\caption{\small{Images of the trajectory  for some irrational and rational values of $\alpha$.
 One can  show in  particular   that $r(\varphi)=r(\varphi+\alpha l \pi)$, $l=1,2,\ldots$. From here 
one deduces that in the case $\alpha=q/k$ the number of maxima/minima of $r$ on the orbit   is 
$\mathscr{N}_{\text{max}/\text{min}}=k(q\,\text{mod}\,2+1)$. } }
\label{figure2Sec4}
\end{figure}

By applying the classical conformal bridge transformation to  generators (\ref{ConeHK}) and (\ref{ConeDPhi}) 
we obtain the $\mathfrak{sl}(2,\R)\oplus\mathfrak{u}(1)$ generators of the harmonic oscillator in 
the geometry defined by (\ref{Metric}), 
\begin{eqnarray}
&
\mathcal{J}_0=\frac{1}{2} \mathfrak{b}_a^+\mathfrak{b}_a^-=\frac{1}{2\omega}H_{\text{os}}^{(\alpha)}\,,\quad
\mathcal{J}_{\pm}=\mathfrak{b}_1^\pm \mathfrak{b}_2^\pm\,,\quad
\mathcal{L}_2=\frac{1}{2}(\mathfrak{b}_1^+\mathfrak{b}_1^--\mathfrak{b}_2^+\mathfrak{b}_2^-)=\frac{1}{2}\alpha p_\varphi\,, &
\\&
\mathfrak{b}_1^-=\frac{1}{2}e^{i(\omega t-\frac{\varphi}{\alpha})}
\left(
\alpha\sqrt{m\omega}r
+\frac{p_\varphi}{\sqrt{m\omega}r} +\frac{ip_r}{\alpha\sqrt{m\omega}}\right)\,,\qquad 
\mathfrak{b}_1^+=(\mathfrak{b}_1^-)^*\,,
&\\&
\mathfrak{b}_2^-= \frac{1}{2}e^{i(\omega t+\frac{\varphi}{\alpha})}
\left(
\alpha\sqrt{m\omega}r
-\frac{p_\varphi}{\sqrt{m\omega}r} +\frac{ip_r}{\alpha\sqrt{m\omega}}\right)\,,
\qquad 
\mathfrak{b}_2^+=(\mathfrak{b}_2^-)^*\,,&
\end{eqnarray}
where 
\begin{eqnarray}
&
H_{\text{os}}^{(\alpha)}=\frac{1}{2m}\left(\frac{p_r^2}{\alpha^2}+
\frac{p_{\varphi}^2}{r^2}\right) +\frac{m\omega^2\alpha^2}{2}r^2\,,
&
\end{eqnarray}
is the Hamiltonian of the isotropic harmonic oscillator in a cosmic string background, 
and the formal dynamical integrals $\mathfrak{b}_i^\pm$ correspond to the mapping 
\begin{eqnarray}
&  \label{frakmapping}
  \hat{\mathfrak{S}}: (\hat{\Xi}_+,\hat{\Xi}_-,\hat{\Pi}_+,\hat{\Pi}_-)\,\,\,
 \rightarrow\,\,\,\left(\sqrt{\frac{2m\hbar}{\omega}}\hat{\mathfrak{b}}^+_{1},
 \sqrt{\frac{2m\hbar}{\omega}}\hat{\mathfrak{b}}^+_{2},-i\sqrt{2m\omega\hbar}\hat{\mathfrak{b}}^-_{2},
 -i\sqrt{2m\omega\hbar}\hat{\mathfrak{b}}^-_{1}\right).
 &
 \end{eqnarray}
By solving the associated  equations of motion of the system one gets
\begin{eqnarray}
\label{CosmicStrTra}
&r^2(\varphi)=\frac{p_\varphi^2}{m H_{\text{os}}^{(\alpha)}}\left(1+\delta\cos(\frac{2}{\alpha}(\varphi-\varphi_*))\right)^{-1}\,,\quad 
\delta=\sqrt{1-\left(\frac{\omega\alpha p_{\varphi}}{ H_{\text{os}}^{(\alpha)}}\right)^2}\,,&
\end{eqnarray}
from where one finds  that the closed trajectories are possible only in the rational case 
$\alpha=q/k$, see Fig. \ref{figure2Sec4}.

In correspondence with (\ref{CosmicStrTra}), there are  globally well defined in the phase space integrals 
of motion that control the periodic behaviour of the trajectory  iff $\alpha$ is rational. 
To find these integrals we use the relations (\ref{frakmapping}) to transform the quantities (\ref{O})
and (\ref{S}) (up to inessential multiplicative constant factors) into 
\begin{eqnarray}
\mathcal{G}_{\mu,\nu}^+=(\mathfrak{b}_1^+)^\mu(\mathfrak{b}_2^-)^\nu\,,\quad (\mathcal{G}_{\mu,\nu}^+)^*=\mathcal{G}_{\nu,\mu}^-
\,,\quad 
\mathcal{F}_{\mu',\nu'}^+= (\mathfrak{b}_1^+)^{\mu'}(\mathfrak{b}_2^-)^{\nu'}\,,\quad
(\mathcal{F}_{\mu',\nu'}^+)^*=\mathcal{F}_{\nu',\mu'}^-\,.
\end{eqnarray}

Since the classical CBT is a canonical transformation, the algebraic properties of the free particle algebra are inherit 
by the harmonic oscillator algebra (with  $2i\omega D$ as a pre-image of the harmonic oscillator Hamiltonian).
This implies that in the case $q=2n$ with $n=1,2,\ldots$, 
the true integrals of the harmonic oscillator system in the cosmic string background are 
$
(\mathcal{G}_{n,n}^\pm,\mathcal{F}_{2n,2n}^\pm=(\mathcal{G}_{n,n}^\pm)^2)$ and  
in the case $q=2n+1$, the true integrals are $\mathcal{F}_{2n+1,2n+1}^\pm$. 
This is due to the dilatation invariance of their corresponding pre-images,  see (\ref{commutingDOS}) and comments below.

At the quantum level, the corresponding Hamiltonian operator, the eigenstates and 
the spectrum are given by 
\begin{eqnarray}
\label{HarmonicOcilator}
&\hat{H}_{\text{os}}^{(\alpha)}=
-\frac{\hbar^2}{2m}
\left(\frac{1}{\alpha^2r}\frac{\partial}{\partial r}\left(r\frac{\partial}{\partial r}\right) +
\frac{1}{r^2}\frac{\partial^2}{\partial\varphi}\right)+\frac{\alpha^2 m \omega^2}{2}r^2\,,&\\&
\label{EigenstatesHar}
\psi_{n_r,l}^\pm(r,\varphi) =\left(\frac{m\omega\alpha^2}{\hbar}\right)^{\frac{1}{2}}
\sqrt{\frac{n_r!}{2\pi\alpha\Gamma(n_r+\alpha l+1)}}\,
\zeta^{\alpha l}L_{n_r}^{(\alpha l)}(\zeta^2)
e^{-\frac{\zeta^{2}}{2} \pm i l\varphi}\,,\quad \zeta=\sqrt{\frac{m\alpha^2\omega}{\hbar}}r\,,&\\&
E_{n,l}=\hbar\omega(2n_r+\alpha l+1)\,,\quad
n_r\,,l=0,1,\ldots\,.&
\end{eqnarray} 
Eigenstates (\ref{EigenstatesHar}) and the spectrum can be obtained by applying the corresponding realization 
of the operator $\hat{\mathfrak{S}}$ on the zero energy Jordan states 
of the free particle Hamiltonian (\ref{free-Hamiltonian}), which are simultaneously 
eigenstates of $2i\omega\hat{D}$. In this case these Jordan states 
 are given by 
$\Omega_{n_r,l}^\pm(r,\varphi) =r^{2n_r+\alpha l }e^{\pm l \varphi}$.
In the same vein, the application of    $\hat{\mathfrak{S}}$ to functions (\ref{FreeParticleE})
gives us the coherent states of the system. 
 Due to this connection 
one deduces that the quantum anomaly mentioned above for the free system 
is also present in the  harmonically confined one. We refer for the details to   \cite{InzPly7}.

In correspondence with the spectral properties of the confined system, 
one notes that it acquires a special degeneracy when 
$\alpha=q/k$, however, due to the presence of the quantum anomaly, only in the case $\alpha=q$ 
one can construct well defined  operators that correctly reflect the degeneracy in the spectrum  \cite{InzPly7}. 
When $\alpha=2n$ ($\alpha=2n+1$),
  these operators are $\hat{\mathcal{G}}_{n,n}^\pm$  ($\hat{\mathcal{F}}_{2n+1,2n+1}^\pm$).

\subsection{CBT in a Dirac monopole background}

The non-trivial three dimensional example  we present here corresponds 
to the dynamics of a particle with electric charge $e$, which is coupled to a 
Dirac magnetic monopole
with magnetic charge $g$ and  is subjected to the
central potential $V(r)=\frac{m\omega^2 r^2}{2}+\frac{\alpha}{2mr^2}$. 
Parameter $\alpha$ is a real numerical constant and $\omega>0$ is a
 frequency associated to the harmonic trap. 
The model and its supersymmetric extensions were extensively studied in \cite{InzPlyWipf2} 
and here we just consider the system in relation to CBT. 

The Hamiltonian of the system is  
\begin{eqnarray}
&
\label{DiracMonoH}
H_\omega=\frac{\vpi^2}{2m}+\frac{m\omega^2 r^2}{2}+\frac{\alpha}{2mr^2}\,,
\quad \vpi=\vp-e\vA\,,\quad 
\nabla\cross\vA=\vB=g\frac{\vr}{r^3}\,.
&
\end{eqnarray}
By considering 
the Poincar\'e integral of the system 
\be
\vJ=\vr\cross\vpi-\nu\vn\,,\qquad J^2=\vJ^2\,,\qquad  \vn\cdot\vJ=-\nu\,,\qquad eg=\nu\,,
\ee
we note that the 
Hamiltonian (\ref{DiracMonoH}) in spherical coordinates admits an 
`AFF model representation',
\begin{eqnarray}\label{defEll}
&
H_\omega=
\frac{\pi_r^2}{2m}+\frac{\mathscr{L}^2}{2mr^2}+\frac{m\omega^2r^2}{2}\,,
\qquad \mathscr{L}^2:=\vJ^2-\nu^2+\alpha\,,
&
\end{eqnarray}
and by using the fact that $\{r,\pi_r\}=1$, it is  deduced that the generators 
\begin{eqnarray}
&\label{Sl2Rmonopolar}
\mathcal{J}_0=\frac{1}{2\omega^2}H_\omega\,,\qquad
\mathcal{J}_\pm=-
\frac{1}{2\omega}( H_0- \omega^2 K_0 \pm i2\omega D_0)\,,
&
\end{eqnarray}
produce the $\mathfrak{sl}(2,\R)$ algebra. In these last relations 
we have introduced the $\mathfrak{so}(2,1)$ symmetry generators 
of the system without the harmonic  trap
\begin{eqnarray}
\label{freemotion}
H_0=\frac{\pi_r^2}{2m}+\frac{\mathscr{L}^2}{2mr^2}\,,\qquad
D_0=\frac{1}{2}r\pi_r-H_0t\,,\qquad
K_0=\frac{mr^2}{2}-Dt-H_0t^2\,.
&
\end{eqnarray}
Both forms of  dynamics are connected 
by the classical conformal bridge transformation (\ref{GenConBr+}). 
The model $\hat{H}_0$ and its supersymmetric extensions were studied in details in \cite{PlyWipf}.
It is worth to mention that the Poincar\'e integral $\vJ$ Poisson commutes with all 
generators (\ref{Sl2Rmonopolar}) and (\ref{freemotion}), and plays the role of angular momentum of the system. 
 
\begin{figure}[hbt!]
\begin{center}
\begin{subfigure}[c]{0.27\linewidth}
\includegraphics[scale=0.3]{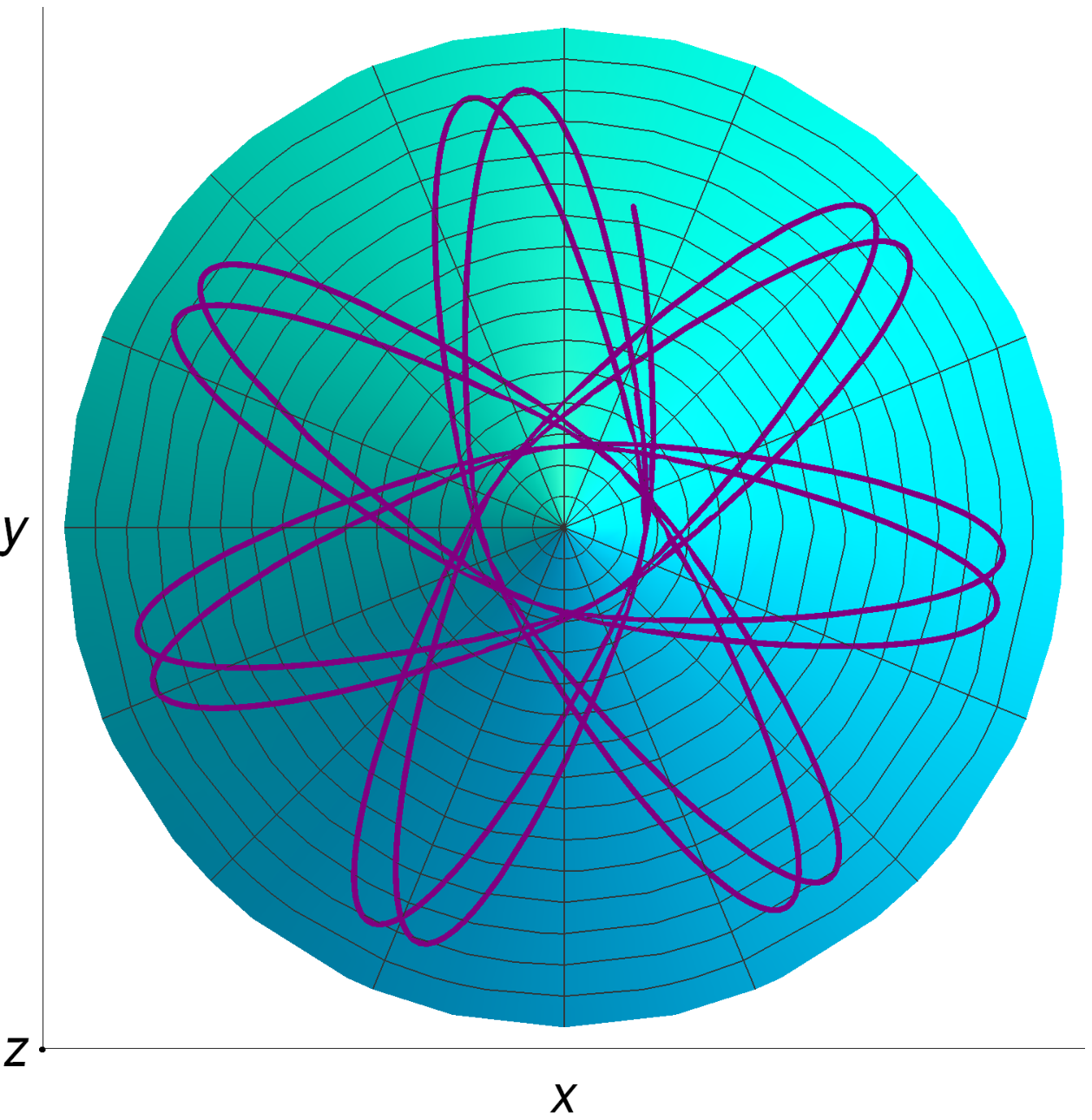}
\caption{\small{$\alpha=1/e$}}
\end{subfigure}
\begin{subfigure}[c]{0.27\linewidth}
\includegraphics[scale=0.3]{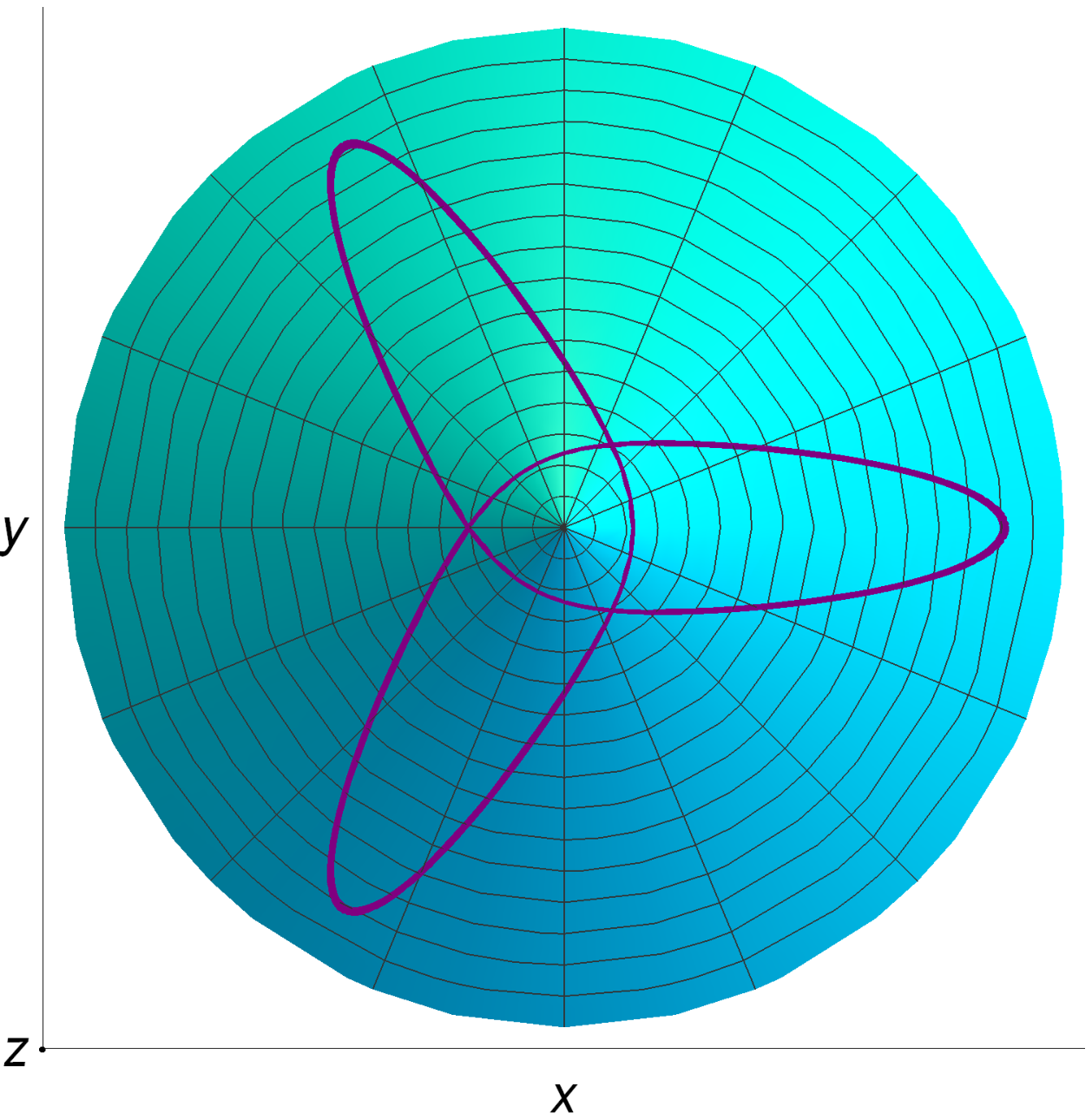}
\caption{\small{$l_a/l_r=3/2$}}
\end{subfigure}
\begin{subfigure}[c]{0.27\linewidth}
\includegraphics[scale=0.3]{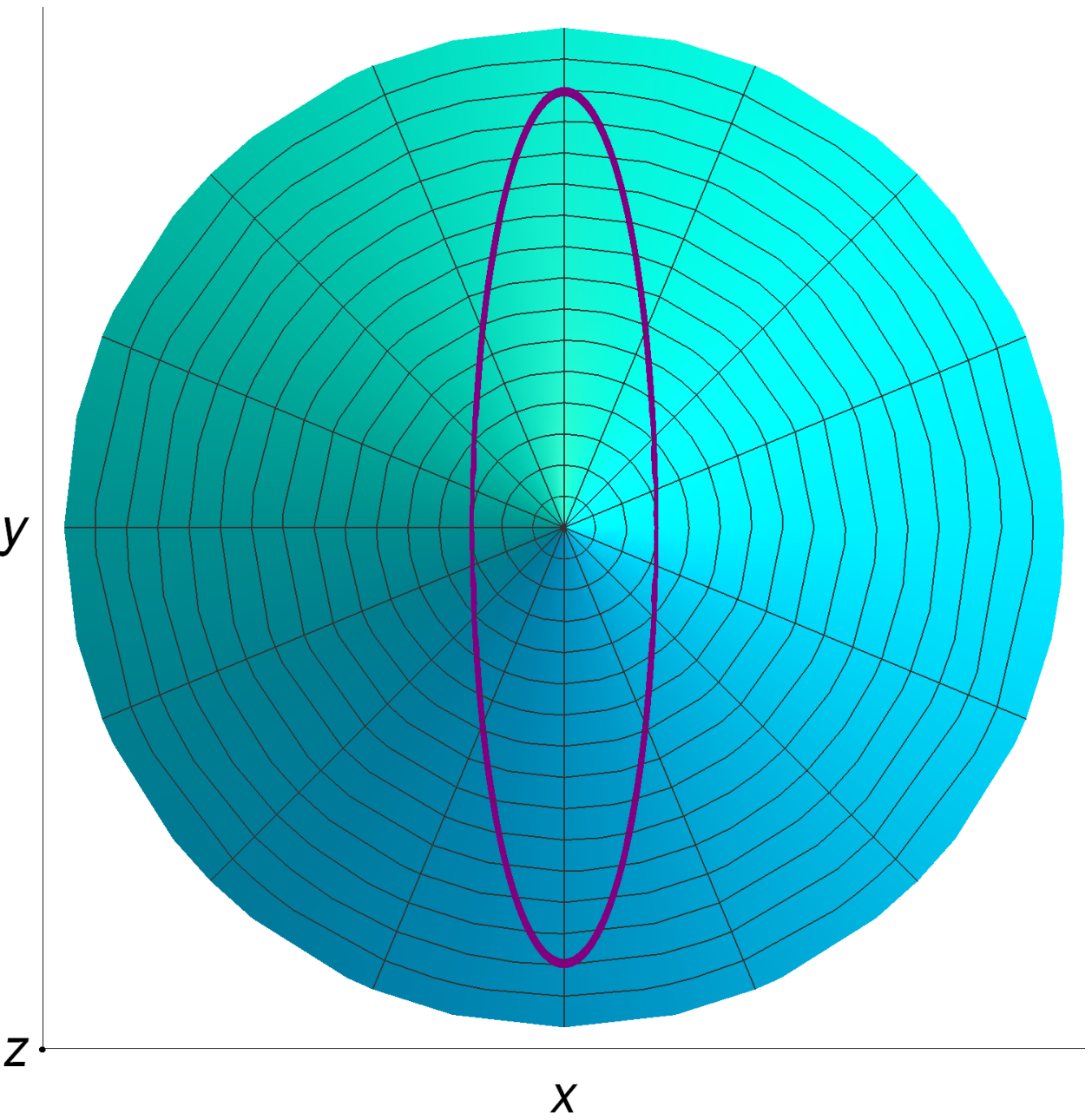}
\caption{\small{$l_a/l_r=2$}}
\label{Eliticinthecone}
\end{subfigure}
\caption[Classical trajectories 1, Sec. 9.1]{\small{
Examples of  some  non-closed and 
closed trajectories   are shown.  
The last
relation  $l_a/l_r=2$ corresponds to  the special case $\alpha=\nu^2$. }}
\label{figure1monopolo}
\end{center} 
\end{figure}

 After solving the trajectory equation for the system (\ref{DiracMonoH}),  one gets
$\vr=r(\varphi) \vn$, where 
\begin{eqnarray}
&
r^2(\varphi)=\frac{\mathscr{L}^2}{mH_\omega}[1-\rho\cos(\frac{2\mathscr{L}}{J}\varphi)]^{-1}\,,
\qquad\rho=\sqrt{1-\frac{\omega^2\mathscr{L}^2}{H_\omega^2}}\,,
&\\& \label{vncono}
\vn=\frac{-\nu\vJ}{J}+\vn_\bot(\varphi)\,,\qquad
\vn_\bot(\varphi)=\vn_\bot(0)\cos(\varphi)+J^{-1}\vJ\cross\vn_\bot\sin(\varphi)\,,
&
\end{eqnarray}
and $\vn_\bot\cdot\vJ=0$. After that, the angle  $\varphi=\varphi(t)$ is obtained by substitution of $r^2(\varphi)$ 
into the 
$\mathfrak{sl}(2,\R)$ generators (\ref{Sl2Rmonopolar}). 
From (\ref{vncono}) one concludes that the dynamics occurs on
 the surface of a dynamical cone defined by the
equation $\vr\cdot\vJ=-\nu r$.
Also from these solutions, with taking into account the definition  (\ref{defEll}) of $\mathscr{L}$,
we find that the trajectories are closed for arbitrary values of angular momentum $J$
 only  when  $ \alpha = \nu ^ 2 $. 
 On the other hand, for $\alpha\neq \nu^2$, the trajectories are closed 
 only for special values of $J$ given by the equation
\begin{eqnarray}
\label{alpha nu J}&
\alpha=\nu^2
+\left(\frac{1}{4}\frac{l_a^2}{l_r^2}-1\right)J^2\,,\qquad  l_a, l_r=1,2,\ldots\,.&
\end{eqnarray}
 The  special properties  at $\alpha=\nu^2$ are expected to be reflected in 
the presence of the hidden symmetry associated with additional non-trivial
 integrals of  motion  in  the system 
 \cite{InzPlyWipf2}. In Fig. \ref{figure1monopolo} 
 examples of the  trajectories are shown for some  irrational and rational values
 of the parameter $\alpha$.

To find the already anticipated hidden integrals of the system we employ the classical CBT. 
For this aim,  let use introduce the  analogs of the Laplace-Runge-Lentz vector and generator of the Galilean boost  
transformations 
for the asymptomatically  free system governed by the conformal generators (\ref{freemotion}), 
that are only available  when $\alpha=\nu^2$ \cite{PlyWipf},
\begin{eqnarray}
\label{LRL-G}
&
\vV=\vpi\cross\vJ\,,\qquad
\vG=(m\vr-\vpi t)\cross\vJ\,.
&
\end{eqnarray}
The  components of these vector quantities satisfy the relations  
\begin{eqnarray}
&
\{H_0,G_i\}=-V_i\,,\qquad
\{K_0,V_i\}=G_i\,,\qquad
\{H_0,V_i\}=\{K_0,G_i\}=0&\\&
\{D_0,V_i\}=\frac{1}{2}V_i\,,\qquad
\{D_0,G_i\}=-\frac{1}{2}G_i\,,\qquad
\{D_0,V_{i}G_{j}\}=0\,,&
\end{eqnarray}
and the classical conformal bridge transformation corresponds to the mapping 
\begin{eqnarray}
&
\mathfrak{S}: (\vV,\vG)\,\,\,\rightarrow \,\,\, (-i \sqrt{m\omega} \va,\sqrt{ \frac{m}{\omega}} \va^{*})\,,\quad 
\va=\sqrt{\frac{m\omega}{2}}(\vr +\frac{i}{m\omega}\vpi )\cross\vJ\, e^{i\omega t}\,,&\\&
 \mathfrak{S}: (-i\omega(G_iV_j+G_jV_i), \omega(G_iV_j-G_jV_i))\,\,\,\rightarrow\,\,\, ( T_{(ij)},, T_{[ij]})\,,
&\\&
\label{ladders1monopole}
\quad
T_{(ij)}=m\omega(a_i^*a_j+a_j^*a_i)\,,\quad
T_{[ij]}=-im\omega(a_i^*a_j-a_j^*a_i)\,.&
\end{eqnarray}
Here,  $T_{(ij)}$ is the symmetric tensor integral of the system (\ref{DiracMonoH}), being the 
 analog of the 
Fradkin tensor integral of the three-dimensional isotropic harmonic oscillator \cite{Frad}, 
while  $ T_{[ij]}$ is the anti-symmetric tensor proportional to the Poincar\'e integral. 
In terms of $\vr$ and $\vpi$, the explicit form of the components of these tensors 
are 
\be
\label{Tensor}
2T_{(ij)}=(\vpi \times \vJ)^i (\vpi \times \vJ)^j+
m^2\omega^2 (\vr \times \vJ)^i (\vr \times \vJ)^j\,,\qquad
2T_{[ij]}=\epsilon_{ijk}m\omega(J^2-\nu^2)J_{k}\,,
\ee
from where we explicitly see that these are  the higher order integrals of motion
of the  hidden symmetries.  It turns out that  the complete geometric information 
on the trajectory that appears in Fig \ref{Eliticinthecone}
is encoded in the symmetric  tensor, see \cite{InzPlyWipf2}.

As in the previous examples, here we can also obtain  all the information for  the quantum version 
of the model by applying the quantum CBT to the asymptotically free version without harmonic potential.
This time it 
is necessary to take into account the quantization prescription for the Dirac monopole, where 
the parameter $ \nu $ has to  be an integer or half integer number \cite{Mono1,Mono2,Mono2+}, 
and the eigenstates are given in terms 
of the monopole harmonics \cite{monoharm,monoharm+}. 
Due to the presence of the integral (\ref{Tensor}), the spectrum, 
which is discrete and bounded from below, 
 has a special degeneracy depending on the choice of  the quantized parameter $ \nu $. 
Furthermore, the corresponding spectrum generation operators can be constructed from
the quantum version of the complex vectors $\va$ and $\va^*$ introduced in (\ref{ladders1monopole}). 
For more details,  see  ref. \cite{InzPlyWipf2}.

\section{$\mathcal{PT}$-symmetric systems and extreme waves} 
\label{SecDar}
The conformal bridge transformation presented in all the previous sections 
shows how to derive the properties of harmonically confined systems from 
the associated  
model whose dynamics is asymptotically  free. 
In this section, with the help of the generalized 
Darboux transformations \cite{MatSal,CooperSUSY}
we construct  reflectionless  $\mathcal {PT} $-symmetric systems 
with rational potentials of the type  we already  considered in section 
\ref{SecNon-HermitianCal}.  Then we promote 
the obtained stationary  potentials  
to the  complex $\mathcal {PT} $-symmetric solutions 
to the  KdV equation  and higher equations of its hierarchy,
whose peculiar evolution reveals the  properties  typical for  extreme waves. 
We follow here refs.   \cite{JuanMP1,JuanMP2}, and 
work in the units $\hbar=1$, $m=1/2$. 

First, we remind that  the usual Darboux transformation 
and its generalizations allow ones to generate from a given 
one-dimensional  quantum system $\hat{H}_{0}=-\frac{d^2}{dx^2}+V(x)$ 
another system described by 
the Hamiltonian  
\begin{eqnarray}
\label{H0Dar}&
\hat{H}_{[n]}=-\frac{d^2}{dx^2}+V(x)-2\frac{d^2}{dx^2}\ln(W(\phi_{1},\ldots,\phi_{n}))\,.&
\end{eqnarray}
Here  functions $(\phi_{1},\ldots,\phi_{n})$ are the so-called  seed states, which  
are physical or formal, non-physical  eigenfunctions  of $\hat{H}_{0}$ with 
different  eigenvalues $\lambda_{j}$,
and  $W(\ldots)$
is the  Wronskian. 
If the seed states are chosen so that 
$W(\phi_{1},\ldots,\phi_{n})\neq 0$ in the domain 
where  the potential $V(x)$ of the 
initial system $\hat{H}_{0}$ is regular, then the potential of the generated system
$\hat{H}_{[n]}$
will be nonsingular in the same domain.
The Darboux transformation ensures that 
any, physical or non-physical, 
 eigenfunction   $\psi$  of $\hat{H}_0$  of eigenvalue $E$ not included in the set of the seed states 
can be mapped into the corresponding eigenfunction $\Psi_{[n]}$  of $\hat{H}_{[n]}$, 
\begin{eqnarray}
&\label{EigenHnDar}
\Psi_{[n]}=\frac{W(\phi_{1}\,\ldots,\phi_{n},\psi)}{W(\phi_{1},\ldots\phi_{n})}\,,\qquad
\hat{H}_{[n]}\Psi_{[n]}=E\Psi_{[n]}\,.
&
\end{eqnarray}
These relations can be verified by employing the intertwining  relations 
$
\hat{\A}_{n}^-\hat{H}_{0}=\hat{H}_{[n]}\hat{\A}_{n}^-$ and $
\hat{\A}_{n}^\dagger\hat{H}_{[n]}=\hat{H}_{0}\hat{\A}_{n}^\dagger\,,
$
where the operator 
\begin{eqnarray}
&\label{Am operaorDar}
\hat{\A}_{n}^-=\hat{A}_{n}^-\ldots \hat{A}_{1}^-\,,\qquad
\hat{A}_{k}^-=\hat{\A}_{k-1}\phi_{i}\frac{d}{dx}\left(\frac{1}{\hat{\A}_{k-1}^-\phi_{k}}\right)\,,\qquad
\hat{\A}_{0}^-=1\,,
&
\end{eqnarray}
 is constructed iteratively.
 With the help of this operator, 
eigenstate (\ref{EigenHnDar}) can  be presented in the form   
$\Psi_{[n]}=\hat{\A}_{n}^-\psi$. In particular case in which $n=1$, the function 
$W(x)=-(\ln(\phi_{1}(x))'$ is called  super-potential,   
and the systems $\hat{H}_{0}$ and   
$\hat{H}_{[n]}$ 
can also be presented  in the equivalent, up to an additive common  shift, 
 form 
\begin{eqnarray}
&\label{superpartners}
\hat{H}_\pm=-\frac{d^2}{dx^2}+V_\pm\,,\qquad
V_\pm=W^2 \pm W'
\,.&
\end{eqnarray}
The confluent Darboux transformation follows the same rules but now 
Jordan states of $\hat{H}_0$ can appear in the set of the seed 
states,   see \cite{JuanMP1,JuanMP2,CJP,CarPlyJ,InzPly3}. 
In Sec. \ref{SecNon-HermitianCal} we showed  how the 
$\mathcal{PT}$-regularized  Calogero system with  integer coupling constant $\nu=m$
can be related with   the free particle by means of  an
appropriate Darboux transformation. 

It worth to note here a similarity of the non-unitary CBT we described  in the previous sections 
with the  Darboux transformations. As in the CBT construction, the system 
$\hat{H}_{[n]}$  produced from the 
initial system $\hat{H}_0$  in general case is not  completely isospectral to it. However,  the generated system 
$\hat{H}_{[n]}$  inherits some important properties of $\hat{H}_0$. This happens, for instance,  
in the case when $\hat{H}_0$ corresponds to the free particle, from which reflectionless quantum systems
are produced. They  represent  snapshots of the soliton  solutions to  the KdV equation.
This can be compared with the CBT that transforms an asymptotically free system 
possessing  conformal symmetry into the harmonically trapped system with 
the same conformal symmetry but realized  in another form.  The essential difference between
the two transformations is that the Darboux transformation, being generated by an  operator
of finite differential order, is local. The generator of the CBT $\hat{\mathfrak{S}}$ is, however,  essentially 
non-local since it includes in its structure the exponent of the second order differential Hamiltonian
operator of the initial conformally invariant system. In the next, concluding  section, we will 
return to some aspect of similarity between the CBT, based on the conformal symmetry,  and 
supersymmetry generated by the Darboux transformations of the second order with $n=2$.

\vskip0.1cm
Let us start with the $\mathcal{PT}$-regularized Calogero type Hamiltonian
\begin{eqnarray}
&
\label{HDar}
2\hat{H}_{\alpha,1}=-\frac{d^2}{dx^2}+\frac{2}{\xi^2}\,,\qquad \xi=x+i\alpha \,.
&
\end{eqnarray}
Putting  $\nu=1$ in Eqs. 
(\ref{Am operaor})  and (\ref{AmIntert}),  
one deduces that the system (\ref{HDar}) can be obtained 
from the free particle via the first order Darboux transformation 
 by selecting  its  formal zero energy eigenfunction  
$\Omega_{0,0}^{\alpha}=\xi=x+i\alpha x$ as the seed state. 
Eq.  (\ref{LacNoviIN}) yields  us then 
the Lax-Novikov integral of the system (\ref{HDar}),  
\begin{eqnarray}\label{LaxNov1}
&
2\hat{\mathcal{P}}_{\alpha,1}=-\left(\frac{d}{dx}-\frac{1}{\xi}\right)\hat{p}\left(\frac{d}{dx}+
\frac{1}{\xi}\right)=-i\frac{1}{4}\hat{M}\,,\quad
\hat{M}=-4\frac{d^3}{dx^3}+6 u \frac{d}{dx}+3u\,,&
\end{eqnarray}
where 
$u(\xi)=\frac{2}{\xi^2}$.
The condition of commutativity of the third order operator (\ref{LaxNov1})
with Hamiltonian (\ref{HDar}) means that the potential $u(\xi)=\frac{2}{\xi^2}$ satisfies 
the stationary KdV equation $- 6u u_x+u_{xxx}=0$. 
Using the Galilean invariance of the KdV equation,
one finds then that the function
\begin{eqnarray}\label{Uxtau}
&U(x,\tau)=-\frac{1}{6}c+\frac{2}{(x+i\alpha-c\tau)^2}\,,\quad c\in \R\,,  &
\end{eqnarray}
will satisfy the 
dynamical KdV equation,
which can be presented equivalently in the Lax form 
\begin{eqnarray}
\label{Lax-Eq}
&\partial_{\tau}\hat{L}=[\hat{L},\hat{M}] \quad \Leftrightarrow\quad
U_\tau- 6U U_x+U_{xxx}=0\,,&
\end{eqnarray}
where  $L=-\frac{d^2}{dx^2}+U(x,\tau)$ and $\hat{M}$ is given by Eq.  
(\ref{LaxNov1}) with $u$ there changed for $U(x,\tau)$. 
Notice that if we extend the definition of the time reflection operator $\mathcal{T}$  by requiring 
additionally
$\mathcal{T} \tau = -\tau \mathcal{T}$, the time-dependent KdV 
equation will be invariant under the $\mathcal{PT}$ transformation
if $U(x, \tau)$ is $\mathcal{PT}$-symmetric: $[U(x, \tau), \mathcal{PT}] = 0$. 
The real and imaginary part of such  a field $U(x,\tau)=v(x,\tau)+iw(x,\tau)$, like this happens in the 
particular simplest case  (\ref{Uxtau}),
will satisfy the system of coupled non-linear dynamical equations  
\begin{eqnarray}
&
v_\tau-3(v^2-w^2)_x+v_{xxx}=0\,,\qquad
w_\tau-6(vw)_x+w_{xxx}=0\,.
&
\end{eqnarray}
   
\begin{figure}[hbt!]
\begin{subfigure}[c]{0.5\linewidth}
\includegraphics[scale=0.5]{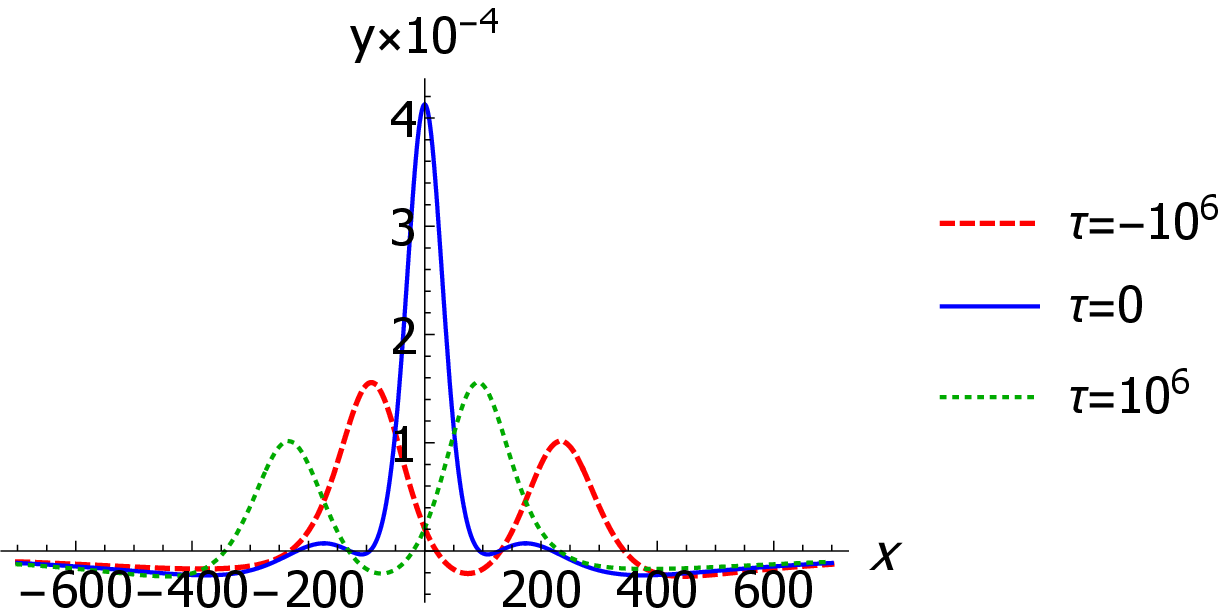}
\caption{\small{Snapshots of the real part,}}
\end{subfigure}
\begin{subfigure}[c]{0.5\linewidth}
\includegraphics[scale=0.55]{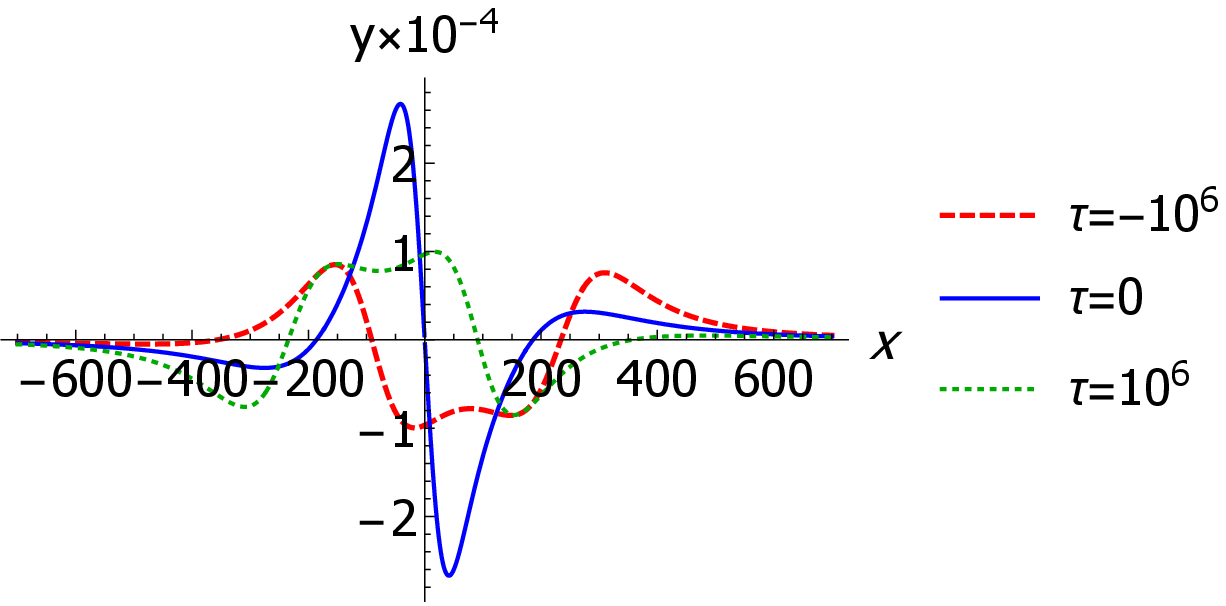}
\caption{\small{Snapshots of the imaginary part.}}
\end{subfigure}

\begin{subfigure}[c]{0.5\linewidth}
\includegraphics[scale=0.4]{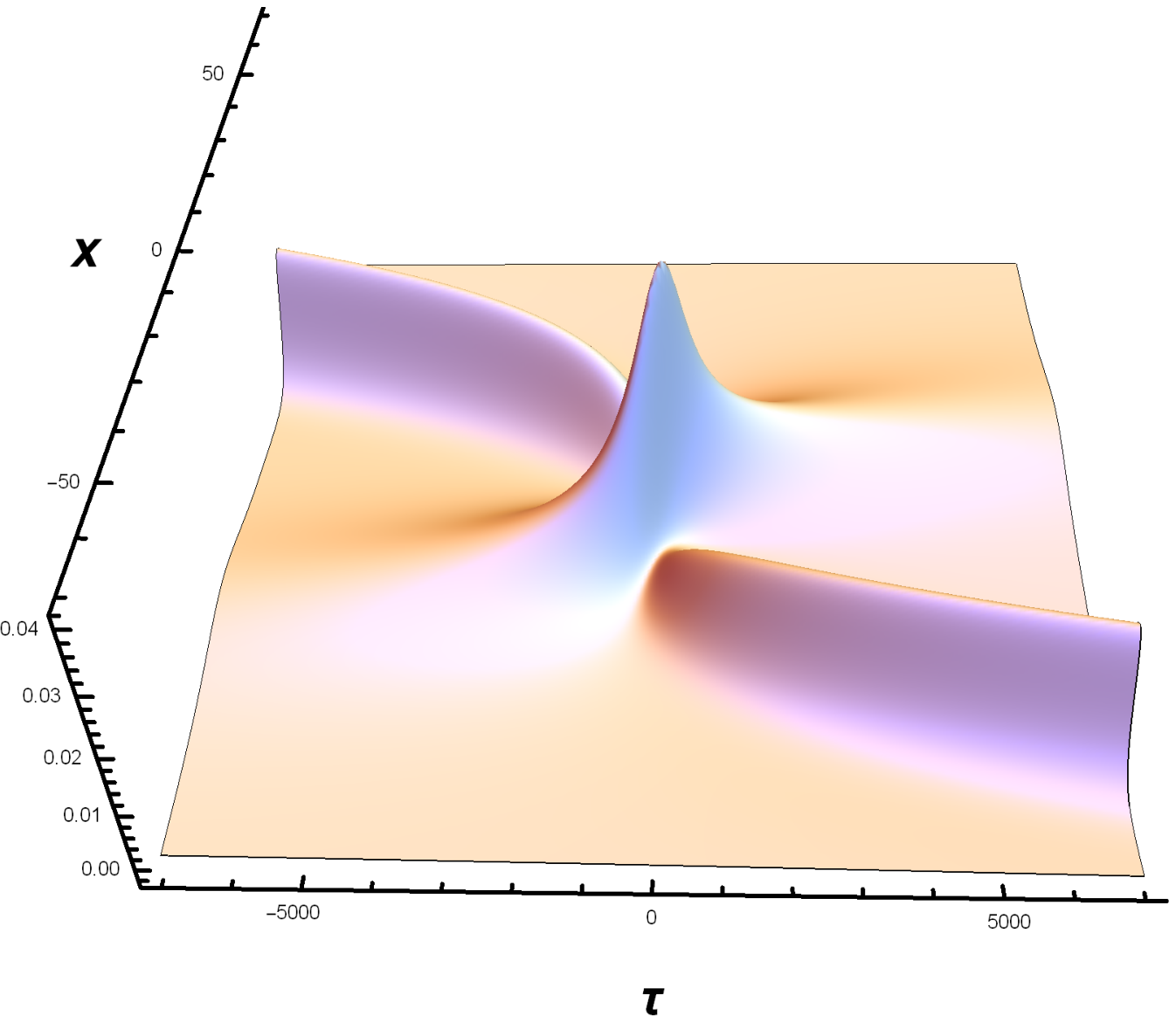}
\caption{\small{The $x-\tau$ plot of the real part. }}
\end{subfigure}
\begin{subfigure}[c]{0.5\linewidth}
\includegraphics[scale=0.4]{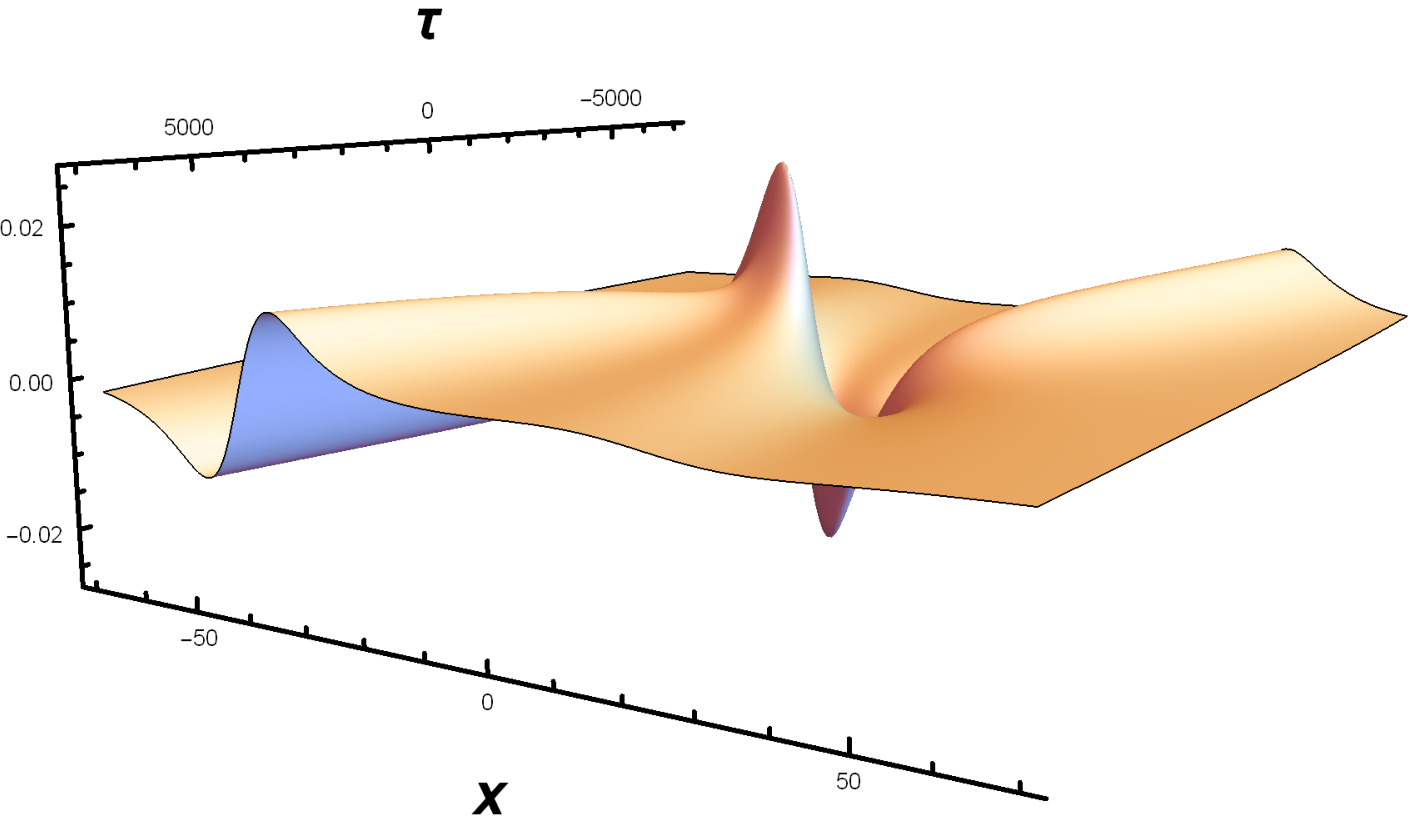}
\caption{\small{The  $x-\tau$ plot of the imaginary part.}}
\end{subfigure}
\caption{\small{The plots of the real and imaginary parts of   $-V_{+}^{(1)}(x,\tau;\alpha,\varrho)$ 
with $\alpha=10$ and $\varrho=5$.
}}
\label{figurePot1} 
\end{figure}

To construct a more interesting    $\mathcal{PT}$-symmetric solution to the KdV equation, 
 let us use  system (\ref{HDar}) as a starting point for a new  Darboux transformation.
To this aim   we select  as the seed state the function 
$
\psi_{\alpha,\gamma}^{1}=\gamma \Xi_{0,1}^{\alpha}+\Omega_{0,1}^{\alpha}=\gamma \xi^{-1}+\xi^{2}$.
This is  a zero energy eigenfunction of (\ref{HDar}). With pure imaginary parameter  $\gamma$, 
function $\psi_{\alpha,\gamma}^{1}$  is $\mathcal{PT}$-invariant. A further restriction 
$\gamma=i\varrho\alpha^3$ with $\varrho\in \R$, $\varrho\neq -8,1$, 
guarantees that $\mathcal{PT}$-odd  superpotential 
$
 W=-\frac{d}{dx}\ln(\psi_{\alpha,\gamma}^{1})=\frac{1}{\xi}-\frac{3\xi^2}{\xi^3+\gamma}
$
does nor take zero value. 
The generated potentials of   the corresponding supersymmetric partner  systems 
(\ref{superpartners}),
\begin{eqnarray}
&
\label{V+-1}
V_+=\frac{6}{\xi^2}-\frac{6\gamma(4\xi^3+\gamma)}{\xi^2(\xi^3+\gamma)^2}:=V_{+}^{(1)}(x;\alpha,\gamma)
\,,\qquad
V_-=\frac{2}{\xi^2}\,,
&
\end{eqnarray}
are $\mathcal{PT}$-symmetric  non-singular functions.
Potential  $V_+$ is a stationary solution to the higher order equation of the KdV hierarchy
\begin{eqnarray}
\label{HeKdV}
&U_\tau+30U^2U_x-20U_xU_{xx}-10U U_{xxx}+U_{xxxxx}=0\,.&
\end{eqnarray}
The substitution  $\gamma\rightarrow \gamma(\tau) = 12 \tau+i\varrho\alpha^2$ with 
$\varrho>1$ transforms function $V_+$ into the dynamical field 
 $V_{+}^{(1)}(x,\tau;\alpha,\varrho)=V_{+}^{(1)}(x;\alpha,\gamma(\tau))$ 
 which is a  $\mathcal{PT}$-symmetric solution of the
KdV equation  to be non-singular for $\tau\in(-\infty,\infty)$ \cite{JuanMP1}.
Some plots of the real and imaginary parts of the inverted field 
 $-V_{+}^{(1)}(x,\tau;\alpha,\varrho)$ are shown  on Fig. \ref{figurePot1}.

In the same way we provide a second example based on 
the $\mathcal{PT}$-symmetric system 
\begin{eqnarray}
&\label{HkdVcosas}
2\hat{H}_{\alpha,2}=-\frac{d^2}{dx^2}+\frac{6}{\xi^2}\,.
&
\end{eqnarray}
This system can be obtained via a second order confluent  Darboux transformation from the free particle system, 
where the corresponding seed states are $\Omega_{0,0}=\xi$ and  $\Omega_{1,0}=\xi^3$
 \cite{JuanMP1}. 
The  Lax-Novikov 
integral of this model   
is given by the five order differential operator 
\begin{eqnarray}&
4\hat{\mathcal{P}}_{\alpha,2}=\left(\frac{d}{dx}-\frac{2}{\xi}\right)
\left(\frac{d}{dx}-\frac{1}{\xi}\right)\hat{p}\left(\frac{d}{dx}+\frac{1}{\xi}\right)\left(\frac{d}{dx}+\frac{2}{\xi}\right)\,.&
\end{eqnarray}
As in the previous example, one can show that this operator is written in terms of derivatives and 
the  potential $u=\frac{6}{x^2}$ only,
which is a stationary solution to the higher order equation of the KdV hierarchy 
(\ref {HeKdV}).

\begin{figure}[hbt!]
\begin{subfigure}[c]{0.5\linewidth}
\includegraphics[scale=0.5]{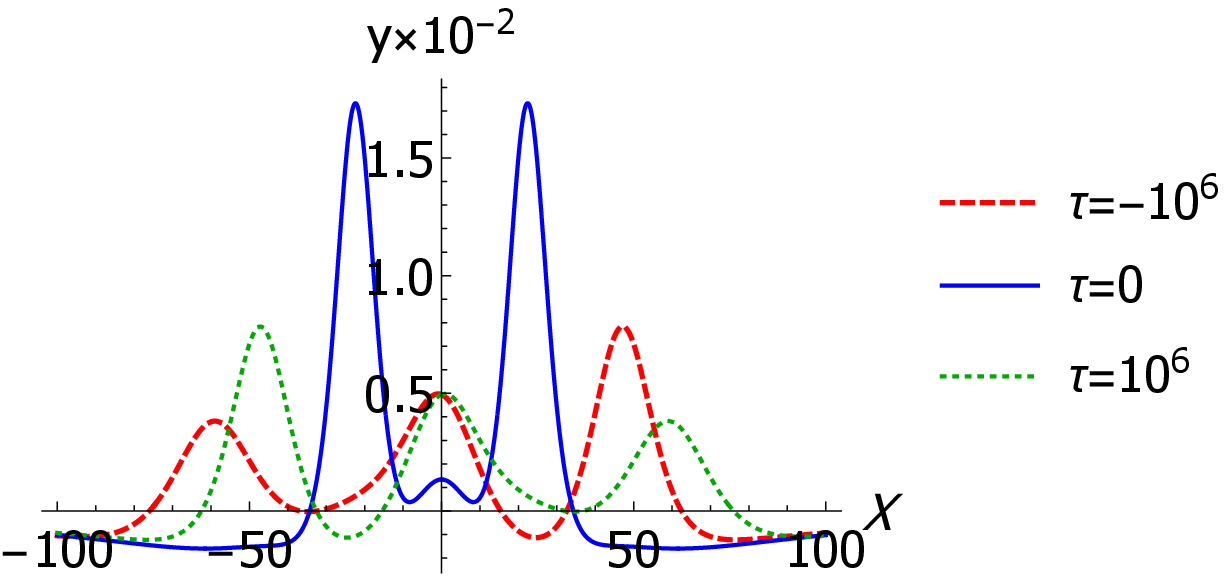}
\caption{\small{Snapshot of the real part.}}
\end{subfigure}
\begin{subfigure}[c]{0.5\linewidth}
\includegraphics[scale=0.5]{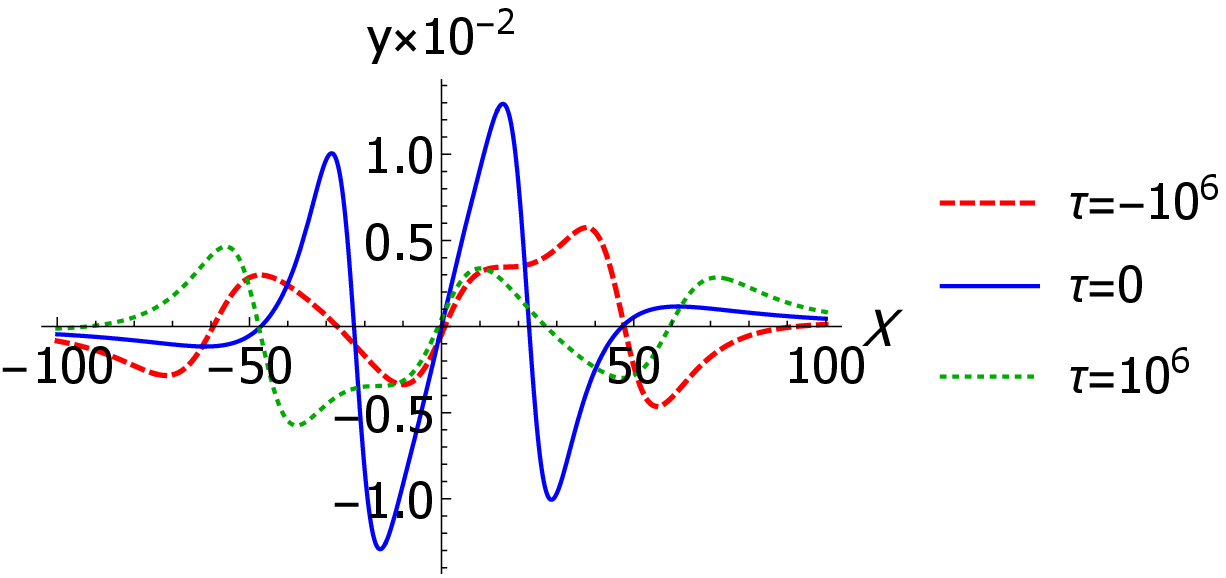}
\caption{\small{Snapshot of the imaginary part.}}
\end{subfigure}

\begin{subfigure}[c]{0.5\linewidth}
\includegraphics[scale=0.4]{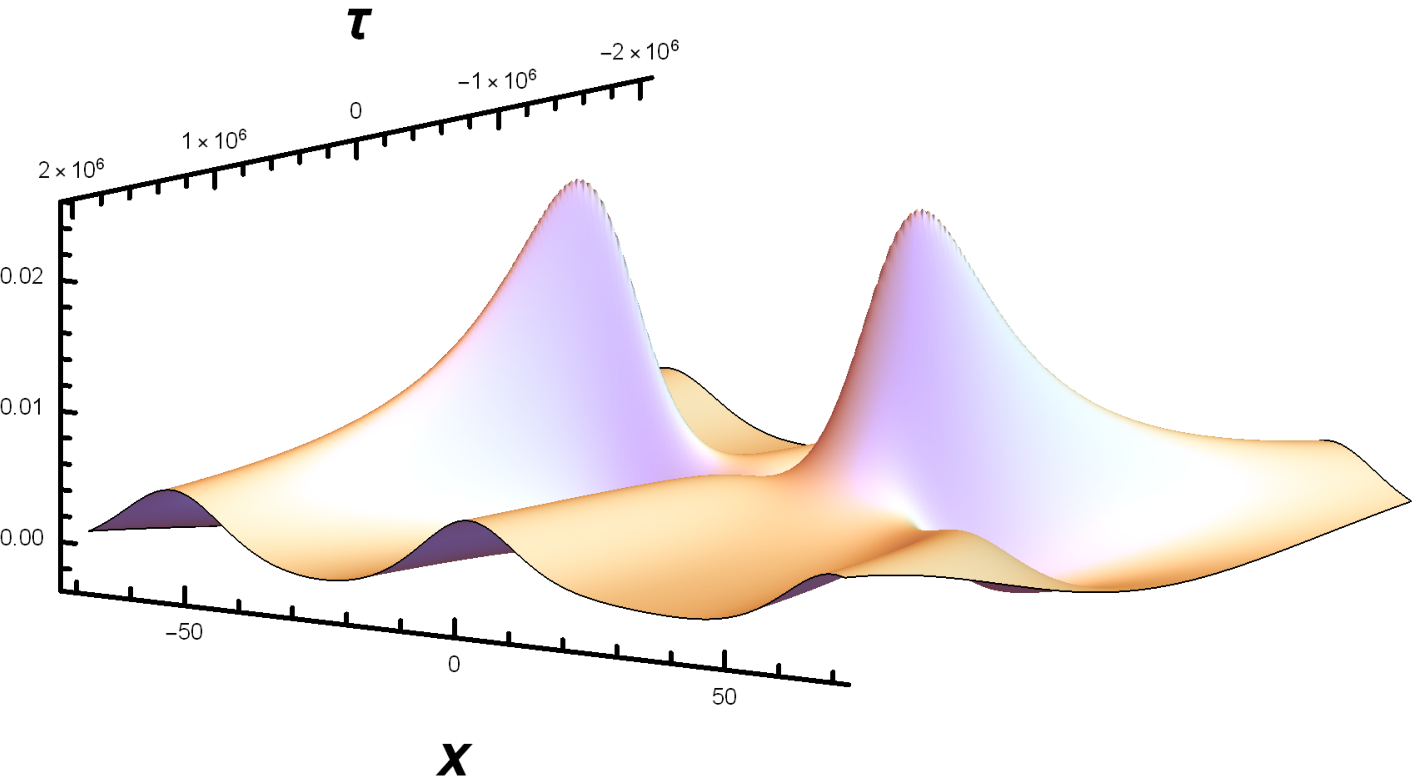}
\caption{\small{The $x-\tau$ plot of the real part.}}
\end{subfigure}
\begin{subfigure}[c]{0.5\linewidth}
\includegraphics[scale=0.4]{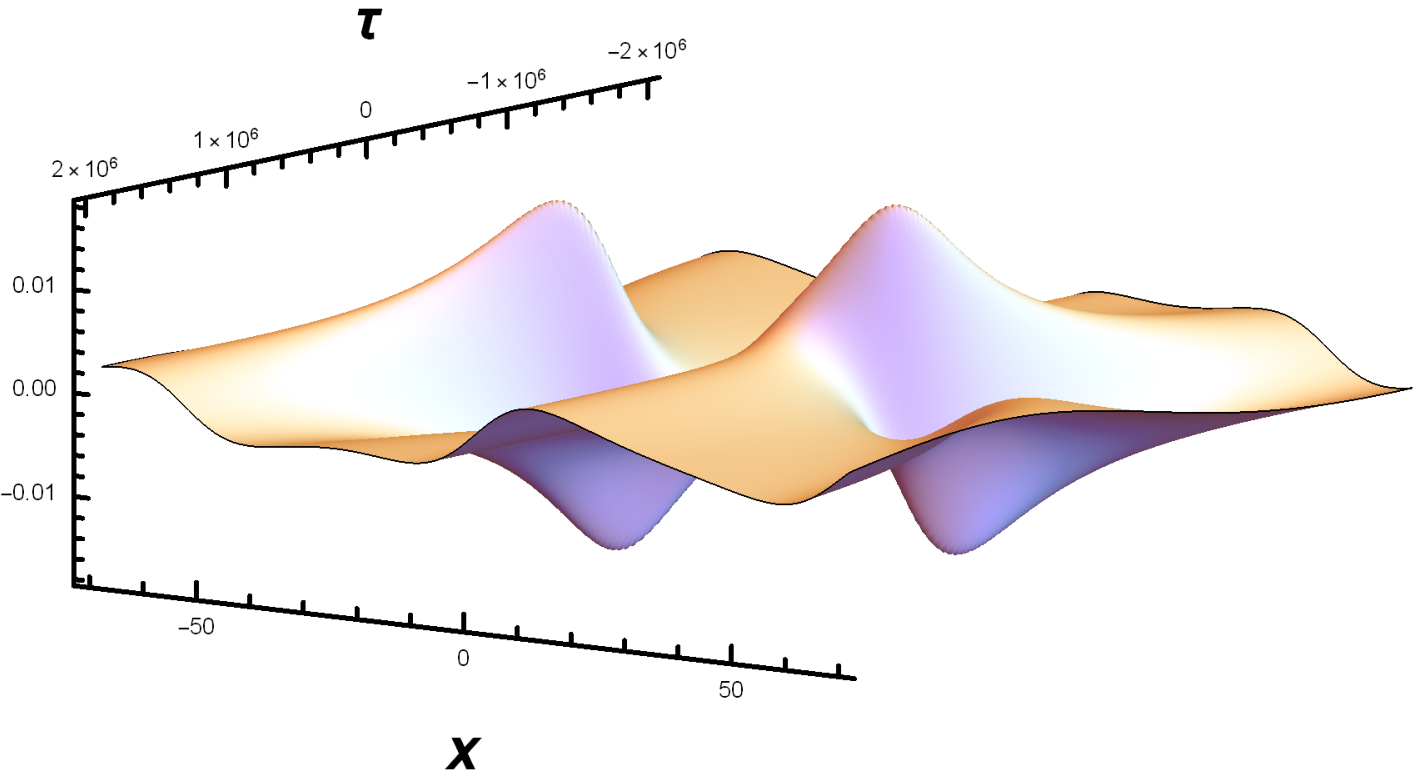}
\caption{\small{The  $x-\tau$ plot of the imaginary part.}}
\end{subfigure}
\caption[Classical trajectories 1, Sec. 9.1]{\small{
The plots of the real and imaginary parts of  $-V_{+}^{(2)}(x;\alpha,\gamma(\tau))$  with  
$\alpha=20$ and $\varrho=25$.
}}
\label{figurePot2} 
\end{figure}

As a seed state we choose  the following  nodeless zero energy eigenfunction of (\ref{HkdVcosas}),
$
\psi_{\alpha,\gamma}^{2}=\gamma \Xi_{0,2}^{\alpha}+\Omega_{0,2}^{\alpha}=\gamma \xi^{-2}+\xi^{3}$,
with $
\gamma=i\varrho \alpha^5$, ans now $\varrho$ is a real numerical parameter 
 different from  $-1$ and $-4$. 
With this choice  one gets the $\mathcal{PT}$-odd superpotential 
$W=-\frac{d}{dx}\ln(\psi_{\alpha,\gamma}^{2})=\frac{2}{\xi}-\frac{5\xi^4}{\xi^5+\gamma}
$, 
and the $\mathcal{PT}$-symmetric potentials of the super-partner systems are  given by 
\begin{eqnarray}
&
\label{V+-2}
V_+=\frac{12}{\xi^2}-\frac{10\gamma(6\xi^5+\gamma)}{\xi^2(\xi^5+\gamma)^2}:=V_{+}^{(2)}(x;\alpha,\gamma)\,,\qquad
V_-=\frac{6}{\xi^2}\,.
&
\end{eqnarray}
Similarly to  the previous example,  after the substitution 
 $\gamma\rightarrow\gamma(\tau)=-720\tau +i\varrho\alpha^5$ with $\varrho>24$, 
potential $V_{+}^{(2)}(x;\alpha,\gamma(\tau))$
satisfies the higher order non-linear field equation  (\ref{HeKdV}).  
The real and imaginary parts of the inverted function $-V_{+}^{(2)}(x;\alpha,\gamma(\tau))$
 are shown in Fig \ref{figurePot2}.

Finally, we  note  that near a critical value of the parameter $\varrho$, 
which in the first example is $\varrho=1$ and 
in the second case it is $\varrho=24$,  
the real part of the potentials,  defined respectively in (\ref{V+-1}) and (\ref{V+-2}), 
have a $\delta$-function like behaviour, 
while  the corresponding imaginary parts have a form 
similar to  $\delta'$-function \cite{JuanMP1,JuanMP2}. 
This behaviour is typical for  the  extreme (or, the so-called rogue)  waves, that corresponds to 
soliton type waves with extreme values of the amplitude emerging
in the process of their evolution.

\section{Discussion and outlook}
\label{SecDis}
In conclusion, we indicate some open  questions and problems  related to the considered
 topics  that deserve a further attention.

{\bf 1.}
 The two systems presented in Sec. \ref{SecNon-HermitianCal} 
have interesting spectral properties. The first one  is asymptotically  free,  and
corresponds to a $\mathcal{PT}$-regularized  Calogero type  model. It
is a perfectly  invisible zero-gap system with the unique $L^2(\R)$ integrable
eigenfunction of zero energy when the parameter  $\nu$  in the potential term $\nu(\nu+1)(x+i\alpha)^{-2}$ 
is an integer number, $\nu=m$,  and the system can be related to a free particle by 
the Darboux transformation.  These spectral 
properties are  coherently reflected by the presence (available only at $\nu=m$)
 of a well defined on all the  real line Lax-Novikov integral \cite{JuanMP1,JuanMP2}. 
On the other hand,  the 
harmonically confined $\mathcal{PT}$-regularized AFF model  
has two spectral towers that do not touch each other for any value of  the parameter $\nu$,
and  for which 
the complete set of the spectrum generating ladder operators 
can only be constructed when, again, the parameter $\nu$ is integer.
 As it was indicated at the end of that section, 
the well defined indefinite scalar  product
for those $\mathcal{PT}$-symmetric systems is unknown for us. 
We also do not  know the equivalent Hermitian systems
into which they can be transformed.
However, due to the similarity of the spectrum of the
$\mathcal{PT}$-regularized AFF model 
to   the spectrum of a  non-local model presented in ref. \cite{hidsusypara}, 
one can expect that those  two families  of the 
$\mathcal{PT}$-regularized models can be 
 related somehow to the non-local models with Hamiltonians
 that include in the structure the spatial reflection 
 operator $\mathcal{P}$.

\vskip0.15cm

{\bf 2.} 
The higher order confluent Darboux transformations, which appeared in
the $\mathcal{PT}$-regularized  Calogero type systems with $\nu=m$, 
are directly related  to the construction of higher order quantum supersymmetry 
\cite{AMGP,CJNP,InzPly3,CooperSUSY,Veronique} since the higher 
order intertwining operators can be promoted to higher order supercharges. 
In general this kind of systems,  including the non-Hermitian ones,  
are described by non-linear  superalgebras \cite{AMGP,CJNP,JuanMP1,JuanMP2,Veronique}. 
The generators of the higher order Darboux transformations can be factorized 
into generators of Darboux transformations of the corresponding lower orders,
and such factorization is non-unique \cite{Klishevich,PlySch}.
In dependence on the choice of the factorization,
the initial and final Darboux-related Hermitian systems 
can be related via ``virtual" systems which can be non-Hermitian.
This happens, in particular, in the case of  the second order supersymmetry 
with Darboux generators to be differential operators of the second order
 \cite{PlySch}. In some examples considered there, 
 the $\mathcal{PT}$-symmetric systems  like (\ref{HDar}) 
 do appear in the form of non-Hermitian virtual systems.. 
 An interesting question is whether other non-Hermitian systems 
 obtained in this way can  
fit into the scheme of the $\mathcal{PT}$-symmetric models
and CBT. 

 \vskip0.15cm
       
{\bf 3.}
Consider the following similarity transformation of the  operator $i\hat{D}$,
\begin{eqnarray}
\label{TransDdis1}
&\hat{\mathfrak{S}}_{a,b}(i\hat{D})\hat{\mathfrak{S}}_{a,b}^{-1}=
i(1+2ab)\hat{D}+b\hat{H}-a(1+ab)\hat{K}:=i\hat{D}_{a,b}\,,\quad\,\,
\hat{\mathfrak{S}}_{a,b}=e^{\frac{a}{\hbar}\hat{K}}e^{\frac{b}{\hbar} \hat{H}}\,,\quad\,\,
&
\end{eqnarray}
where $a=\alpha\omega$,  $b=\beta\omega^{-1}$, 
 and $\alpha$ and $\beta$ can be complex in general case dimensionless parameters.
 Our CBT generator (\ref{QCB}) corresponds here to the particular choice $\alpha=-1$, $\beta=1/2$.
The operator $\hat{\mathfrak{S}}_{a,b}$ is a  generator of the  
internal automorphism of $\mathfrak{so}(2,1)\cong\mathfrak{sl}(2,\R)$ of the most general form
since the inclusion of  the operator $e^{\frac{c}{\hbar}\hat{D}}$ in its structure 
reduces just to the change of the parameters $\alpha$ and $\beta$.
The operator $i\hat{D}_{a,b}$ has the structure of the operator  (\ref{Sw-H-g})
with corresponding identification of the parameters. 
From   (\ref{TransDdis1}) we obtain the relation
\be
\hat{\mathfrak{S}} \hat{\mathfrak{S}}_{a,b}^{-1}
(i\hat{D}_{a,b}) \hat{\mathfrak{S}}_{a,b}\hat{\mathfrak{S}}^{-1}=
\hbar\hat{\mathcal{J}}_0 \,,
\ee
where $\hat{\mathfrak{S}} $ is  the $\mathcal{PT}$-symmetric  generator (\ref{QCB}) of our CBT. 
In our CBT scheme, the eigenvectors  $\ket{\lambda}$ that are transformed into  the
physical  eigenstates of $\hbar\hat{\mathcal{J}}_0$
correspond to the zero energy Jordan states of $\hat{H}$,  which, in turn, 
are also eigenstates of $2i\hat{D}$
with real eigenvalues. 
Then, in this extended scheme, the eigenstates of the operator  $i\hat{D}_{a,b}$, 
that have to be transformed into 
the physical eigenstates of $\hbar\hat{\mathcal{J}}_0$, 
are $e^{\frac{a}{\hbar}\hat{K}}e^{\frac{b}{\hbar} \hat{H}}\ket{\lambda}$.
However,  in the general case we cannot say too much about 
the behaviour of the resulting functions in the coordinate 
representation for particular choice of the initial asymptotically free system
described by the Hamiltonian $\hat{H}$.
 One can expect that the detailed analysis of this aspect  
 should restrict the choice of the parameters $\alpha$
and $\beta$. We  just note that in our CBT 
with $\mathcal{PT}$-symmetric generator $\hat{\mathfrak{S}}$, 
the Jordan states of $\hat{H}$,  which  are transformed into eigenstates of $\hat{\mathcal{J}}_0$, 
satisfy the equation $\hat{H}^n\ket{\lambda}=0$ with some integer $n$.
This  is a scale-invariant equation for asymptotically free examples of 
the systems considered by us here.  It is related with a unique peculiarity 
of our CBT:  its generator $\hat{\mathfrak{S}}$, having the property $\hat{\mathfrak{S}}^4=\mathcal{P}$,
and so,  being internal automorphism
of the $\mathfrak{so}(2,1)\cong \mathfrak{sl}(2,\R)$ algebra,
maps the first order scale-invariant differential operator   $2i\hat{D}$ 
into the second order differential operator $\hbar\hat{\mathcal{J}}_0$.

\vskip0.15cm

{\bf 4.}
For our $\mathcal{PT}$-symmetric CBT it  does not matter if the number of degrees 
of freedom is greater  than two, 
or if we are working in some exotic geometry. 
The only  decisive  factor is to have the  generators 
of the  $\mathfrak{so}(2,1)$ conformal symmetry of the initial asymptotically free system 
 to be able to construct the CBT operator.
In this way, the operator $i\hat{D}$, to which we apply the CBT
to get the Hamiltonian operator of the associated harmonically trapped system,  
can be extended  by  a $\mathcal{PT}$-symmetric
`Zeeman type' term   $\hat{\mathcal{Z}}$ that 
 commutes with the $\mathfrak{so}(2,1)$ generators.
In this case  the operator $(2i\hat{D}+g\hat{\mathcal{Z}})$ will be mapped into 
$2\hbar\hat{\mathcal{J}}_0+g\hat{\mathcal{Z}}$.  For example 
if we consider a free particle  in the cosmic string background,  and take 
the combination 
$2\omega(i\hat{D}+g\hat{J}_0)$, with  $\hat{D}$ and $\hat{J}_0$
to be 
the corresponding quantum analogues of the quantities (\ref{HDkuku}), (\ref{ConeDPhi}), 
the application of the corresponding CBT yields us 
\begin{eqnarray}
&
\hat{H}_{g}^{(\alpha)}=
-\frac{\hbar^2}{2m}
\left(\frac{1}{\alpha^2r}\frac{\partial}{\partial r}\left(r\frac{\partial}{\partial r}\right) +
\frac{1}{r^2}\frac{\partial^2}{\partial\varphi}\right)+\frac{\alpha^2 m \omega^2}{2}r^2
-i\hbar\omega g\alpha \frac{\partial}{\partial \varphi}\,.
&
\end{eqnarray}
This is a  direct analog of the ERIHO quantum Hamiltonian (see Sec. \ref{SecCBT2d})
 in the cosmic string metric (conical background from the viewpoint 
 of condensed matter physics \cite{KatVol,Volovik}).
Results related to  this particular system will be presented by us soon.

In the  case of two dimensions we also can add fermionic 
degrees of freedom by  taking  a term of the form $ \hat{\mathcal{Z}}=\omega g \sigma_3 \hat{p}_\varphi $. 
In particular, the application of the isotropic CBT  to the generator
 $\omega (2 i\hat {D} +  \sigma_3 \hat{p}_\varphi)$ 
 produces Hamiltonian of the supersymmetric Landau  problem
 \cite{CooperSUSY}. 
 This is an important indication how the CBT can be generalized for 
 supersymmetric case. In the same vein, 
 the scheme of the Swanson model can be extended. 
 The indicated generalizations  can be interesting, particularly,   from the point 
 of view of physics of Bose-Einstein condensates \cite{Cooper,ALF,Goldman,BEC-ani} 
 and  physics of anyons \cite{NonCom2,NonComPLB,AGKP}. 

\vskip0.2cm

{\bf {\large 
Acknowledgements}}
\vskip0.1cm

The work was partially supported by the FONDECYT Project 1190842 
and the DICYT Project 042131P\_POSTDOC.

\end{document}